\documentclass[a4paper,11pt]{article}
\pdfoutput=1

\usepackage[utf8]{inputenc}
\usepackage{a4wide}
\usepackage{graphicx}
\graphicspath{{Figures/}} % Attention to the path!!!
\usepackage[small,bf]{caption}
\usepackage{amssymb}
\usepackage{amsmath}
\usepackage{slashed}
\usepackage{mathtools}
\usepackage{cite} 
\usepackage{hyperref}
\usepackage{xcolor}
\DeclareMathAlphabet{\mathpzc}{OT1}{pzc}{m}{it}

\newcommand{\be}{\begin{equation}}
\newcommand{\ee}{\end{equation}}
\newcommand{\ben}{\begin{eqnarray}}
\newcommand{\een}{\end{eqnarray}}
\newcommand{\bi}{\begin{itemize}}
\newcommand{\ei}{\end{itemize}}

\numberwithin{equation}{section}
\makeatletter
\g@addto@macro\bfseries{\boldmath}
\makeatother
\usepackage{tikz}
\usetikzlibrary{shapes.geometric, arrows}

\tikzstyle{startstop} = [rectangle, rounded corners, minimum width=3cm,
minimum height=1cm,text centered, draw=black, fill=red!30]

\tikzstyle{io} = [trapezium, trapezium left angle=70, trapezium right
angle=110, minimum width=3cm, minimum height=1cm, text centered,
draw=black, fill=blue!30]

\tikzstyle{process} = [rectangle, minimum width=3cm, minimum height=1cm,
text centered, draw=black, fill=orange!30]

\tikzstyle{decision} = [diamond, minimum width=3cm, minimum height=1cm,
text centered, draw=black, fill=green!30]

\tikzstyle{myprocess} = [rectangle, minimum width=3cm, minimum height=1cm, text centered, draw=black]

\tikzstyle{mydecision} = [diamond, aspect=2, minimum width=3cm, minimum height=1cm, text centered, draw=black]

\tikzstyle{arrow} = [thick,->,>=stealth]

\newcommand{\Slash}[1]{\ooalign{\hfil$ \diagup $\hfil\crcr$#1$}}
\begin{document}
%==================================================

%==================================================
\begin{titlepage}
%==================================================
\begin{flushright}
IFT–UAM/CSIC–22-135

FTUV-22-1026.0666

IFIC/22-31
\end{flushright}
\vspace*{0.8cm}

\begin{center}
%{\Large\bf $\nu$ EWBG: The Scalar Potential Strikes Back}\\[0.8cm]
{\Large\bf $\nu$ Electroweak Baryogenesis: The Scalar Singlet Strikes Back}\\[0.8cm]
E.~Fern\'andez-Mart\'inez,$^a$
J.~L\'opez-Pav\'on,$^b$
J.~M.~No,$^a$
T.~Ota$^a$
and S.~Rosauro-Alcaraz$^c$\\[0.4cm]
$^{a}$\,{\it Departamento de F\'isica Te\'orica and Instituto de F\'{\i}sica Te\'orica, IFT-UAM/CSIC,\\
Universidad Aut\'onoma de Madrid, Cantoblanco, 28049, Madrid, Spain} \\
$^{b}$\,{\it Instituto de F\'isica Corpuscular, Universidad de Valencia and CSIC, \\
Edificio Insitutos Investigaci\'on, Catedr\'atico Jos\'e Beltr\'an 2, 46980, Paterna, Spain}\\
$^{c}$\,{\it P\^ole Th\'eorie, Laboratoire de Physique des 2 Infinis Irène Joliot Curie (UMR 9012) CNRS/IN2P3, 15 rue Georges Clemenceau, 91400 Orsay, France}
\end{center}
\vspace{0.8cm}

\abstract{  
We perform a comprehensive scan of the parameter space of a general singlet scalar extension of the Standard Model to identify the regions which can lead to a strong first-order phase transition, as required by the electroweak baryogenesis mechanism. We find that taking into account bubble nucleation is a fundamental constraint on the parameter space and present a conservative and fast estimate for it so as to enable efficient parameter space scanning. The allowed regions turn out to be already significantly probed by constraints on the scalar mixing from Higgs signal strength measurements. We also consider the addition of new neutrino singlet fields with Yukawa couplings to both scalars and forming heavy (pseudo)-Dirac pairs, as in the linear or inverse Seesaw mechanisms for neutrino mass generation. We find that their inclusion does not alter the allowed parameter space from early universe phenomenology in a significant way. Conversely, there are allowed regions of the parameter space where the presence of the neutrino singlets would remarkably modify the collider phenomenology, yielding interesting new signatures in Higgs and singlet scalar decays.
}

\end{titlepage}
\section{Introduction}

The origin of the observed Baryon Asymmetry of the Universe (BAU) is one of the fundamental open problems of the Standard Model (SM) of particle physics and one of the few precious pieces of experimental evidence for physics beyond the SM together with the existence of neutrino masses and dark matter.  
The generation of the BAU in the early Universe requires satisfying the three Sakharov conditions~\cite{Sakharov:1967dj}: baryon number violation, $C$ and $CP$ violation, and departure from thermal equilibrium. In principle the SM itself could address the origin of the BAU via the electroweak baryogenesis (EWBG) mechanism~\cite{Shaposhnikov:1986jp,Shaposhnikov:1987tw, Cohen:1987vi, Cohen:1990it, Cohen:1990py, Nelson:1991ab}. 
However, the SM three-family quark mixing encoded in the Cabibbo-Kobayashi-Maskawa (CKM) matrix does not provide enough $CP$ violation to generate a sufficient asymmetry~\cite{Gavela:1993ts, Gavela:1994ds, Gavela:1994dt}, and the early Universe transition from the electroweak (EW) symmetric phase to the EW broken phase in the SM is a smooth crossover~\cite{Kajantie:1996mn,Degrassi:2012ry}, rather than the strongly first-order transition required by the out-of-equilibrium Sakharov condition.

Nevertheless, simple extensions of the SM could solve these issues and make EWBG viable. In particular, extending the scalar sector of the SM by just a real singlet field could allow for a first-order EW phase transition (see e.g.~\cite{Espinosa:1993bs,Profumo:2007wc,Espinosa:2011ax,Profumo:2014opa,Chen:2017qcz}). 
This new scalar singlet may not be alone, but rather be part of an extended dark sector to which it couples. A particularly motivated such scenario is the extension of the SM by (heavy) fermion singlets, i.e. right-handed neutrinos, able to account for the observed pattern of neutrino masses and mixings in Nature. 
%A Yukawa coupling between the fermion and scalar singlets is then allowed at the renormalizable level and should be considered.
Remarkably, it was shown in Refs.~\cite{Hernandez:1996bu,Fernandez-Martinez:2020szk} that the new sources of CP violation that arise in this extension of the SM, from the simultaneous presence of 
Yukawa interactions of the heavy neutrinos with the singlet scalar and with the Higgs doublet and SM neutrinos,
could lead to successful EWBG depending on the evolution of the scalar sector 
%for some configurations of the scalar sector 
during the phase transition (a scenario referred to as $\nu$-EWBG in~\cite{Fernandez-Martinez:2020szk}).

In this work we aim to clarify the conditions on the singlet scalar dynamics during the EW phase transition that possibly allows for EWBG in the above setup. 
Our scope is however more general, and we study the regions of parameter space of the real singlet scalar extension of the SM yielding a strong first-order phase transition (SFOPT), exploring the correlations among different model parameters, and emphasizing those that might arise between measurable quantities. The aim here is not a high-precision computation of the various thermodynamic quantities of the phase transition, which would be numerically challenging if combined with a thorough scan of the model parameter space. Rather, we focus on exploring the parameter space as efficiently as possible, covering broad areas of the multidimensional space via a number of approximations. Even if these approximations are not suitable to obtain highly-accurate results for the SFOPT quantities, they allow to pinpoint the regions of the parameter space with the desired features and test whether they are presently allowed, for subsequent analyses to concentrate in these regions.
%\EFM{While a full and very accurate scan would be technically and numerically very demanding, we instead focus on exploring the parameter space as efficiently as possible. The aim is to cover very broad areas of the multidimensional parameter space with a number of approximations that, even if not suitable to obtain precision results, allow to pinpoint the interesting regions of the parameter space and test whether they are presently allowed for subsequent precision analyses to concentrate in.} 
We pay particular attention to the SFOPT requirement of bubble nucleation for a successful completion of the EW phase transition, for which we provide conservative and fast estimate for assessing if nucleation would take place. While the nucleation dynamics has been studied previously~\cite{Carena:2019une} in the context of a $\mathbb{Z}_2$-symmetric singlet scalar extension of the SM, here we aim at a more general study without the additional constraint of additional symmetries. To the best of our knowledge, this is the first full scan exploring all the different correlations of the parameter space of the scalar singlet extension of the SM aiming to identify the regions where a SFOPT could take place. In addition, we investigate the impact of the heavy neutrinos on the SFOPT dynamics: 
while previous studies indicate that sizable values of the neutrino Yukawa interactions with the scalar singlet can strengthen the first-order phase transition~\cite{Cline:2009sn}, we find that large values of the neutrino Yukawas, unless compensated by other parameters, can also have the effect of destabilizing the EW broken minimum and are thus generally disfavoured. However, their inclusion does not alter significantly the allowed regions of the parameter space as compared to the singlet-only case.

Finally, we also discuss the phenomenological impact of the existence of such heavy neutrinos as compared to the minimal singlet scalar extension of the SM, finding that the phenomenology can be altered dramatically with respect to the latter model. 
Specifically, we find that the singlet-like scalar will dominantly decay into right-handed neutrinos (if allowed by phase space), instead of directly decaying into SM particles. 
These heavy neutrinos may then subsequently decay into SM particles either promptly or via displaced vertexes, depending on the size of their mixing $\theta$ with the active SM neutrinos. Since the production of the right-handed neutrinos from the scalar singlet decay is unrelated to the strongly constrained mixing $\theta$, this heavy neutrino production process could well be the dominant one at colliders like the LHC. Besides, the mixing between the scalar singlet and the Higgs would also lead to exotic Higgs decays into right-handed neutrinos (if these are light enough), which can be probed via Higgs signal strength measurements and also in direct searches for such exotic Higgs decays (see e.g.~\cite{Hernandez:2016kel}) at the LHC.

%Finally, we explore here the phenomenological impact of the existence of such heavy neutrinos as compared to the minimal singlet scalar extension of the SM, finding that the phenomenology can be altered dramatically with respect to the latter model. Specifically, we find that the singlet-like scalar may dominantly decay into right-handed neutrinos, instead of directly decaying into SM particles which could then subsequently decay into SM particles either promptly or leading to displaced vertexes, depending on the size of their mixing with the SM neutrinos. Since the production of the right-handed neutrinos from the scalar singlet decay is unrelated to their strongly constrained mixing with their SM counterparts, this production process could be the dominant one. Interestingly, the mixing between the Higgs and the scalar singlet will also lead to exotic Higgs decays to right-handed neutrinos, if light enough. We briefly discuss the associated phenomenology in this scenario. 

This paper is organized as follows. In Section~\ref{sec:singlet_model} we introduce the real singlet scalar extension of the SM with the addition of heavy neutrinos, and discuss the details of the scalar potential in the early Universe relevant for our SFOPT analysis. Then, in Section~\ref{sec:pheno_singlet} we analyze the experimental constraints on the model, as well as the possible new combined probes of the existence of the singlet and the heavy neutrinos. In Section~\ref{sec:param_scan} we give details of our model parameter scan, and we discuss its results in Section~\ref{sec:results}. We finally conclude in Section~\ref{sec:summary}.

\section{The scalar singlet extension of the SM with heavy neutrinos}
\label{sec:singlet_model}

The simplest extension of the SM scalar sector is the inclusion of a real scalar singlet $s$ that may mix with the Higgs boson. This small addition to the SM may however significantly alter the scalar sector phenomenology. In particular, it can allow for a SFOPT even at tree-level~\cite{Espinosa:1993bs,Espinosa:2011ax}, re-opening the possibility of explaining the origin of the observed matter-antimatter asymmetry of the Universe in the context of EWBG %~\cite{Shaposhnikov:1986jp} 
if new sources of CP violation beyond the SM are also present. 

More interestingly, the scalar singlet field $s$ could be a window to a dark sector capable of addressing some of the other open problems of the SM. Indeed, given its singlet nature, renormalizable (and therefore less suppressed) couplings are expected between the scalar and both the SM and the extended dark sector. Such scenarios could for example account for the observed dark matter of the Universe~\cite{Arcadi:2019lka} (see also~\cite{Falkowski:2009yz,Lindner:2010rr,Bertoni:2014mva,GonzalezMacias:2015rxl,Batell:2017cmf,Blennow:2019fhy,Baker:2019ndr,Baker:2021zsf}) or the simultaneous origin of neutrino masses and the BAU~\cite{Hernandez:1996bu,Fernandez-Martinez:2020szk}.   
Motivated by the latter, but easily generalizable, in this work we will consider a dark sector that comprises the real scalar field $s$ and $n$ new Dirac neutrinos, $N'=(N'_{L}, N'_{R})$, singlets under the SM gauge group and with lepton number $+1$.
In order to have an accessible extended neutrino sector with masses around the EW scale that may play a non-trivial role in the SFOPT and the baryogenesys process, we consider low-scale realizations of the Seesaw mechanism~\cite{Minkowski:1977sc,Mohapatra:1979ia,Yanagida:1979as,Gell-Mann:1979vob} with an approximate lepton number conservation so as to protect and ensure the lightness of neutrino masses~\cite{Branco:1988ex,Kersten:2007vk, Abada:2007ux}. Indeed in the so-called inverse~\cite{Mohapatra:1986aw,Mohapatra:1986bd} or linear~\cite{Akhmedov:1995ip,Malinsky:2005bi} Seesaw mechanisms the new heavy neutrinos arrange in Dirac pairs while the SM neutrinos remain massless if the lepton number symmetry is exact. Upon softly breaking this symmetry, the SM neutrinos will acquire small masses and the mass degeneracy of the two members of each Dirac pair will be slightly broken~\cite{Fernandez-Martinez:2022gsu}. 
Neglecting the small lepton-number-violating terms (which are suppressed by the tiny SM neutrino masses), the most general lepton-number-conserving interaction Lagrangian among the dark sector fields $s$, $N'$ and the SM fields is:
\begin{equation}
    \mathcal{L}\supset \left(-\overline{L_L}\tilde{\Phi}Y_{\nu}N'_R-\overline{N'_L}\,s Y_N N'_R+h.c.\right)+V\left(\Phi^{\dagger}\Phi,s\right),
    \label{eq:Lag_1}
\end{equation}
where $\Phi$ is the $SU(2)_L$ doublet Higgs field, $\tilde{\Phi} = i\sigma_{2} \Phi^{*}$, $L_L$ is the $SU(2)_L$ lepton doublet and $Y_{\nu}$ and $Y_{N}$ 
are general $3\times n$ and $n\times n$ Yukawa matrices, respectively. Without loss of generality we will work in the basis where $Y_N$ is diagonal.

\vspace{1mm}

The most general Lagrangian scalar potential for the Higgs doublet $\Phi$ and the singlet scalar $s$ 
%(at $T=0$)\SR{We haven't talked about $T$ yet, is this clarification necessary?} Good ppoint! :)
is given by (see e.g.~\cite{Espinosa:2011ax})
\begin{equation}
\begin{split}   V\left(\Phi^{\dagger}\Phi,s\right)=&-\tilde{\mu}_h^2 \Phi^{\dagger}\Phi+\lambda_h (\Phi^{\dagger}\Phi)^2+\frac{1}{2}\tilde{\mu}_s^2 s^2+
    \frac{1}{4}\lambda_s s^4\\
    &+ \frac{1}{2} \mu_m s \Phi^{\dagger}\Phi+\frac{1}{2} \lambda_m s^2 \Phi^{\dagger}\Phi+\tilde{\mu}_1^3 s+\frac{1}{3}\mu_3 s^3.
    \label{eq:ScalarPot0}
\end{split}
\end{equation}
By writing $\Phi = (h^{+}, (h + i \chi)/\sqrt{2})$, the scalar potential for the neutral fields $h$ and $s$, relevant for EW symmetry breaking, is found to be 
\begin{align}
    V (h,s) =&-\frac{1}{2} \tilde{\mu}_h^2 h^{2} +\frac{1}{4} \lambda_h h^{4}+\frac{1}{2}\tilde{\mu}_s^2 s^{2}+\frac{1}{4}\lambda_s s^{4}+ \frac{1}{4} \mu_m s h^{2}+\frac{1}{4} \lambda_m s^2 h^{2}+\tilde{\mu}_1^3 s+\frac{1}{3}\mu_3 s^3\, .
    \label{eq:ScalarPot1}
\end{align}
In the rest of this work, we will denote the (zero-temperature) vacuum expectation values (vevs) of the Higgs and singlet fields stemming from the potential~\eqref{eq:ScalarPot1} by $v_{EW}$ and $\omega_{EW}$, respectively.
We note that all parameters from $V(h,s)$ are real, which means that the only sources of CP violation beyond the SM would arise from the Yukawa couplings in Eq.~(\ref{eq:Lag_1}), in the absence of further new physics contributions.

\vspace{2mm}
%\SR{This vertical space? It seems we have many of them all around. It's ok, just the first time I see their use.}
In general, to study the early Universe dynamics of the scalar sector and the possibility to have a SFOPT, zero-temperature loop corrections (at 1-loop, this corresponds to the so-called Coleman-Weinberg contribution~\cite{Coleman:1973jx}) as well as finite-temperature contributions to the scalar potential~\cite{Dolan:1973qd,Weinberg:1974hy}, should be taken into account. 
These corrections, however, introduce gauge dependence~\cite{Patel:2011th} and renormalization scale dependence in the effective potential of the theory, leading to important theoretical uncertainties~\cite{Croon:2020cgk,Papaefstathiou:2020iag}.\footnote{A possibility to alleviate these problems consists of performing dimensional reduction, working with a 3-dimensional effective theory~\cite{Kajantie:1995dw,Brauner:2016fla,Schicho:2021gca,Niemi:2021qvp,Schicho:2022wty}. This procedure consists in practice on successively integrating-out all the heavy energy scales of the system (see e.g.~\cite{Croon:2020cgk} for a recent discussion on the topic).} 
Nevertheless, the scalar potential of the singlet scalar extension of the SM may already lead to the generation of a tree-level barrier between the EW symmetric and broken minima~\cite{Espinosa:2011ax} and, in such a case, an analysis based on the tree-level potential~\eqref{eq:ScalarPot1} supplemented by the leading ($\sim T^2$) thermal corrections in a high-temperature 
%expansion of the finite-temperature effective potential
approximation, which do not depend on the choice of the gauge, captures the most relevant features needed for the study of the SFOPT. 
%
%In order to study the early Universe dynamics of the scalar sector and the possibility to have a SFOPT, both temperature dependent contributions to the scalar potential~\cite{Dolan:1973qd,Weinberg:1974hy}, as well as the Coleman-Weinberg correction~\cite{Coleman:1973jx} should be included. These corrections lead to gauge and renormalization scale dependent results, introducing a series of theoretical uncertainties~\cite{Croon:2020cgk,Papaefstathiou:2020iag}. One possibility to alleviate these problems is performing dimensional reduction, working with a 3-dimensional effective theory~\cite{Kajantie:1995dw,Brauner:2016fla}. In practice, this procedure consists on successively integrating out all the heavy energy scales of the system. 
%
%Nevertheless, the potential under study may already lead to the generation of a tree-level barrier between the symmetric and broken minima~\cite{Espinosa:2011ax} and, in such a case, the high-temperature approximation of the thermal corrections, which does not depend on the choice of the gauge, captures the most relevant features needed for the study of the SFOPT.
%
At the same time, working at this level of approximation allows to study the relevant features of the phase transition analytically, as advocated in Ref.~\cite{Espinosa:2011ax}. This is very 
advantageous in order to efficiently scan the parameter space of the model. We have verified the generic validity of this approximation concerning the results of our global parameter scan, as we discuss in more detail in Section~\ref{sec:param_scan}.
The finite-temperature effective potential $V_T$ can in this case be written as:
\begin{equation}
\begin{split}
    V_T\left(h, s,T\right)=&-\frac{1}{2}\mu_h^2 h^2+\frac{1}{4}\lambda_h h^4+\frac{1}{2}\mu_s^2 s^2+\frac{1}{4}\lambda_s s^4 + \frac{1}{4}\mu_m s h^2+\frac{1}{4}\lambda_m s^2 h^2\\
    &+\mu_1^3 s+\frac{1}{3}\mu_3 s^3+\left[\frac{1}{2}c_h h^2+\frac{1}{2}c_s s^2+m_3 s\right](T^2-T_c^2)\, .
    \label{eq:ScalarPot2}
\end{split}
\end{equation}
The explicit appearance as a free parameter in Eq.~\eqref{eq:ScalarPot2} of the critical temperature $T_c$, 
at which the EW symmetric and broken minima are degenerate in energy, proves very convenient  
in a scan of the model parameter space requiring the presence of a SFOPT.
Indeed, when imposing that at $T=T_c$ the two minima are degenerate, an analytical condition among the other potential parameters in Eq.~(\ref{eq:ScalarPot2}) is obtained, effectively trading its freedom for $T_c$ and allowing to explore only potentials for which the two-degenerate-minima condition is fulfilled (see Appendix~\ref{appendix} for details). 
The %mass-squared 
parameters in Eq.~(\ref{eq:ScalarPot2}), defined at $T=T_c$, are related to those of Eq.~(\ref{eq:ScalarPot1}), defined at $T=0$, by
$
\tilde{\mu}_{h}^{2} 
\equiv 
\mu_{h}^{2} + c_{h}T_{c}^{2}$, 
$\tilde{\mu}_{s}^{2} 
\equiv 
\mu_{s}^{2} - c_{s} T_{c}^{2}$,
and 
$\tilde{\mu}_{1}^{3} 
\equiv 
\mu_{1}^{3} - m_{3}  T_{c}^{2}$.
The constants $c_h$, $c_s$ and $m_3$ are given by
\begin{equation}
\begin{gathered}
    c_h=\frac{1}{48}\left[9g^2+3g^{\prime 2}+2(6 Y_t^2+12\lambda_h+\lambda_m+2 \mathcal{Y}_{\nu}^2)\right],\\
    c_s=\frac{1}{12}\left[2\lambda_m+3\lambda_s+2\mathcal{Y}_{N}^2\right],\\
    m_3=\frac{1}{12}\left[\mu_3+\mu_m\right],
    \label{eq:Running_consts}
\end{gathered}
\end{equation}
where $g$ and $g'$ are respectively the $SU(2)_L$ and $U(1)_Y$ gauge couplings, $Y_t$ is the top Yukawa coupling, and
$\mathcal{Y}_{\nu}^{2}$, $\mathcal{Y}_{N}^{2}$ are defined as $\mathcal{Y}_{\nu}^2\equiv\mathrm{tr}\left(Y_{\nu}^{\dagger}Y_{\nu}\right)$,
$\mathcal{Y}_{N}^2\equiv\mathrm{tr}\left(Y_{N}^{\dagger}Y_{N}\right)$.

For the study of the temperature evolution of the scalar potential minima and the SFOPT, it is also convenient to rewrite the potential $V_T$ from Eq.~\eqref{eq:ScalarPot2} in terms of the temperature-dependent vevs $v_T\equiv \langle h\rangle(T)$ and $\omega_T\equiv \langle s \rangle(T)$ in the broken minimum as~\cite{Espinosa:2011ax} 
\begin{equation}
\begin{split}
    V_T\left(h,s,T\right)=&\frac{m_h^2}{8v_T^2}\left(h^2-v_T^2\right)^2+\frac{m_{sh}^2}{2v_T}\left(h^2-v_T^2\right)\left(s-\omega_T\right) \\
    &+\frac{1}{4}\left[2m_s^2+\lambda_m\left(h^2-v_T^2\right)\right]\left(s-\omega_T\right)^2\\
    &+\frac{v_T}{2m_h^2}\left(\lambda_m m_{sh}^2+4m_{*}\right)\left(s-\omega_T\right)^3+\frac{v_T^2}{8m_h^2}
    \left(4\lambda^2+\lambda_m^2\right)\left(s-\omega_T\right)^4,
    \label{eq:ScalarPot_wv}
\end{split}
\end{equation}
%
%with the parameters $\{v,\omega,m_{h}^{2}, m_{s}^{2}, m_{sh}^{2}, \lambda_{m}, \lambda, m_{*}\}$.
where all dimensionful parameters have an implicit dependence on the temperature $T$.
The mass parameters $m_{h}^{2}$, $m_{s}^{2}$, and $m_{sh}^{2}$ are defined as
\begin{equation}
    m_h^2\equiv \frac{\partial^2V}{\partial h\partial h}
    \Biggl|_{\left(v_T,\omega_T\right)},
    \quad
    m_s^2\equiv \frac{\partial^2V}{\partial s\partial s}
     \Biggl|_{\left(v_T,\omega_T\right)},
    \quad 
    m_{sh}^2\equiv \frac{\partial^2V}{\partial h\partial s}
     \Biggl|_{\left(v_T,\omega_T\right)},
    \label{eq:ScalarMassElements}
\end{equation}
evaluated at the EW broken minimum at $T$. The effective coupling $\lambda^2$
and mass $m_{*}$ in Eq.~\eqref{eq:ScalarPot_wv} are defined as 
\begin{equation}
    \lambda^2\equiv \lambda_h \lambda_s-\frac{1}{4}\lambda_m^2,
    \quad
    m_{*}\equiv \lambda^2\omega_T+\frac{1}{3}\lambda_h \mu_3-\frac{1}{8}\lambda_m\mu_m\, .
    \label{eq:lambda_m*}
\end{equation}
Furthermore, in the parameter scans in Section~\ref{sec:param_scan} we will eventually trade $m_{sh}^2$ for the quantity $\omega_p$, defined as
\begin{equation}
    \omega_p\equiv \omega_T-\frac{m_{sh}^2}{\lambda_m v_T} = \frac{-\mu_m}{2 \lambda_m}\,, 
    %\quad \JMN{\omega \quad {\rm or} \quad \omega_T \, ?}\SR{{\rm Actually \quad also} \quad v_T} jajajaja yes!
    \label{eq:omegap}
\end{equation}
which has the advantage of being temperature-independent.
The parametrization~\eqref{eq:ScalarPot_wv} explicitly shows that a shift in the field $s\rightarrow s+\sigma$ keeps the finite-temperature scalar potential $V_T$ invariant with a redefinition of $\omega_T\rightarrow \omega_T+\sigma$. 
The relations between the new parameters in Eq.~\eqref{eq:ScalarPot_wv} and the coefficients in Eq.~\eqref{eq:ScalarPot1} are found in Ref.~\cite{Espinosa:2011ax}.

\vspace{2mm}

%The starting point of our parameter scan is the potential of Eq.~(\ref{eq:ScalarPot2}) at $T=T_c$ with two degenerate minima located in general at $(0, \omega_{0})$ and $(v, \omega) \equiv (v_{T_c}, \omega_{T_c})$ in the $(h,s)$ space. \SR{Changed the $(h,s)=(0,\omega_0)$ but still not happy with this. Maybe change $(h,s)$ space for scalar-configuration-space?}Throughout the analysis, we take $\omega_0=0$ at the critical temperature $T_c$ making use of the freedom introduced by the shift symmetry of the potential to rescale $\omega$, which would translate into $\mu_1=0$ at $T_{c}$ in Eq.~(\ref{eq:ScalarPot2}). The potential at the critical temperature can be conveniently parametrized as~\cite{Espinosa:2011ax} 

The starting point of our analysis of SFOPT scenarios is the finite-temperature potential $V_T$ from Eq.~(\ref{eq:ScalarPot2}) at $T = T_c$, with two degenerate minima located in general at $(0, \omega_{0})$ and $(v, \omega) \equiv (v_{T_c}, \omega_{T_c})$ in the two-dimensional field space.
We will require that both the Higgs and the singlet field acquire a vev in the EW broken phase, in order to generate masses for the heavy neutrinos after the phase transition. 
Furthermore, in Ref.~\cite{Fernandez-Martinez:2020szk} it was shown that successful baryogenesis in the present scenario favours the heavy neutrinos to be approximately massless at the onset of the SFOPT. We will thus 
make use of the shift symmetry of the potential via $\omega_T\rightarrow \omega_T+\sigma$ discussed above to set 
$\omega_0=0$ at the critical temperature $T_c$ (this corresponds to setting $\mu_1 = 0$ %at $T = T_{c}$ 
in Eq.~(\ref{eq:ScalarPot2})),
and mainly focus on phase transitions from $(\langle h \rangle, \langle s \rangle) = (0,\,0)\rightarrow(v,\,\omega)$.
For studies on other phase transition scenarios in the singlet scalar extension of the SM, we refer the reader to Refs.~\cite{Fuyuto:2014yia,Kotwal:2016tex,Hashino:2016xoj,Kurup:2017dzf,Chen:2017qcz,Chiang:2018gsn,Carena:2019une,Kozaczuk:2019pet,Papaefstathiou:2020iag,Liu:2021jyc,Carena:2022yvx,Azatov:2022tii}. Following~\cite{Espinosa:2011ax}, we can use Eq.~(\ref{eq:ScalarPot_wv}) to conveniently parametrize the finite-temperature potential at the critical temperature as 
\begin{align}
\begin{split}
    V_T(h,s,T_c) = &\frac{v^{2} m_{h}^{2}}{8}\Biggl\lbrace\left(\frac{h^2}{v^2} - 1\right)^{2}+\left(\frac{s}{\omega} -1\right)^{3} \left(1 + \frac{3 s}{\omega}\right)+2 \frac{\lambda_{m} \omega^2}{m_h^2}\left(\frac{s}{\omega} - 1\right)^{2} \left(\frac{h^2}{v^2} - \frac{s^2}{\omega^2}\right) \\
    &+\frac{4 m_s^2}{ m_h^2} \frac{s^2}{v^2} \left(\frac{s}{\omega}-1\right)^{2}+ \frac{4 m_{sh}^2\omega}{ m_h^2 v}\left(\frac{s}{\omega} - 1\right)\left[ \frac{h^2}{v^2} + \frac{s^2}{\omega^2} \left( \frac{2 s}{\omega} - 3\right) \right]\Biggr\rbrace\,.
\end{split}
\label{eq:V-2degmin-Espinosa}
\end{align}

\vspace{-1mm}

The %rest of the 
set of parameters of Eq.~(\ref{eq:ScalarPot2}) at $T = T_{c}$ can be recovered from the new parametrization given in Eq.~\eqref{eq:V-2degmin-Espinosa} (together with $\mu_1 = 0$) via the following relations:\footnote{These are only valid at $T=T_c$. We refer the reader to Ref.~\cite{Espinosa:2011ax} for the general relation between both sets of parameters.}
\begin{align}
\mu_{h}^{2} =& 
    \frac{1}{2} 
    \left(
    m_{h}^{2} 
    - 
    \lambda_{m} \omega^2
    +
    2 \frac{\omega m^2_{sh}}{v}    
    \right),
    \\
%%%%%
\lambda_{h} =& \frac{1}{2} \frac{m_{h}^{2}}{v^{2}},
\\
%%%%%
\mu_{s}^{2} =&
    \frac{1}{2}
    \frac{ v^{2}}{\omega^{2}}
    \left(
        3 m_{h}^{2}
        -
        \lambda_{m} \omega^2
        +
        2
        \frac{\omega^2 m^2_s}{v^2}
        +
        6
        \frac{\omega m^2_{sh}}{v}
    \right),
    \\
%%%%%
\lambda_{s} =&
    \frac{1}{2}
    \frac{v^{2}}{\omega^{4}}
    \left(
        3 m_{h}^{2}
        -2
        \lambda_{m} \omega^2
        +
        4
        \frac{\omega^2 m^2_s}{v^2}
        +
        8
        \frac{\omega m^2_{sh}}{v}
    \right),
    \\
%%%%%
\mu_{m} =& -2  \left( \omega \lambda_{m} - \frac{m_{sh}^2}{v} \right),
\\
%%%%%
\mu_{3} =& 
    - \frac{3}{2}
    \frac{ v^{2}}{\omega^{3}}
    \left(
        2 m_{h}^{2}
        -
        \lambda_{m} \omega^2
        +
        2
        \frac{\omega^2 m^2_s}{v^2}
        +
        5
        \frac{\omega m^2_{sh}}{v}
    \right)\, .
\end{align}
%
%In summary, the following set of parameters\footnote{Notice that we have already made use of the shift symmetry in the singlet direction and taken $\omega_0=0$ without loss of generality, thus having 7 free parameters to describe the potential.} evaluated at $T_c$ is enough to fully characterize the scalar potential with two degenerate minima~\cite{Espinosa:2011ax} at $(0,0)$ and $(v,\omega)$: $\{\omega,\omega_{p},v,m_{h}^{2}, m_{s}^{2}, \lambda_{m} \}$. Together with $T_{c}$, this set of parameters allows us to specify the potential at any $T$ in the singlet scalar model. When the extension with extra heavy Dirac neutrinos is considered, we will also add $\mathcal{Y}_{\nu}$ and $\mathcal{Y}_{N}$ to free parameters, since they appear in the thermal corrections to the potential.
%
Then, the set of parameters $\{\omega,\omega_{p},v,m_{h}^{2}, m_{s}^{2}, \lambda_{m} \}$ evaluated at $T = T_c$
is enough to fully characterize the scalar potential with two degenerate minima at $(0,0)$ and $(v,\omega)$.\footnote{We re-stress that we have already made use of the shift symmetry in the singlet field direction and taken $\omega_0=0$ without loss of generality, thus having 7 free parameters to describe the potential.}
Together with $T_{c}$, this set of parameters allows us to specify the finite-temperature potential %at any 
as a function of $T$ in the singlet scalar extension of the SM, in the high-$T$ approximation.
%We have verified the generic validity of this approximation concerning the results of our global parameter scan, as we discuss in more detail in Section~\ref{sec:param_scan}.}
%
%\EFM{We have verified the validity of the high temperature approximation we adopted by comparing the value of $T_c$ characterizing our points in the parameter space with the one obtained computing its value through {\tt FindBounce}~\cite{Guada:2020xnz} from a potential with the same parameters at $T=0$ but taking into account the full evolution. This comparison was performed for all points that passed the selection criteria we introduce in Section~\ref{sec:param_scan}. We find that the two values agree within a $5\%$ for the vast majority of the points scanned with a few outliers extending to $\sim 10 \%$. Even though, as we will show in the following sections, the typical values for the scalar singlet mass we find are of the same order as $T_c$, the high temperature approximation remains accurate enough for our purposes (a fast efficient scan of the parameter space) since one of our selection criteria is to keep all couplings below $2$ (see Section~\ref{sec:param_scan}) so that radiative corrections are not very significant.}
%
When considering the addition of extra heavy Dirac neutrinos, we also need to add
$\mathcal{Y}_{\nu}$ and $\mathcal{Y}_{N}$ as free parameters in our analysis, since they appear in the thermal corrections of the potential $V_T$, see Eq.~\eqref{eq:Running_consts}.
%
%Therefore, with a set of parameters $\{\omega, \omega_{p}, v, m_{h}^{2}, m_{s}^{2}, \lambda_{m}, T_{c},\mathcal{Y}_{\nu}, \mathcal{Y}_{N}\}$, we can compute the potential at $T=0$, which is guaranteed to have two degenerate minima at $T=T_{c}$.

\begin{figure}[h!]
    \centering
    \includegraphics[width=0.485\textwidth]{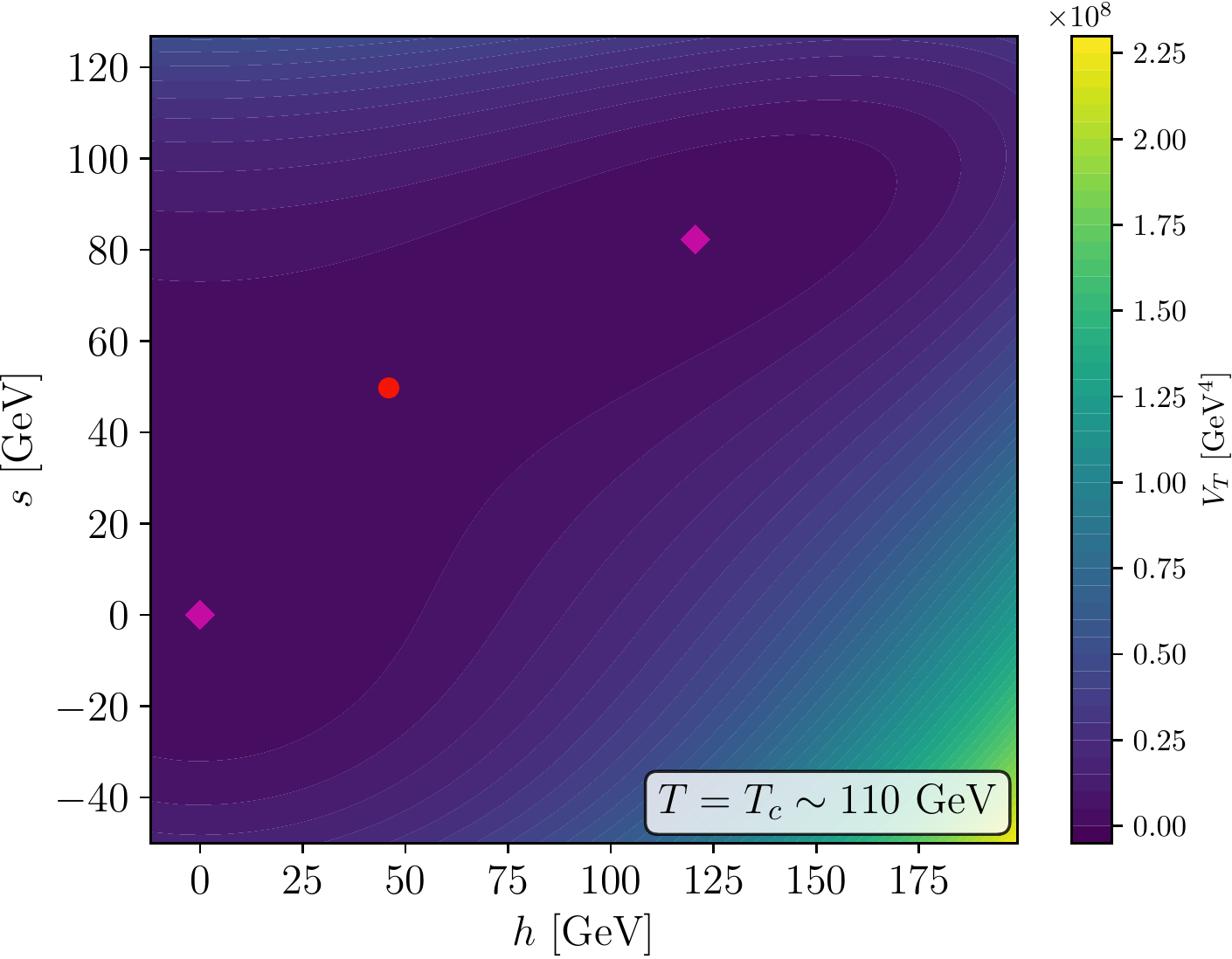}
    \hspace{2mm}
    \includegraphics[width=0.485\textwidth]{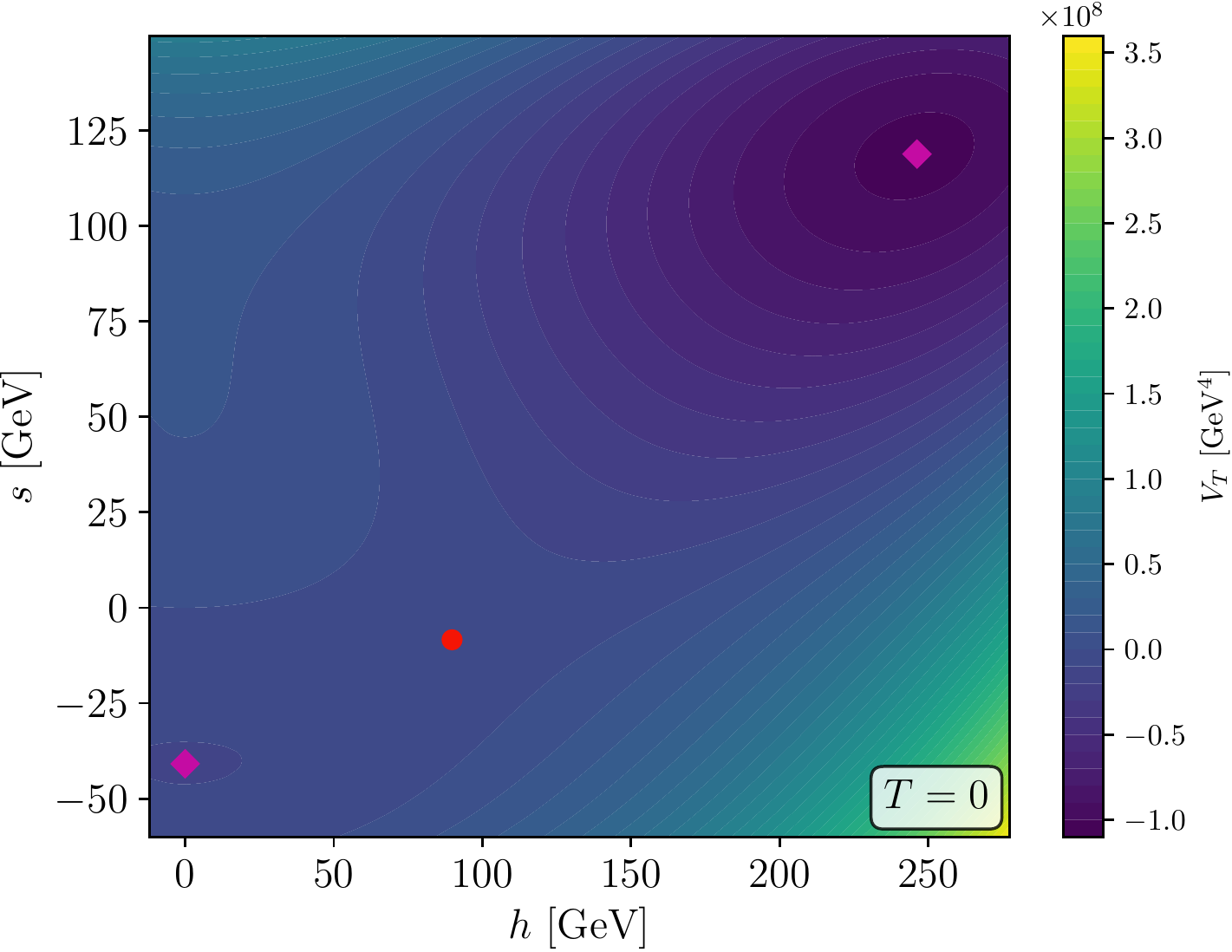}
    \vspace{-2mm}
    \caption{Shape of the scalar potential for the example point in parameter space specified in Table~\ref{tab:parameter_point} at the critical temperature $T_c$ (left panel) and at $T=0$ (right panel). The color bar denotes the value of the scalar potential, $V_T$. The purple diamonds denote the location of the minima, which are degenerate at $T_c$. The red dot denotes the location of the saddle point, which is close to the field trajectory for bubble nucleation (the bounce solution) at the nucleation temperature $T_N$ (see Section~\ref{sec:param_scan} for details).}
    \label{fig:Contour_plots}
\end{figure}

\vspace{2mm}

\begin{table}[h!]
\begin{center}
\begin{tabular}{ |c|c|c|c|c|c|c|c|c|c|c|c| } 
\hline
$\tilde{\mu}_h^2$~[GeV$^2$] & $\lambda_h$ & $\tilde{\mu}_s^2$~[GeV$^2$] & $\lambda_s$ & $\mu_m$~[GeV] & $\lambda_m$\\
\hline
1859.56 & 0.143276 & $-2076.39$ & 1.9975 & $-164.272$ & 0.415282 \\ 
\hline
$\tilde{\mu}_1^3$~[GeV$^3$] & $\mu_3$~[GeV] & $\mathcal{Y}_N$ & $\omega_{EW}$~[GeV] & $M_S$~[GeV] & $\sin{\xi}$\\
\hline
341158 & $-173.445$ & 0.293868 & 118.772 & 236.786 & 0.207\\%0.978341 \\ 
\hline
\end{tabular}
\end{center}
\vspace{-2mm}

\caption{Parameter set at $T=0$ corresponding to the scalar potential shown in Fig.~\ref{fig:Contour_plots}. This parameter set gives rise to a SFOPT with successful nucleation and satisfying all the bounds from Higgs phenomenology, including those on the scalar mixing (see Section~\ref{sec:pheno_singlet} for details).}
\label{tab:parameter_point}
\end{table}

Fig.~\ref{fig:Contour_plots} illustrates an example of the type of scalar potentials that would lead to a SFOPT with the characteristics described above, with the specific values of the corresponding potential parameter set at $T=0$ given in Table~\ref{tab:parameter_point}. In the left panel of Fig.~\ref{fig:Contour_plots}, the potential $V_T(h,s,T)$ is shown at $T = T_c$ with the two degenerate minima, represented by the purple diamonds. The red dot indicates the location of the saddle point yielding a potential barrier between both minima. In the right panel we show the potential at $T=0$ where the EW breaking minimum ($v_{EW},\omega_{EW}$) is now the true vacuum and also the real singlet has an $\mathcal{O}(100\,\,{\rm GeV})$ vev that generates EW-scale masses for the heavy neutrinos. 

\vspace{2mm}

In general, the potential $V_T$ from Eq.~\eqref{eq:V-2degmin-Espinosa} characterized by a random set of parameters $\{\omega,\omega_{p},v,m_{h}^{2}, m_{s}^{2}, \lambda_{m} \}$ at $T_c$ satisfying the conditions in Appendix~\ref{appendix} from Ref.~\cite{Espinosa:2011ax}, despite satisfying the desired property of featuring two degenerate minima at $(0,0)$ and ($v,\omega$), will not reproduce the correct value for the EW symmetry breaking vev at $T=0$, $v_{EW} =246.22$ GeV, obtained from the measurement of the Fermi constant via the muon decay width~\cite{Workman:2022ynf}. 
In addition, upon diagonalization of the scalar mass matrix at $T = 0$,
\begin{equation}
    \mathcal{M}^{0}_{s}=\begin{pmatrix}
    m_h^2 & m_{sh}^2\\
    m_{sh}^2 & m_s^2
    \end{pmatrix}\Bigg|_{T=0} \, ,
\end{equation}
with $m_h^2$, $m_s^2$ and $m_{sh}^2$ defined in Eq.~(\ref{eq:ScalarMassElements}) and evaluated in the $T = 0$ EW broken minimum $\left(v_{EW},\,\omega_{EW}\right)$, the eigenvalue $M_H$ for the mostly-doublet mass eigenstate will generally not reproduce the measured value for the Higgs boson mass $M_{H}=125.10$ GeV.
Satisfying these two physical requirements at $T=0$ is rather non-trivial in our setup, and considerably reduces the allowed %$\{\omega,\omega_{p},v,m_{h}^{2}, m_{s}^{2}, \lambda_{m} \}$
parameter space: 
given the high accuracy of the $v_{EW}$ and $M_H$ measurements, two combinations of the free parameters in the scalar potential are effectively determined.
In Section~\ref{sec:param_scan} we will discuss how these requirements are implemented in our numerical scan of the parameter space of the model. 

Finally, we also need to consider the existing constraints on the mixing $\xi$ between the Higgs doublet and the scalar singlet, arising from the diagonalisation of the $T = 0$ scalar mass matrix, $\mathcal{M}^0_{s}$. 
We have
\begin{equation}
    \begin{split}
        h=&v_{EW}+\cos{\xi}\, H +\sin{\xi}\, S,\\
        s=&\omega_{EW}-\sin{\xi} \, H +\cos{\xi}\, S,
        \label{eq:ScalarMixing}
    \end{split}
\end{equation}
where $S$ ($H$) is the mass eigenstate corresponding to the mostly-singlet (doublet) scalar combination with a mass $M_S$ ($M_H = 125.10$~GeV).
In the next section we will discuss the present experimental constraints on the ($T=0$) model parameters, affecting in particular the possible allowed values of the mixing $\xi$, which will also be applied to our parameter scan in Section~\ref{sec:param_scan}.

\section{Experimental constraints and phenomenological probes}
\label{sec:pheno_singlet}

In this section we discuss the relevant experimental limits on the singlet scalar extension of the SM, making emphasis on how the possible presence of the extra heavy singlet neutrinos can affect them. 
These experimental constraints will translate into bounds on the parameters of the potential from Eq.~(\ref{eq:ScalarPot2}) at $T = 0$. 
We also discuss the main phenomenological probes of the model, particularly in connection with both the structure of the scalar potential and the presence of the heavy neutrinos in comparison with the minimal singlet scalar extension of the SM.

\subsection{SM-heavy neutrino mixing}
\label{sec:theta-mixing}

The new Dirac neutrinos with components $N'_R$ and $N'_L$ introduced in Eq.~(\ref{eq:Lag_1}) mix with the SM neutrinos after spontaneous symmetry breaking (SSB) and may participate in the generation of light neutrino masses. The mixing matrix between the SM active and the heavy sterile neutrinos is given by
\begin{equation}
    \theta\equiv \frac{v_{EW}}{\sqrt{2}\,\omega_{EW}}Y_{\nu}Y_N^{-1},
    \label{eq:nu_mixing}
\end{equation}
with the Yukawa matrices $Y_N$ and $Y_\nu$ for singlet and SM neutrinos (see Eq.~\eqref{eq:Lag_1} for details), respectively. 
Thus, neglecting all small $L$-violating parameters that would eventually lead to the generation of the masses of the mostly SM-like light neutrinos, the heavy neutrinos have Dirac masses $M_{N_i}\simeq \omega_{EW} Y_{Ni}$ and the two chiralities of the mass eigenstates $N_i$ are given by
\begin{equation}
   N_R = N'_R, \quad N_L \simeq N'_L - \theta^\dagger \nu_L.
    \label{eq:nu_mixing2}
\end{equation}
For sterile neutrinos %heavier than the EW scale 
with masses $M_{N_i} > M_W$, the active-sterile neutrino mixing is bounded from above by a combination of EW precision tests\footnote{Notice that the recent anomalous measurement of $M_W$ by the CDF II collaboration~\cite{CDF:2022hxs} could be potentially explained through a non-zero neutrino mixing~\cite{Blennow:2022yfm}. However, this result is in tension with the other observables and we conservatively do not take it into account here.} and flavour observables~\cite{Fernandez-Martinez:2016lgt,Dani}
\begin{equation}
    \mathrm{tr}\left(\theta\theta^{\dagger}\right) < 0.0048 \quad (2 \sigma) .
    \label{eq:bound_mixing}
\end{equation}
For lighter sterile neutrinos ($M_{N_i} < M_W$), direct searches at colliders and beam dump experiments as well as searches for peaks and distortions in the decay products of mesons, leptons and beta decays set much more stringent constraints on the active-sterile neutrino mixing, and we refer the reader to Refs.~\cite{Bolton:2019pcu,MatheusRepository} for a comprehensive list of these limits.

The bound~\eqref{eq:bound_mixing} will be used as a $\chi^2$ contribution added to the weight function that we construct to guide our scan of the parameter space (see Appendix~\ref{appendix}). 
To constrain $\mathcal{Y}^2_{\nu} = \mathrm{tr}\left(Y_{\nu}^{\dagger}Y_{\nu}\right)$ in our parameter scan, we use that 
\begin{equation}
    \mathrm{tr}\left(\theta\theta^{\dagger}\right)\leq \frac{v_{EW}^2}{2\,\omega_{EW}^2}\mathcal{Y}_{\nu}^2\,\mathrm{tr}\left(Y_N^{-2}\right),
    %\leq \frac{n}{2} \left(v_{EW}/M^{\rm min}_{N_i}\right)^2    \mathcal{Y}_{\nu}^2\leq 0.0085\, .
    \label{eq:bound_nuyuk}
\end{equation}
%
%where $n$ is the number of singlet fermions, %$M^{\rm min}_{N_i}$ is the mass of the lightest (Dirac) sterile neutrino $N_i$ %(we recall the masses of the sterile neutrinos are given by $M_{N_i}\simeq Y_{N_i}\,\omega_{EW}$ in our setup)
%and 
since $\mathrm{tr}\left(A B\right)\leq \mathrm{tr}\left(A\right)\mathrm{tr}\left(B\right)$. Thus, when imposing the present bound on $ \mathrm{tr}\left(\theta\theta^{\dagger}\right)$ to the right-hand side of Eq~\eqref{eq:bound_nuyuk}, a conservative bound is implemented since $ \mathrm{tr}\left(\theta\theta^{\dagger}\right)$ will always be smaller than this quantity. %\JMN{I am confused by this! Apparently this would give a lower bound on $\mathcal{Y}_{\nu}^2$??} \EFM{I was confused by this too at some point. Check the sentence I added, should hopefully be clearer now.}
In practice, because we only have access to $\mathcal{Y}_N$ and not the individual values of the Yukawas, when constraining $\mathcal{Y}_\nu$ we assume a degenerate spectrum such that $\mathrm{tr}\left(Y_N^{-2}\right)=n^2\mathcal{Y}_N^{-2}$. Any other choice would translate into smaller values for $\mathcal{Y}_\nu$.
%Additionally we assume that the mass of the heavy neutrinos should be larger than some scale $M_{min}\sim M_W$ for the bounds from Ref.~\cite{Fernandez-Martinez:2016lgt} to apply. 

\subsection{LHC Higgs signal strengths}
\label{sec:Higgs_Signal_Strengths}

\begin{figure}
    \centering
    \includegraphics[width=0.75\textwidth]{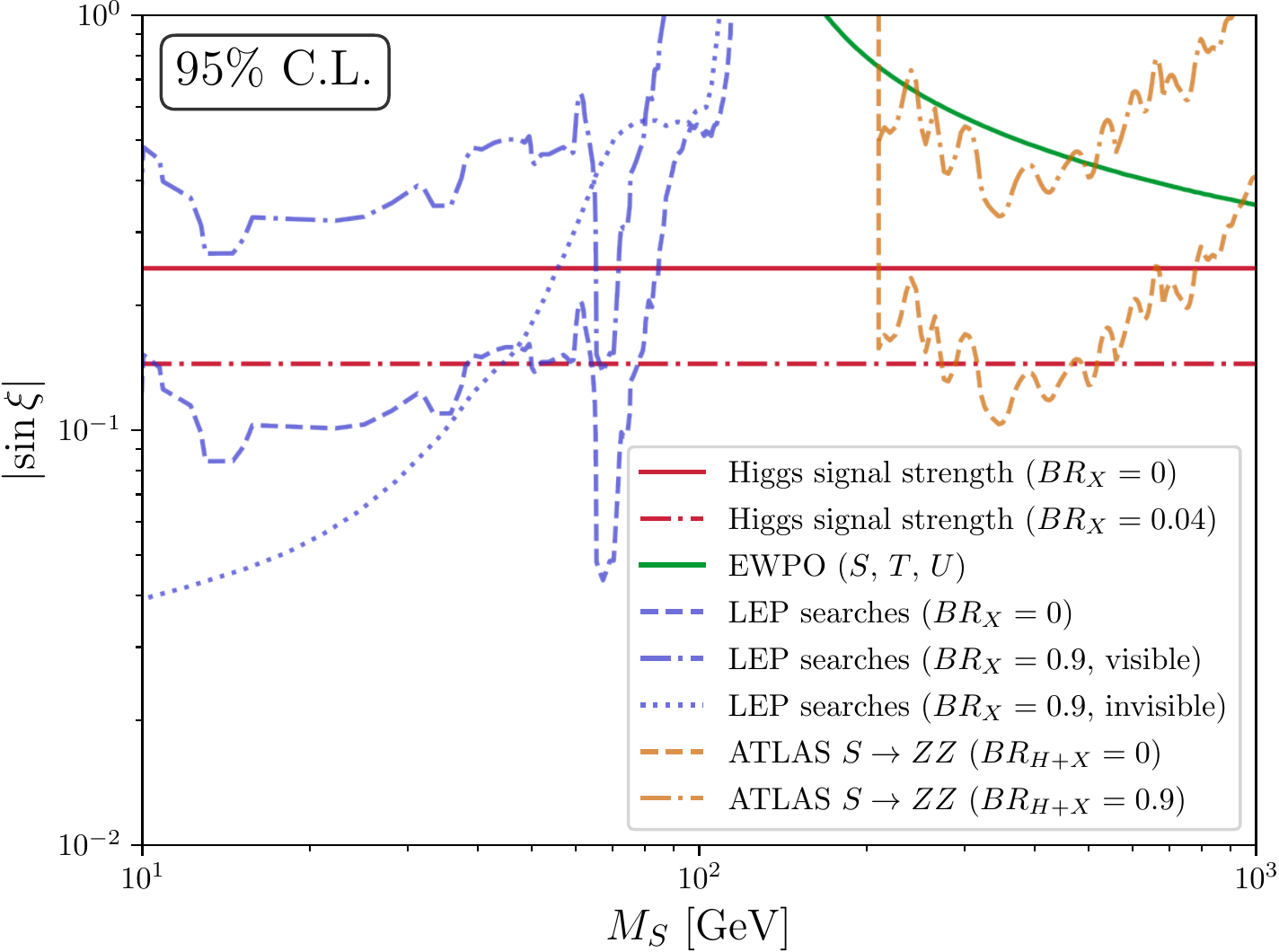}
    \caption{Existing 95\% C.L. constraints on the singlet-doublet scalar mixing $\sin{\xi}$ as a function of $M_S$ from EWPO (solid green), from LHC Higgs signal strength measurements (red) with $\mathrm{BR}_{H \to \mathrm{BSM}} \equiv \mathrm{BR}_X = 0$ (solid) and $\mathrm{BR}_X = 0.04$ (dash-dotted), from direct searches for $p p \to S \to Z Z$ by ATLAS~\cite{ATLAS:2020tlo} (ocher), with $\mathrm{BR}_{S\to H H} + \mathrm{BR}_{S\to \mathrm{BSM}} \equiv \mathrm{BR}_{H+X}  = 0$ (dashed) %, $\mathrm{BR}_X = 0.5$ (dashed) 
    and $\mathrm{BR}_{H+X} = 0.9$ (dash-dotted), 
    and from LEP searches for light scalars (blue) respectively assuming $\mathrm{BR}_{S \to \mathrm{BSM}} \equiv \mathrm{BR}_{X} = 0$ (dashed), $\mathrm{BR}_{X} = 0.9$ with visible BSM  decays (dash-dotted) and $\mathrm{BR}_{X} = 0.9$ with invisible BSM decays (dotted).
    %\TO{We should state: LEP bound is obtained with $\text{Br}(S \rightarrow \text{BSM}) = 0$. Higgs signal strengths bound is with $\text{Br}(H \rightarrow \text{BSM}) = 0$. Even if we do not have extra neutrinos, the signal strength bound in $M_{S} < M_{H}/2$ is weakened due to $H \rightarrow SS$ as discussed in Sec.~\ref{sec:Higgs_Signal_Strengths}}
    }
    \label{fig:Singlet_mass_vs_mixing_Bounds}
\end{figure}

The latest measurements of the 125 GeV Higgs boson signal strength by the ATLAS~\cite{ATLAS:2019nkf} and CMS~\cite{CMS:2020gsy} collaborations provide an important constraint on deviations of Higgs couplings from their SM values. In the singlet scalar extension of the SM, all SM couplings to the Higgs-like mass eigenstate $H$ become rescaled relative to the SM values by $\cos{\xi}$, which, in the absence of exotic Higgs decays (see discussion below), yields an overall suppression of Higgs signal strength given by 
\begin{equation}
\label{Higgs_Signal_Mu}
    \mu\equiv \frac{\sigma\cdot BR}{\left(\sigma\cdot BR\right)_{SM}}=\cos^2{\xi}
    %\quad     \text{(without $H$-BSM couplings)}
    \, .
\end{equation}
This allows to constrain the singlet-doublet scalar mixing via Higgs measurements (see~\cite{Robens:2015gla,Buttazzo:2015bka,Fuchs:2020cmm,Dawson:2021jcl}).
We use the latest measurements of the Higgs signal strength from ATLAS, $\mu = 1.05 \pm 0.06$~\cite{ATLAS:2022vkf} and CMS, $\mu = 1.002 \pm 0.057$~\cite{CMS:2022dwd}, and combine them to derive a bound on $\cos{\xi}$ following the Feldman-Cousins~\cite{Feldman:1997qc} prescription.\footnote{We adopt the Feldman-Cousins method to incorporate in a consistent way the fact that the best fit value from~\cite{ATLAS:2022vkf,CMS:2022dwd}, corresponding to $\mu > 1$, is not achievable in the scalar singlet extension of the SM.} 
We find $\mu \geq 0.94$ at the $95\%$~C.L. which translates into $\left|\sin{\xi}\right| < 0.245$,  as shown in Fig.~\ref{fig:Singlet_mass_vs_mixing_Bounds}. This is in fairly good agreement with other recent analyses (see e.g.~\cite{Dawson:2021jcl}) where Feldman-Cousins is however not applied in general. 
%\SR{How do we know they are not using Feldman-Cousins?}

\vspace{1mm}

In the presence of an exotic Higgs branching fraction into beyond-the-SM (BSM) states $\mathrm{BR}_{H\to\mathrm{BSM}}$, 
the Higgs signal strength modifier $\mu$ in Eq.~\eqref{Higgs_Signal_Mu} becomes
\begin{align}
\mu = \Bigl(1 - \mathrm{BR}_{H\to\mathrm{BSM}} \Bigr) \cos^2{\xi}
\label{eq:H_BSM_Higgs_Signal}
%\quad \text{(with $H$-BSM couplings)},
\end{align}
since $\cos{\xi}$ and $\mathrm{BR}_{H\to\mathrm{BSM}}$ now yield a combined dilution of the global Higgs signal strength with respect to the SM. 
%In particular, for $M_H > 2 M_S$ the presence of the decay $H \to S S$ makes the Higgs signal strength constraint on $\cos^2{\xi}$ tighter, depending on the specific value of $\mathrm{BR}_{H\to S S}$, as compared to the $M_H < 2 M_S$ scenario.
%
Thus, the presence of exotic Higgs decays yield a tighter bound on $\cos^2{\xi}$ from Higgs signal strengths, as shown in Fig.~\ref{fig:Singlet_mass_vs_mixing_Bounds} for 
the specific value $\mathrm{BR}_{H\to \mathrm{BSM}} \equiv \mathrm{BR}_X = 0.04$\footnote{It is clear from~\eqref{eq:H_BSM_Higgs_Signal} that an exotic branching fraction $\mathrm{BR}_{H\to \mathrm{BSM}} \geq 0.06$ is by itself ruled out at 95\% C.L.}. 
In particular, the interactions in Eq.~\eqref{eq:Lag_1} between the Higgs boson $H$ and the extra heavy neutrino states $N_i$ could lead to $H\rightarrow N_i\bar{N}_i$ if $M_{N_i}<M_{H}/2$, with the heavy neutrino masses $M_{N_i} \simeq Y_{N_i}\,\omega_{EW}$.
This occurs via $Y_{N}$ in Eq.~\eqref{eq:Lag_1}, through the singlet-doublet scalar mixing, or via $Y_{\nu}$, through the active-sterile neutrino mixing. 
In addition, the interaction $Y_{\nu}$ may also mediate $H \to \nu \bar{N}_i, \, \bar{\nu} N_i$ decays~\cite{Das:2017rsu,Das:2017zjc}.     
%
%
%We will discuss Higgs decays into sterile neutrinos in more detail in Sec.~\ref{sec:HNN-SNN}.
%
%Compared to the minimal scalar singlet extension of the SM, the presence of heavy neutrinos $N_i$ would yield new interactions for both the Higgs boson $H$ and the singlet-like scalar $S$ that may have an important impact on the model's phenomenology. 
%Besides the deviation of the coupling of the Higgs-like boson $H$ induced by the presence of the scalar singlet, it is interesting to consider more exotic signatures  also associated to the singlet fermions. 
%
Since the singlet-doublet scalar mixing is much more weakly constrained than the active-sterile neutrino mixing, the leading interaction (assuming $\theta^2 \ll \sin^2{\xi}$) after EW symmetry breaking between the scalar states and the $N_i$ ($i = 1,\, ...,\,  n$), %with masses $M_i\simeq Y_{N_i}\,\omega_{EW}$, 
which is induced by the Lagrangian from Eq.~(\ref{eq:Lag_1}), would be
%Higss and heavy neutrinos is induced after SSB:
\begin{equation}
\label{eq:Interaction_H_NN}
    \mathcal{L} \supset \frac{1}{\omega_{EW}} (\cos{\xi}\, S - \sin{\xi}\, H )\sum_i \bar{N}_i M_{N_i} N_i \, .
\end{equation}
The $H\rightarrow N_i\bar{N}_i$ decay channel is then driven by the scalar mixing, $\Gamma_{H \to N_i \bar{N}_i} \propto \sin^2{\xi}$, while in minimal seesaw scenarios the heavy neutrinos are produced via mixing with the SM neutrinos, leading to $\Gamma_{H \to N_i \bar{N}_i} \propto \theta^4$ and $\Gamma_{H \to \nu N_i} \propto \theta^2$. In our scenario these   generally correspond to subleading effects %(we expect $\mathcal{O}(\theta^2)$ corrections for both $H$ and $S$ interactions in Eq.~(\ref{eq:Interaction_H_NN}))
(the corrections for both $H$ and $S$ interactions in Eq.~(\ref{eq:Interaction_H_NN}) are $\mathcal{O}(\theta^2)$), and we concentrate in the following on the leading interaction from Eq.~\eqref{eq:Interaction_H_NN}.
%\footnote{Given the experimental limits on $\theta^2$ (see e.g~\cite{MatheusRepository}), the effect of the existence of a $H \to \nu N_i$ branching fraction for $M_{N_i} < M_W$ on Higgs signal strength is negligible.\SR{I got the impression the other day that it might not be negligible no?}\EFM{Indeed, I would remove this footnote..}}
%Since $Y_{\nu}$ is constrained from the active-sterile mixing as discussed in Sec.~\ref{sec:theta-mixing}, the decay channel $H \rightarrow \nu \bar{N}/\bar{\nu} N$ can only give a sub-dominant effect on the signal strengths.
% 
%The interaction~\eqref{eq:Interaction_H_NN} is directly linked to the coupling among the heavy neutrinos and the singlet scalar field $s$ in~(\ref{eq:Lag_1}).

%and mix with the SM neutrinos and Higgs, respectively.\TO{Well..., "mix with the SM neutrinos" ($=Y_{\nu}$) is not necessary for this, no? I think, this comes from $Y_{N}$ + $s$-$h$ mixing.} 
%
\vspace{1mm}

The decay $H\rightarrow N_i \bar{N}_i$ can have a significant impact on the LHC bounds on the Higgs signal strength. 
From the latest $\mu$ measurements performed by ATLAS~\cite{ATLAS:2022vkf} and CMS~\cite{CMS:2022dwd} and discussed above, we set the bound\footnote{We neglect $\mathrm{BR}_{H\to \nu N_i}$ as discussed above. In addition, we neglect a possible branching fraction $\mathrm{BR}_{H\to S S}$, potentially present only for very light singlets.}
%(considering for simplicity $\mathrm{BR}_{H\to S S} = 0$ and neglecting $\mathrm{BR}_{H\to \nu N_i}$)
%
\begin{equation}
    \Bigl( 1- \sum_{i} {\rm BR}_{H\rightarrow N_i\bar{N}_i} \Bigr) \cos^2\xi \geq 0.94\, .
    \label{eq:Higgs_Signal_H_NN}
\end{equation}
The total rate of the Higgs-like boson with a mass of 125 GeV decaying into $N_i$ states %$H\rightarrow N\bar{N}$ decay rate 
is given by
\begin{equation}
    \Gamma_{H\rightarrow N\bar{N}}
    \equiv 
    \sum_{i=1}^{k}
    \Gamma_{H\rightarrow N_{i}\bar{N}_{i}}
    \simeq\frac{\sin^2{\xi}}{8\pi}M_{H}
    \sum_{i=1}^{k} Y_{N_i}^2\left(1-\frac{4\,\omega^2_{EW}}{M^2_{H}}Y_{N_i}^2\right)^{3/2},
    \label{eq:total_DR_hN}
\end{equation}
where $k \leq n$ is the number of kinematically accessible heavy neutrinos. The maximum possible value of $\Gamma_{H\rightarrow N\bar{N}}$ (occurring for $k=n$) is given by 
\begin{equation}
\Gamma_{\text{max}} = n\,\frac{\sin^{2}\xi}{80\, \pi} \left(\frac{3}{5}\right)^{3/2} \frac{M_{H}^{3}}{\omega_{EW}^{2}} \simeq 
10^{-2}\,{\rm GeV} \times \left(\frac{n}{3}\right)\, \left(\frac{\sin^2{\xi}}{0.05}\right)\,\left(\frac{200\,{\rm GeV}}{\omega_{EW}}\right)^2 \, ,
\label{eq:Gamma_max-Case1}
\end{equation}    
which is achieved for $Y_{N_i}^2=M_H^2/(10\,\omega_{EW}^2)$
%=\mathcal{Y}^2_{N}/n$ 
with $i=1, 2,...n$, and may well be comparable to the SM Higgs boson total width $\Gamma_{\rm SM} = 0.00412$ GeV~\cite{LHC_Higgs_XS_WG} if the mixing $\sin{\xi}$ is not too suppressed.
For a given value of $\mathcal{Y}^2_{N} = \sum_{i=1}^{n} Y_{N_i}^2$, which is the relevant combination of neutrino Yukawa couplings affecting the thermal history of the scalar sector, the following two ``Cases'' are possible:
\begin{enumerate}

\item {\boldmath $\mathcal{Y}_N^2\, \omega_{EW}^2 < (M_H/2)^2$}: Then all neutrinos are kinematically accessible and $k = n$ in Eq.~\eqref{eq:total_DR_hN} so that the 125~GeV Higgs boson decays into all $n$ heavy neutrinos. 
For a fixed $\mathcal{Y}^2_{N}$ the maximum value of $\Gamma_{H\rightarrow N\bar{N}}$ is achieved when all Yukawa couplings are equal ($Y_{N_i}^2=\mathcal{Y}_N^2/n$ with $i=1, 2,...n$), while the minimum value is obtained when the rate is dominated by a single heavy neutrino contribution ($Y_{N_1}^2\approx\mathcal{Y}_N^2 \gg Y_{N_i}^2$ with $i=2,...n$). 
 
%\TO{As Jacobo mentioned, we cannot put $n \geq 3$ in Eq.~\eqref{eq:Gamma_max-Case1}.
%Then, the maximum possible value is achieved with $n=2$.}

\item {\boldmath $\mathcal{Y}_N^2\, \omega_{EW}^2 > (M_H/2)^2$}:  The decay of the 125 GeV Higgs boson into at least one heavy neutrino may be kinematically forbidden.\footnote{Writing $\mathcal{Y}_N^2 = R \, (M_H/2\, \omega_{EW})^2$ (with $R > 1$), for $R/n > 1$ the Higgs decay into at least one heavy neutrino $N_i$ must be kinematically forbidden, while for $R/n < 1$ it is still possible to have $k = n$ in Eq.~\eqref{eq:total_DR_hN}.} 
Therefore, for a given value of $\mathcal{Y}_N^2$, the decay rate can be arbitrarily suppressed depending on the value of the individual Yukawas $Y_{N_i}$ (e.g. in the limit $Y_{N_1}^2 \to \mathcal{Y}_N^2,\,  Y_{N_i}^2 \to 0$ for $i=2,...n$) and no lower bound on $\Gamma_{H\rightarrow N\bar{N}}$ exists.
Still, sizable Higgs boson branching ratios into sterile neutrinos are also possible (even reaching $\Gamma_{H\rightarrow N\bar{N}} = \Gamma_{\text{max}}$) in this case for a fixed value of $\mathcal{Y}_N^2$.

\end{enumerate}

The inclusion of the exotic $H \to N \bar{N}$ decay channel on the Higgs signal strength bound (recall Eq.~(\ref{eq:Higgs_Signal_H_NN})) allows to exclude a significant fraction of the parameter space in which a SFOPT is possible in the present scenario, as we will show explicitly in Section~\ref{sec:results}.
%
%\JLP{At the end of Section~\ref{sec:results}, we will show that }
%     
Finally, we stress that the heavy neutrinos produced in the decays of the 125~GeV Higgs bosons may themselves decay visibly inside the detector via active-heavy neutrino mixing, leaving a prompt or displaced vertex signal in the detector depending on the value of $\theta^2$. In particular, if the heavy neutrinos $N_i$ are long-lived, they can lead to a two-displaced-vertices signal in the LHC detectors, which would be a very powerful probe of the model~\cite{Graesser:2007pc,Caputo:2017pit}.

\subsection{Electroweak Precision observables}
\label{sec:EWPO}

The properties of the singlet field are also constrained by EW precision observables (EWPO), which limit the value of the mixing $\sin{\xi}$ as a function of the scalar mass $M_S$ in the singlet scalar extension of the SM (see e.g.~\cite{Gorbahn:2015gxa,Beniwal:2018hyi}). This is a result of the shift induced by the presence of the singlet scalar on the EW oblique parameters $S$, $T$, $U$\cite{Peskin:1991sw} with respect to the SM.
A global fit to EWPO measurements yields the respective values of the shifts on the oblique parameters with respect to their SM predictions~\cite{Haller:2018nnx}
\begin{equation}  
S \; = \; 0.04 \pm 0.11, \; T \; = \; 0.09 \pm 0.14, \; U = -0.02 \pm 0.11 \, ,
\label{DeltaObliques}
\end{equation}
with the following correlation coefficients: $+0.92$ between $S$ and $T$, $-0.68$ between $S$ and $U$ and $-0.87$ between $T$ and $U$. Explicit expressions for $S$, $T$, $U$ in the singlet scalar extension of the SM are given in~\cite{Beniwal:2018hyi} as a function of $\sin{\xi}$ and $M_S$. Using these, we obtain the 95\% C.L. limits on the ($M_S, \, \left|\sin{\xi}\right|$) plane from a $\chi^2$ fit to the $S$, $T$, $U$ measurements from Eq.~\eqref{DeltaObliques}. These are shown in Fig.~\ref{fig:Singlet_mass_vs_mixing_Bounds}, highlighting that for values of $M_S$ below a TeV, the bound from Higgs signal strength discussed in the previous section is stronger than that of EWPO. 

Notice, however, that the same EWPO used to constrain $S$, $T$ and $U$ and, from there, derive constraints on $\sin{\xi}$ are affected already at tree level and used to derive the bounds on the heavy-active neutrino mixing $\theta$ as outlined above~\cite{Fernandez-Martinez:2016lgt}. In principle, the two contributions should be studied together to derive a consistent set of constraints. The interplay between new physics contributions to the EWPO through $S$, $T$, $U$ and the presence of heavy neutrinos was studied in detail in Refs.~\cite{Loinaz:2002ep,Loinaz:2004qc,Akhmedov:2013hec,Fernandez-Martinez:2015hxa}. In particular, it was realized that most observables depend on the same combination of elements of $\theta$ and $T$ and that, if a cancellation between these two contributions is present, the bounds on both sources of new physics would weaken significantly. Nevertheless, for this situation to take place, negative and sizable values of $T$ are required~\cite{Fernandez-Martinez:2015hxa}. The scalar singlet contribution to $T$ does indeed become negative for masses above the mass of the Higgs (see e.g.~\cite{Beniwal:2018hyi}). For lighter singlet masses, no cancellation is possible and the two effects would rather reinforce each other, leading to slightly stronger constraints. Nevertheless, since the bounds from Higgs signal strength are more stringent, the potential contribution of the singlet is small and does not alter significantly the constraints on heavy-active neutrino mixing derived in~\cite{Fernandez-Martinez:2015hxa}. Conversely, for a scalar singlet heavier than the Higgs, the bound $\mathrm{tr}\left(\theta\theta^{\dagger}\right)\leq 0.0048$ would weaken if $-2 \alpha T \sim 0.0048$. However, given the bounds on $\sin{\xi}$ from the LHC Higgs signal strength measurements (see Section~\ref{sec:Higgs_Signal_Strengths}), this is never achieved for sub-Planckian scalar masses.
Thus, for the parameter space under study, the possible interplay between the heavy neutrino and scalar singlet contributions to EWPO can be safely neglected.  

\subsection{Searches for singlet-like scalars at LEP and LHC}
\label{Direct_Bounds_scalars}

Under the assumption that the singlet-like scalar decays into SM particles (i.e. its decay is driven by the singlet-doublet mixing), the null results from LEP searches for Higgs-like particles yield strong upper limits on $\left|\sin{\xi}\right|$ for singlet-like scalar masses below $M_S \simeq 115$ GeV (see e.g.~\cite{Robens:2015gla}). These limits are at the level of $\left|\sin{\xi}\right| \lesssim 0.2$ (or below) for masses $M_S < 100$ GeV.
At the same time, LHC searches for BSM scalars decaying to $W W$, $Z Z$ or $H H$ pairs also constrain the doublet admixture of the singlet-like scalar $S$ for $M_S > M_H$. For $M_S > 200$ GeV the strongest such limits are obtained by ATLAS in the $Z Z \to 4\ell$ and $ Z Z \to 2\ell \, 2\nu$ final states~\cite{ATLAS:2020tlo}.~\footnote{For masses $M_H < M_S < 200$ GeV, only $\sqrt{s} = 7$ and $8$ TeV LHC limits exist~\cite{CMS:2013zmy,ATLAS:2015pre}. These are not competitive with present Higgs signal strength bounds on $\sin{\xi}$, and we disregard them.} 
In Fig.~\ref{fig:Singlet_mass_vs_mixing_Bounds} we show the corresponding bounds on the ($M_S,\,\left|\sin{\xi}\right|$) plane from both LEP and LHC searches for new scalars, under the assumption $\mathrm{BR}_{S\to H H} + \mathrm{BR}_{S\to \mathrm{BSM}} \equiv \mathrm{BR}_{H+X} = 0$ (dashed lines). 

\vspace{1mm}

Nevertheless, compared to the minimal singlet extension of the SM, here the presence of the heavy neutrinos may lead to much less stringent bounds on $\sin{\xi}$ from direct scalar searches.
%, if the singlet-scalar decays into the new $N_i$ states.
%
Indeed, the interactions of $S$ with the heavy neutrinos $N_i$ in Eq.~\eqref{eq:Interaction_H_NN} will induce the decay $S \to N_i \bar{N}_i$ if available by phase space. The corresponding partial width $\Gamma^S_{N \bar{N}} \propto \cos^2{\xi}$, in contrast to the partial decay widths of $S$ into SM states, $\Gamma^S_{\rm SM} \propto \sin^2{\xi}$. Thus, $S \to N_i \bar{N}_i$ will generally be the dominant decay channel for the singlet-like scalar in the limit $\left|\sin{\xi}\right| \ll 1$ (as favoured by LHC Higgs signal strength measurements, see Sec.~\ref{sec:Higgs_Signal_Strengths}):
%. As a result, the above limits from direct searches for $S$ at colliders would be significantly weakened:
%
\begin{itemize}
    \item For light singlets ($M_S \lesssim 100$ GeV), the $S\to N_i \bar{N}_i$ decay channel would significantly relax constraints on $\sin{\xi}$ from LEP searches for Higgs bosons decaying visibly (into SM particles), and we show the corresponding dilution of the limits when $\mathrm{BR}_X = \mathrm{BR}_{S\to N_i \bar{N}_i} = 0.9$ in Fig.~\ref{fig:Singlet_mass_vs_mixing_Bounds}. Nonetheless, if the heavy neutrinos $N_i$ are long-lived (e.g. for very small neutrino mixing) and would have escaped the LEP detectors, limits from LEP searches for invisibly decaying Higgses~\cite{ALEPH:2001roc,DELPHI:2003azm,L3:2004svb,OPAL:2007qwz} would apply. 
    We also depict the bounds from such searches on $\sin{\xi}$ in  Fig.~\ref{fig:Singlet_mass_vs_mixing_Bounds}, showing that they become very strong for rather light scalars. We nevertheless re-stress that these only apply under specific conditions (very long-lived $N_i$, leading to invisible $S$ decays), which depend on the details of the neutrino sector of the model. 
    %, softening the dilution of the LEP limits discussed above. 

%    (it is also possible to reinterpret LEP searches for generic scalars $\phi$ (decaying to SM particles) recoiling against invisible $Z$ bosons, $e^+ e^- \to \phi\, Z, \, Z \to \nu \bar{\nu}$~\cite{Lopez-Fernandez:1993bnm,deCampos:1996bg}). This also yields limits on $\sin{\xi}$, yet much weaker than those quoted in the previous section. \EFM{But the previous ones are relaxed... how do we know they are weaker?}
    %

    \item For $M_S \gtrsim 200$ GeV, the presence of the $S \to N_i \bar{N}_i$ decay would weaken the LHC limits on $\sin{\xi}$ from $p p \to S \to Z Z$ searches, as shown explicitly in Fig.~\ref{fig:Singlet_mass_vs_mixing_Bounds} for $\mathrm{BR}_{H+X} = 0.9$.\footnote{The presence of a non-zero $S \to H H$ partial width $\Gamma^S_{H H}$ would also weaken the limits on $\sin{\xi}$ from $p p \to S \to Z Z$ searches, allowing at the same time to search for $S$ via resonant di-Higgs production (see e.g.~\cite{ATLAS:2021ifb,ATLAS:2022xzm}). Yet, di-Higgs searches are generally less sensitive than $Z Z$ ones for equal branching fractions, and the equivalence theorem~\cite{Lee:1977eg} naively yields $\Gamma^S_{H H} \sim \Gamma^S_{Z Z}$ in the $M_S \gg v$ limit (since also $\Gamma^S_{H H} \propto \sin^2{\xi}$). We have thus not considered here the would-be limits from resonant di-Higgs searches for $\Gamma^S_{H H} \neq 0$ for simplicity.} At the same time, this BSM decay would open a new avenue to probe the existence of $S$ and $N_i$ at the LHC, either when the $N_i$ decay products are resolved in the ATLAS/CMS detector or merge into a single reconstructed object (for $M_S \gg M_{N_i}$, producing a ``neutrino jet"~\cite{Mitra:2016kov,Mattelaer:2016ynf}). 
    Yet, current LHC searches for heavy neutrinos generally consider $N_i$ production modes (e.g. Drell-Yan or $W\gamma$ fusion, see~\cite{Degrande:2016aje} for a discussion) which yield kinematic properties of the $N_i$ rather different from those of $S \to N_i \bar{N}_i$ decays,\footnote{An exception is given by LHC searches for $Z'$ gauge bosons decaying to heavy neutrinos~\cite{CMS:2022irq}, which feature similar kinematics and could allow for a reinterpretation in our setup. We defer this for future work.} and as such present LHC limits (see~\cite{Cai:2017mow,Han:2022qgg} for reviews) are difficult to extrapolate to our scenario. 
    Moreover, the possibility that the $N_i$ yield displaced decays (for $\theta^2 \ll 1$) would dramatically reduce the sensitivity of those existing searches, providing at the same time a new avenue for discovery in long-lived particle searches to be explored in the future.  
\end{itemize}

\subsection{Higgs self-coupling}

Finally, the existence of the singlet scalar would induce a deviation on the Higgs boson trilinear self-coupling $\lambda_{HHH}$ from its SM value. This is currently being searched for at the LHC~\cite{CMS:2018ipl,ATLAS:2019qdc,ATLAS-CONF-2021-052} via non-resonant di-Higgs production, albeit with limited precision at present. 
At tree-level, we find
\begin{equation}
    \lambda_{HHH}=
    \lambda_h\,v_{EW}\, c_{\xi}^3-\frac{2\,\lambda_m \,\omega_{EW} + \mu_m}{4}c_{\xi}^2s_\xi+\frac{\lambda_m\,v_{EW}}{2}c_{\xi}s_{\xi}^2-\frac{3\,\lambda_s \,\omega_{EW}+ \mu_3}{3}s_{\xi}^3 \, ,
    \label{eq:Trilinear}
\end{equation}
with $\cos{\xi}\,(\sin{\xi})=c_{\xi}\,(s_{\xi})$. Additionally, and particularly relevant in the $\left|\sin{\xi}\right| \ll 1$ limit, the one-loop corrections to the trilinear self-coupling coupling should be taken into account, as they contain terms that do not vanish even when $\sin{\xi} \rightarrow 0$. The one-loop contribution reads, in the $\left|\sin{\xi}\right| \ll 1$ limit~\cite{Chen:2017qcz}
\begin{equation}
    \Delta \lambda_{HHH}^{\mathrm{1-loop}}=\frac{1}{16\pi^2}\left(\lambda_m^3\frac{v_{EW}^3}{2\,M_S^2}+27\frac{M_H^4}{v_{EW}^3}+3\,\lambda_m^2\frac{\mu_3\,v_{EW}^2}{M_S^2}s_{\xi}\right)\, .
    \label{eq:Trilinear1loop}
\end{equation}
We parametrize the deviation with respect to the SM as
\begin{equation}
    \kappa_\lambda\equiv \frac{\lambda_{HHH}+\Delta\lambda_{HHH}^{\mathrm{1-loop}}}{\lambda_{HHH}^{SM}} \, ,
    \label{eq:Kappa_lambda}
\end{equation}
with $\lambda_{HHH}^{SM}=M_H^2/(2v_{EW})$ the tree-level value of the SM Higgs boson self-coupling.\footnote{We remark that our definition of $\kappa_\lambda$ matches that of the ATLAS and CMS experimental collaborations, yet $\kappa_\lambda = 1$ corresponds to the SM prediction only if one-loop corrections to $\lambda_{HHH}$ in the SM (which amount to $9\%$ of the tree-level value~\cite{Dorsch:2017nza}) are neglected.} The most stringent constraint on $\kappa_\lambda$ has been recently set by the ATLAS Collaboration~\cite{ATLAS-CONF-2021-052} to be
\begin{equation}
    -1.0 \leq\kappa_\lambda\leq 6.6\;\;(95\%\,\rm{C.L.}).
    \label{eq:Kappa_lambda_bound}
\end{equation}
We note that the measurements of $\lambda_{HHH}$ at the LHC via non-resonant di-Higgs production can be significantly altered by the presence of a resonant contribution to the di-Higgs signature (see~\cite{Carena:2018vpt,Arco:2022lai} for a discussion), appearing in the singlet scalar extension of the SM via the $p p \to S \to H H$ process. Still, depending on the singlet-like scalar mass $M_S$, it should be possible to exploit the di-Higgs invariant mass distribution $m_{HH}$ to measure the self-coupling $\lambda_{HHH}$~\cite{Carena:2018vpt} from the non-resonant part of the distribution (with the resonant part properly identified and subtracted), achieving comparable precision to the scenario with no resonant $S \to H H$ contribution. A detailed analysis of this issue is however beyond the scope of the present work. Moreover, as we will see in Section~\ref{sec:results} this observable barely deviates from its SM value in the interesting regions of the parameter space and is therefore not a relevant probe of the scenario under study.

\section{Parameter scan}\label{sec:param_scan}

In this section we describe our procedure to scan the parameter space of the model and collect the sets of parameters which fulfill the various necessary conditions for a SFOPT.
As described in Section~\ref{sec:singlet_model},  in general a potential characterized by a random set of parameters $\{\omega, \omega_{p}, v, m_{h}^{2}, m_{s}^{2}, \lambda_{m}, T_{c},
\mathcal{Y}_{\nu}, \mathcal{Y}_{N}\}$ at $T = T_{c}$ will not 
yield the correct values of the Higgs vev and Higgs mass at $T=0$, namely $v_{EW} = v_{EW}^\text{exp} \equiv  246.22$ GeV and $M_{H} = M_{H}^\text{exp} \equiv 125.10$ GeV.
%satisfy that $v=v_{EW}$ and $M_{H} = M_{H,\text{exp}}$ at $T=0$.
%
The first condition 
%$v = v_{EW}$ 
$v_{EW} = 246.22$ GeV can always be imposed starting from any given set of parameters by shifting all the 
%mass dimensionful parameters 
parameters $\eta$ with dimension of mass (including $T_{c}$)
as $\eta \rightarrow (v_{EW}^\text{exp}/v_{EW})\, \eta $.
To satisfy the second condition, $M_{H}=M_{H}^{\text{exp}}$,
we solve for $m_{h}^{2}$ for each generated set $\vec{w} = \{\omega, \omega_{p}, v, m_{s}^{2}, \lambda_{m}, T_{c},
\mathcal{Y}_{\nu}, \mathcal{Y}_{N}\}$
to find values which reproduce
the correct Higgs boson mass: in practice, for a given set $\vec{w}$  
we scan $m_{h}^{2}$ imposing $v_{EW}/M_{H} = v_{EW}^\text{exp} / M_{H}^\text{exp}$ before the aforementioned shift of the mass dimensionful parameters $\eta$, which guarantees $M_{H} = M_{H}^\text{exp}$ after it.
%In practice, we generate a $\vec{w}$ and we scan $m_{h}^{2}$ imposing that $v/M_{H}|_{T=0} = v_{EW} / M_{H,\text{exp}}$ before the shift of the mass dimensionful parameters, which guarantees $M_{H} = M_{H,\text{exp}}$ after the shift. 
%
A solution does not always exist depending on the actual values of $\vec{w}$.
%
%In order to have a complete description of the potential, one should also scan over $m_h^2=\partial^2_h V_T$. However, as already pointed out in Section~\ref{sec:pheno_singlet}, the precise determination of both $v_{EW}$ and $M_{H,exp}$ effectively reduces the parameter space by fixing one parameter, which we choose to be $m_h^2$. Thus, when exploring the parameter space though the set of parameters $\vec{w}$, the first step is to fix the value for $m_h^2$ which allows to reproduce the experimental ratio of $v^2/M_H^2|_{T=0}$. If such a value does not exist, the point is discarded and a new set of parameters is generated. 
%
%Once a potentially good set of parameters $\vec{w}$ is found, all mass parameters are rescaled by a factor $A$, such that $v|_{T=0}=v_{EW}$, which automatically sets as well $M_H|_{T=0}=M_{H,exp}$ given that the ratio $v/M_H|_{T=0}$ does not depend on this rescaling. Additionally, the structure of the potential remains the same such that we still have two degenerate minima at $T_c$. 
%
In this way, we obtain the sets of parameters which have two degenerate minima at $T=T_{c}$ and also reproduce the correct Higgs vev and mass 
%the correct EW phenomenology, $M_{H,\text{exp}}$ and $v_{EW}$ 
at $T=0$.

\begin{figure}
\centering
\begin{tikzpicture}[node distance=2cm]
%%%%%%%%%%%%%%%%%%%%%%%%%%%%%%%%%%%%%%%%%%%%%%%%%%%%%%%%%%%%%%%%%%%%%%
\node (genparamTc) [myprocess] 
{\begin{minipage}{4.5cm}
\begin{center}
Generate parameters at $T_c$\\
\bigskip
$\{\omega,\omega_{p},v,m_{s}^{2},\lambda_{m},\mathcal{Y}_{\nu}, \mathcal{Y}_N\}$
\end{center}
\end{minipage}};

\node (scanmhSqTc) [myprocess,below of=genparamTc] 
{Scan $m_{h}^{2}$ at $T_c$};

\node (checkatT0) [mydecision, below of=scanmhSqTc, yshift=-0.8cm] 
{\begin{minipage}{4.5cm}
\vspace{-0.3cm}
\begin{center}
Does $m_h^2$ satisfying\\
%\medskip
$v_{EW}/M_{H} = v_{EW}^\text{exp} / M_{H}^\text{exp}$
%$\left.v/M_H\right|_{T=0}=v_{EW}/M_{H,\text{exp}}$ 
exist?
\vspace{-0.5cm}
\end{center}
\end{minipage}
};

\node (shiftV) [myprocess, below of=checkatT0, yshift=-1.1cm] 
{\begin{minipage}{6.5cm}
\begin{center}
Re-scale dimensionful parameters to fix
$M_{H} = M_{H}^\mathrm{exp}$ \& $v_{EW} = v_{EW}^\mathrm{exp}$
%$\left.M_{H}\right|_{T=0} = M_{H,\mathrm{exp}}$ \& $\left.v\right|_{T=0}=v_{EW}$
\end{center}
\end{minipage}};

\node (checkEspinosa) [mydecision, below of=shiftV, yshift=-0.9cm] 
{
\begin{minipage}{4.5cm}
\vspace{-0.1cm}
\begin{center}
Weighted MCMC SFOPT conditions in Appendix~\ref{appendix}.
\end{center}
\vspace{-0.7cm}
\end{minipage}};

\node (success) [startstop, below of=checkEspinosa, yshift = -0.8cm]
{
\begin{minipage}{4.5cm}
\begin{center}
Keep point in\\ parameter space
\end{center}
\end{minipage}};
%%%%%%%%%%%%%%%%%%%%%%%%%%%%%%%%%%%%%%%%%%%%%%%%%%%%%%%%%%%%%%%%%%%%%%
\draw [arrow] (genparamTc) -- (scanmhSqTc);
\draw [arrow] (scanmhSqTc) -- (checkatT0);
\draw [arrow] (checkatT0.west) node [left,yshift=-0.3cm,xshift=+0.1cm] {No} -- +(-0.5,0) |- (genparamTc.west);
\draw [arrow] (checkatT0.south) node [left, yshift=-0.3cm] {Yes} -- (shiftV);
\draw [arrow] (shiftV) -- (checkEspinosa);
\draw [arrow] (checkEspinosa.east) node [right, yshift=-0.3cm,xshift=-0.1cm] {Rejected} -- +(0.5,0) |- (genparamTc.east);
\draw [arrow] (checkEspinosa.south) node [right, yshift=-0.3cm] {Accepted} -- (success);
%%%%%%%%%%%%%%%%%%%%%%%%%%%%%%%%%%%%%%%%%%%%%%%%%%%%%%%%%%%%%%%%%%%%%%
\end{tikzpicture}
\caption{Flowchart for the selection of parameter sets generating the correct Higgs vev and mass at $T=0$ and satisfying all necessary conditions to potentially have a SFOPT from Ref.~\cite{Espinosa:2011ax}.}
\label{fig:flowchart}
\end{figure}
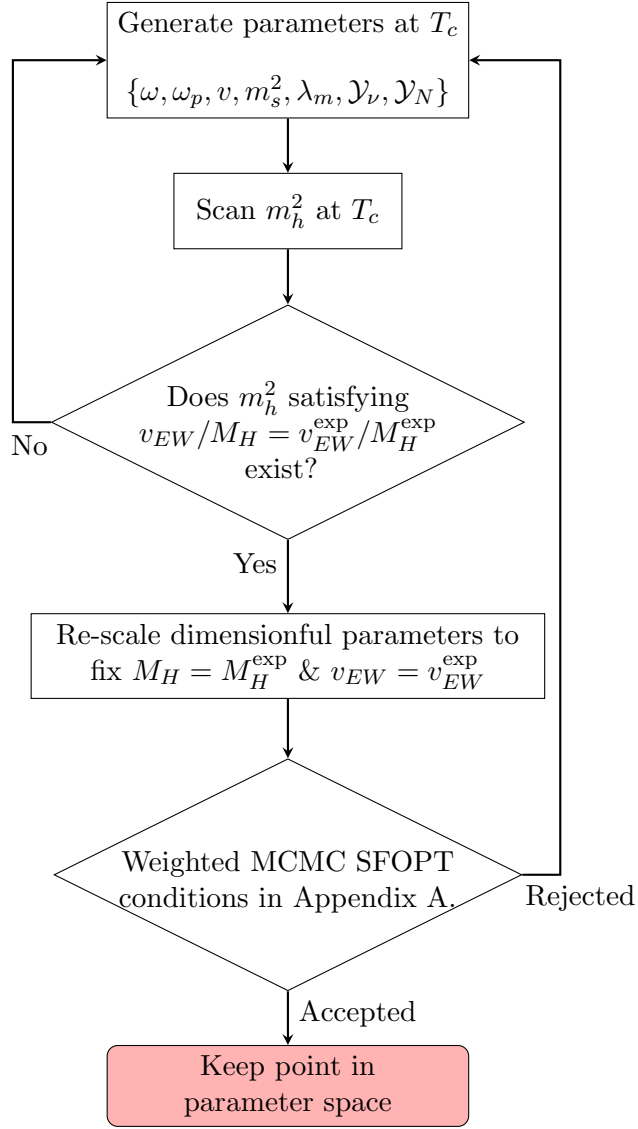

Following Ref.~\cite{Espinosa:2011ax},
we bias our scan towards the parameter sets that satisfy the necessary conditions for a SFOPT.\footnote{Among the different conditions, we look for potentials that are bounded from below for which the EW minimum is the global one. While the EW minimum could be metastable, such setups are beyond the scope of this work.}
For this purpose, we have defined an ad-hoc weight function to rate how well the selected points satisfy these conditions, in order to prioritize the parameter regions to which the points belong in our scan.
We then use this weight function in place of the log-likelihood for a Markov Chain MonteCarlo (MCMC) using the standard Metropolis Hastings algorithm to sample the interesting regions of the parameter space with MonteCUBES~\cite{Blennow:2009pk}. The conditions for a SFOPT and the weight function used in the MCMC are explicitly defined in Appendix \ref{appendix}.
Our procedure of the parameter scan is summarized in Fig.~\ref{fig:flowchart}.

Finally, we have verified the validity of the high-$T$ approximation adopted in Section~\ref{sec:singlet_model} through a comparison of the value of $T_c$ obtained in this approximation with the one obtained from a potential with the same parameters at $T=0$ but implementing the temperature evolution with the full 1-loop thermal potential (see e.g.\cite{Quiros:1999jp}). The comparison has been performed for points passing all viability criteria, as discussed below. Both values of $T_c$ agree within $5\%$ accuracy for the vast majority of the points scanned (a few outliers extend to $\sim 10 \%$). Then, even though in some cases the value for the scalar singlet mass $M_S$ found in the scan is of the same order as $T_c$, the high-$T$ approximation can remain suitable for our purposes (a fast efficient scan of the parameter space). In addition, we impose a stringent perturbativity condition on the scalar quartic couplings, $\lambda_i\leq2$ (see below), which favours that radiative corrections are not significant.

\vspace{2mm}

The parameter sets output of our MCMC scan are further classified according to the following viability criteria:
\begin{itemize}
    \item Points with the scalar potential quartic couplings $\lambda_i\leq2$ to ensure perturbativity.
    %for which the inclusion of higher orders
    %would be necessary, rendering the analysis inaccurate.
    
    \item Points that lead to a sufficiently strong first-order phase transition
    (if the phase transition occurs).
    As a rough estimate, we ask for the ratio $v/T_c>1$.
    This is required to make EW baryogenesis possible
    by decoupling sphaleron processes in the EW broken phase.
    
    \item Points for which the bubbles of the EW broken phase can 
    %potentially 
    actually nucleate and the phase transition does take place. 
    Although the conditions summarized
    %that we already took account 
    in Fig.~\ref{fig:flowchart} are needed to realize a SFOPT, they are not sufficient to guarantee it.
    It is important to study
    whether a nucleation temperature $T_N < T_c$ exists for which the bubbles of the EW broken phase (the true vacuum for $T<T_{c}$) successfully grow~\cite{Carena:2022yvx,Biekotter:2021ysx,Baum:2020vfl,Kozaczuk:2019pet,Chen:2017qcz} (and the Universe does not become trapped in the false vacuum).
    In our scan of parameters, we gauge the nucleation of EW bubbles as follows:
    
    %In the presence of two almost degenerate minima, 
    The transition probability from the false to the true vacuum is proportional to $e^{-S_3/T}$, with $S_3$ the three-dimensional bounce action.
    %and $T_N$ is the so-called nucleation temperature at which bubbles of true vacuum form. 
    At temperatures slightly below $T_c$, in the so-called thin-wall regime for which the two minima are almost degenerate, the action $S_3$ diverges for $T \to T_c$~\cite{Coleman:1980aw,Ivanov:2022osf} and thus no transition is possible in this regime. As the Universe cools down from $T_c$ to the nucleation temperature $T_N$, away from the thin-wall limit, the computation of the bounce action becomes more involved. It requires to solve the equations of motion for the background fields to find the bounce solution.
    This is usually done via numerical solvers like {\tt CosmoTransitions}~\cite{Wainwright:2011kj}, {\tt BubbleProfiler}~\cite{Athron:2019nbd} or {\tt FindBounce}~\cite{Guada:2020xnz}.
    Nucleation will be possible if there exists a temperature $T_N > 0$ at which $S_3/T_N\sim 140$~\cite{Linde:1980tt,Anderson:1991zb,McLerran:1990zh,Dine:1991ck}, for which the nucleation rate is comparable to the Hubble expansion rate during radiation domination.
   
 \begin{figure}
    \centering
    \includegraphics[width=0.6\textwidth]{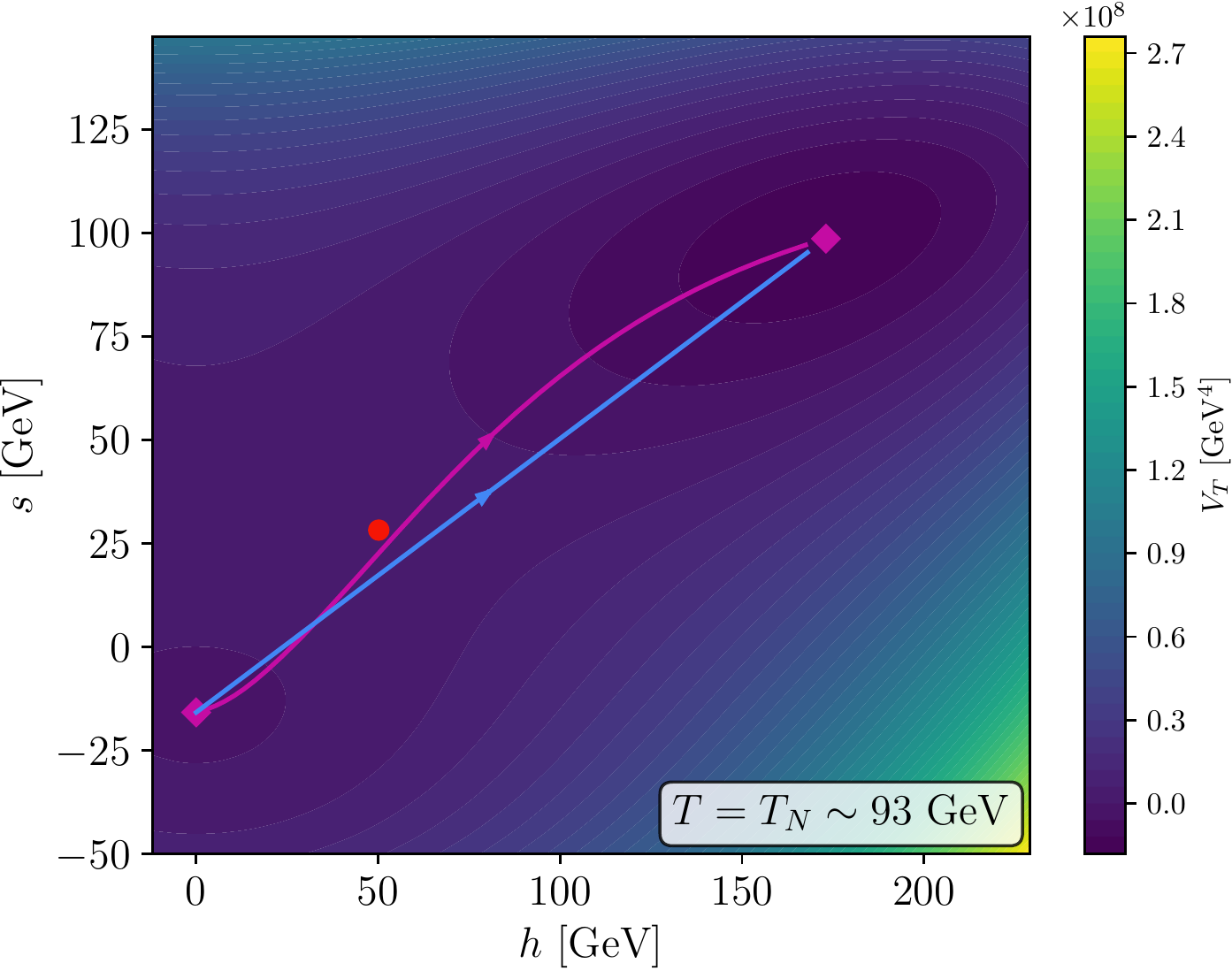}
    \caption{The same as Fig.~\ref{tab:parameter_point} but at the nucleation temperature $T_N$. The blue line corresponds to the straight path in the field configuration space which we use as an approximation to estimate the action and judge if nucleation may happen, while the purple curve corresponds to the actual bounce solution passing closer to the saddle point (red dot) between the two minima (purple diamonds).
    %\TO{$T_{N}\sim 93$ GeV, but how different is $T_{N}$ with $S_{3}=140$ from that with $S_{3}^{\text{app}}=140$ in this concrete example?}
    %\SR{With the proxy it is 84 GeV which also make sense, assuming we have a monotonically decreasing action $S_3$ with $T$, at the same temperature $S_3<S_3^{app}$. I don't know though if we really want to go into these details, also because anyone can take the point and check it themselves.}
    }
    \label{fig:Contour_plotsTN}
\end{figure}  
   
    In order to estimate the bounce action and thus the nucleation temperature $T_N$, instead of computing the bounce solution along the path that minimizes the tunneling action, we approximate the  solution by calculating the bounce action along a straight path in field space, which connects both minima at $T_N$. The action for such a field configuration, $S_3^{\text{app}}$, will by construction be larger than the tunneling solution~\cite{Coleman:1987rm,Cline:1999wi,Espinosa:2018szu}, $S_3\leq S_3^{\text{app}}$. 
    We find that there is good agreement between the true action $S_3$ and our estimation for the cases of interest, and thus successful nucleation for points in parameter space is expected to occur when $S_3^{\text{app}}/T_N\lesssim140$, which in turn represents a conservative estimate. In practice, we rewrite the scalar potential in terms of a linear combination of $h$ and $s$ along the straight line connecting the minima at a given $T$, $\phi_{\parallel}$, and the orthogonal one, $\phi_{\perp}$, as prescribed in Ref.~\cite{Akula:2016gpl}. By taking $\phi_{\perp}\rightarrow 0$ one can quickly find the bounce solution along the straight line using the overshoot-undershoot method in one dimension with, for example, {\tt FindBounce}. In this manner, the action $S_3^{\text{app}}$ can be computed at the temperature $T$. The temperature at which $S_3^{\text{app}}/T\sim140$ defines the nucleation temperature $T_N$. 
    As an example of our approximation, we compare in Fig.~\ref{fig:Contour_plotsTN} the approximated and the actual bounce trajectories in field space with the set of parameters given in  Table~\ref{tab:parameter_point}. 
    This approximation allows us to efficiently scan the parameter space. 
    In the results shown in the following section, the boundaries between the nucleating and non-nucleating points should therefore be understood as an approximated result and somewhat conservative. In Section~\ref{sec:results} we will perform a comparison between the regions of parameter space selected by our 
    nucleation criteria and those found using
    %the results from 
    {\tt CosmoTransitions}, finding good agreement between both methods.
    %the two. 
    
    %the distinction between nucleating and non-nucleating points represents interesting regions of parameter space, but the boundaries of those regions should not be taken as precise and are somewhat conservative. 
    
    \item Points with the scalar mixing $\xi$ at $T=0$ allowed 
    by collider searches as described in the previous section. 
\end{itemize}
In the next section, we will show the impact of each of these conditions in the parameter space to reveal the correlations among the parameters and the preference for particular parameter regions.
As we will see, the condition of the bubble nucleation will prove to be the most constraining one~\cite{Carena:2022yvx,Baum:2020vfl,Kozaczuk:2019pet,Chen:2017qcz}, which greatly reduces the allowed parameter space. 
%Moreover, we find that when considering large neutrino Yukawa couplings, $\mathcal{Y}_N$, the EW broken minimum may be destabilized unless this effect is compensated by other parameters. \JLP{The following is not entirely true right? I would just remove the next sentences or state somewhere that what we mean by large $\mathcal{Y}_N$ is larger than $2$}. This option however turns out to be ruled out by the constraints on the singlet-doublet scalar mixing. Thus, their preferred values are not large enough to open new regions in the allowed parameter space. 

\section{Results}
\label{sec:results}

\begin{figure}
    \centering
    \includegraphics[width=0.49\textwidth]{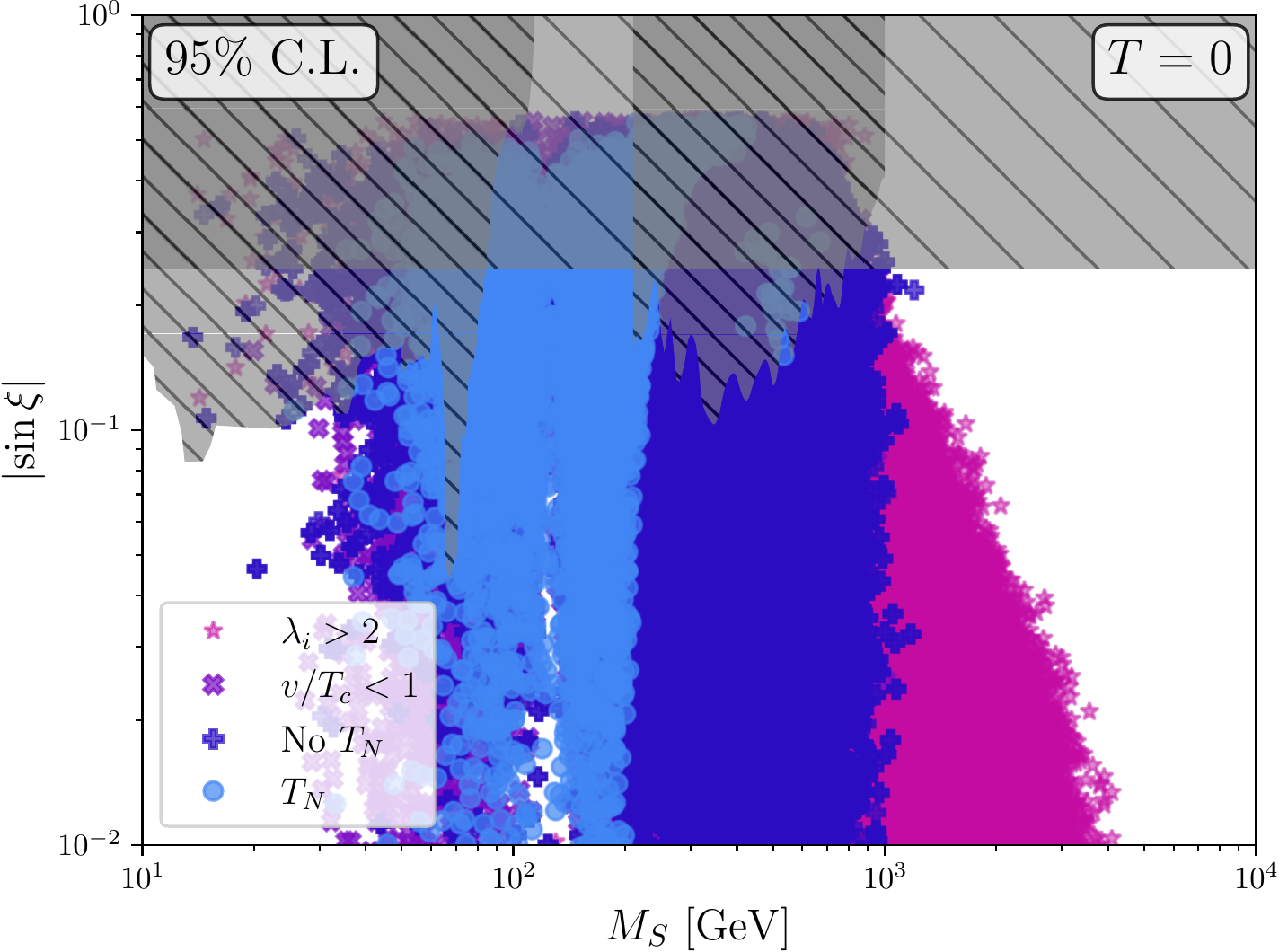}
    \includegraphics[width=0.49\textwidth]{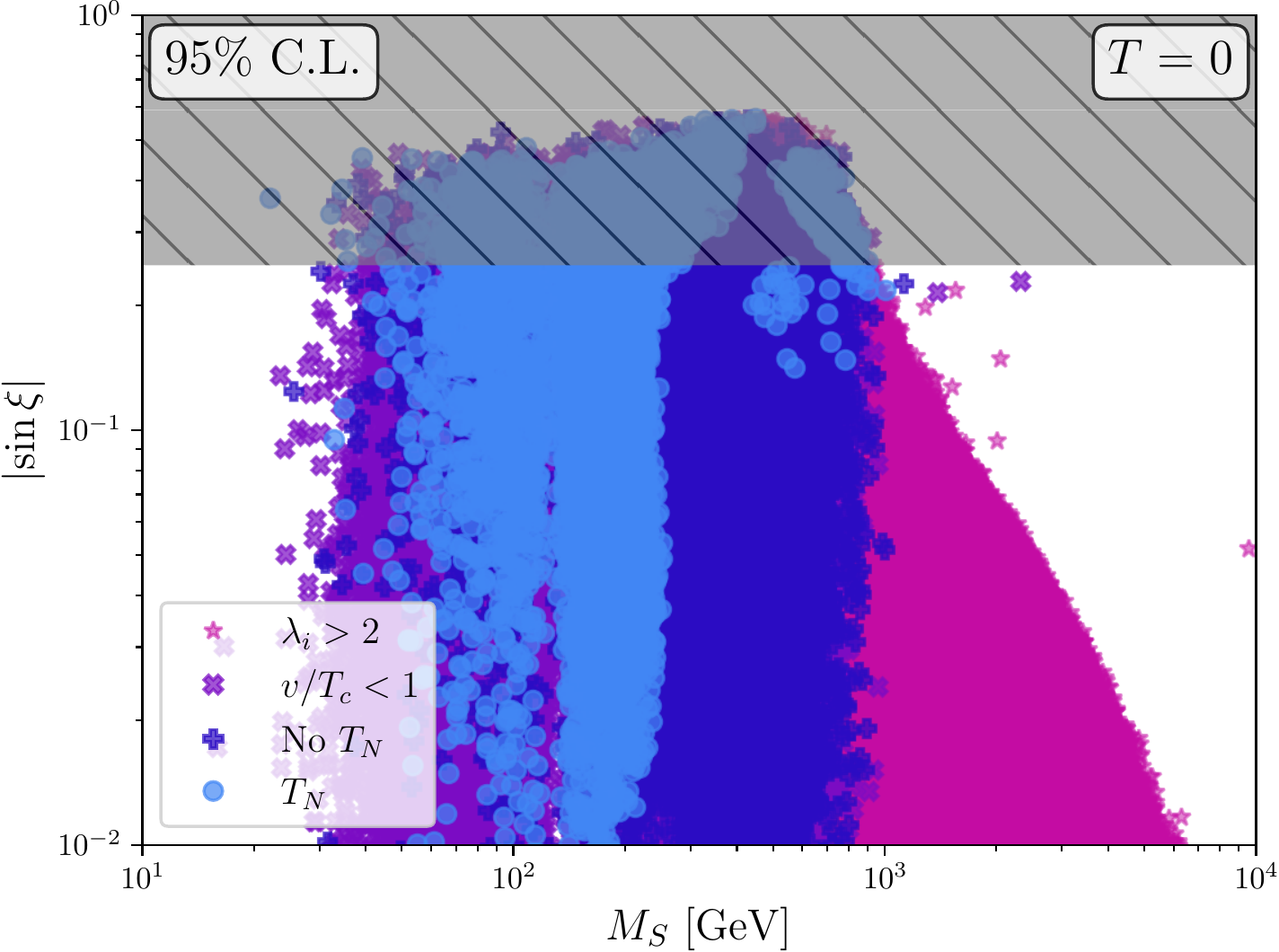}
    \caption{Results of the parameter scan in the scalar mass, $M_S$, and mixing, $\sin{\xi}$, plane. In the right (left) panel the scan was performed with(out) the addition of the heavy Dirac neutrinos. The pink stars correspond to points with non-perturbative couplings, the purple crosses to points for which sphaleron transitions would not decouple in the broken phase. The light blue dots (dark blue pluses) additionally do (not) satisfy the nucleation condition as described in the text. The grey-shaded region corresponds to the bounds on the scalar mixing described in Section~\ref{sec:pheno_singlet}. In the right panel, we do not show the bounds that depend on the parameters in each point.}
    \label{fig:Singlet_mass_vs_mixing}
\end{figure}

In this section we present and analyze the results from the parameter scan described in Section~\ref{sec:param_scan} where the different constraints and conditions described in the previous section have been implemented. While the scan is performed over all parameters (at $T = T_c$) in $\vec{w} = \{\omega, \omega_{p}, v, m_{s}^{2}, \lambda_{m}, T_{c},
\mathcal{Y}_{\nu}, \mathcal{Y}_{N}\}$ with $m^2_h$ fixed so as to reproduce the correct Higgs mass, the constraint in the active-heavy neutrino mixing $\mathrm{tr}\left(\theta\theta^{\dagger}\right)|_{\text{exp}}\leq 0.0048$~\cite{Fernandez-Martinez:2016lgt} implies $\mathcal{Y}_{\nu}^2$ will have a negligible impact on the scalar potential. We will therefore not show this parameter in the following. 
Instead we will mainly focus on parameters which have a direct connection to experimental observables, and thus refer the reader to Figs.~\ref{fig:Triangle_plot} and \ref{fig:Triangle_plot_Comparison} in Sec.~\ref{sec:summary} for results on the complete set of parameters at $T_c$, $\left\lbrace\vec{w},\,m_{h}^{2}\right\rbrace$, and $T=0$ respectively, as a summary of our results.

In Fig.~\ref{fig:Singlet_mass_vs_mixing} we show the points collected in our parameter scan in the plane of the mass and mixing of the scalar singlet ($M_S$, $\left|{\rm sin} \xi \right|$). 
We study and compare the scenarios with(out) the addition of the heavy Dirac neutrinos in the right (left) panels. All the points have been selected according to the algorithm summarized in Fig.~\ref{fig:flowchart} and, therefore, satisfy the conditions from Ref.~\cite{Espinosa:2011ax} for a SFOPT and have the correct Higgs mass and vev at $T=0$.
The points with different colours and symbols are classified by the conditions listed in Section~\ref{sec:param_scan}.
The pink stars are discarded since they have at least one very large scalar coupling ($\lambda_{i} > 2$).~\footnote{Such large couplings will drive the model into a non-perturbative regime at scales very close to the EW scale, and thus we disregard those points in our scan.}
For the purple crosses, this perturbativity condition is satisfied, but the first-order phase transition is not strong enough to decouple the sphaleron process in the EW broken phase ($v/T_{c}<1$), even if the bubbles of the broken phase may nucleate.
The dark blue crosses labeled with ``No $T_{N}$'' have $\lambda_{i}<2$ and $v/T_{c} >1$ but the nucleation condition  $S_{3}^{\text{app}}/T \lesssim 140$ is not satisfied at any $T<T_{c}$ (and therefore, there is no $T_{N}$). 
Finally, the light blue dots labeled with ``$T_{N}$'' have $\lambda_{i} <2$ and $v/T_{c} >1$ and also fulfill the nucleation condition. 
Grey-shaded areas in the left panel represent the values of the scalar mixing ruled out by LHC Higgs signal strength measurements %for $M_S> M_H/2$, 
(assuming $\mathrm{BR}_{X} = 0$) as described in Section~\ref{sec:Higgs_Signal_Strengths}, or by direct searches for Higgs-like particles at LEP for $M_S\lesssim 100$~GeV and at ATLAS for $M_S>200$~GeV (assuming $\mathrm{BR}_{S\to H H} = 0$). On the right panel we only display the conservative Higgs signal strength bound in the absence of exotic Higgs decays, since the bounds from direct scalar searches at LEP and LHC may be diluted when heavy Dirac neutrinos are included, depending on the values of neutrino couplings as discussed in Section~\ref{Direct_Bounds_scalars}. As can be seen from the plots, these constraints are quite relevant and a big portion of the parameter space is ruled out by them, so that only small values of $\sin{\xi}$ are still allowed. 
Moreover, we also find that the condition of successful bubble nucleation considerably reduces the size of the viable parameter space, as pointed out in Refs.~\cite{Baum:2020vfl,Biekotter:2021ysx} for other scenarios. 
Thus, only the light blue dots below the grey-shaded regions are successful candidates for a SFOPT satisfying all phenomenological constraints listed in the previous section. From Fig.~\ref{fig:Singlet_mass_vs_mixing} we can also see that the Universe 
%can
may
undergo an EW SFOPT only if the mass of the singlet scalar $S$ is $M_{S} \lesssim 300$~GeV. Generally speaking, higher values of $M_S$ would also imply larger $\omega$ and hence a significant distance between the two minima, in general too large to allow for bubble nucleation. The apparent exception to this rule by the few points clustered around $M_S \sim 500-1000$~GeV can be understood through a closer inspection of their thermal evolution. Indeed, in these cases we find a SFOPT only in the singlet direction at $T\gg \mathcal{O}(100)$~GeV. After this transition, both the Higgs and the singlet vevs roll towards their values at $T=0$, $v_{EW}$ and $\omega_{EW}$, respectively, with $\omega_{EW}\gg v_{EW}$.

\begin{figure}[t]
    \centering
    \includegraphics[width=0.49\textwidth]{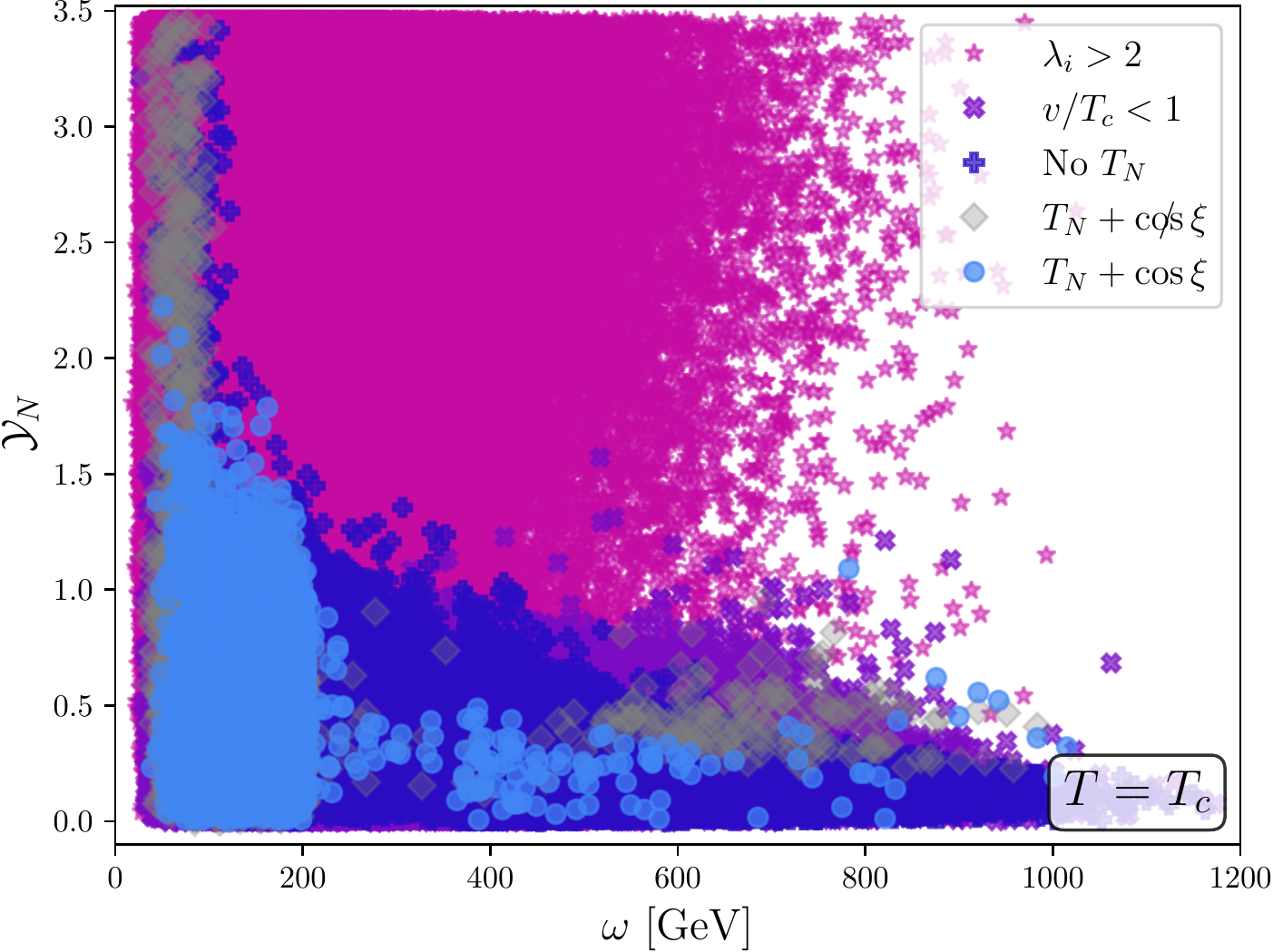}
    \includegraphics[width=0.49\textwidth]{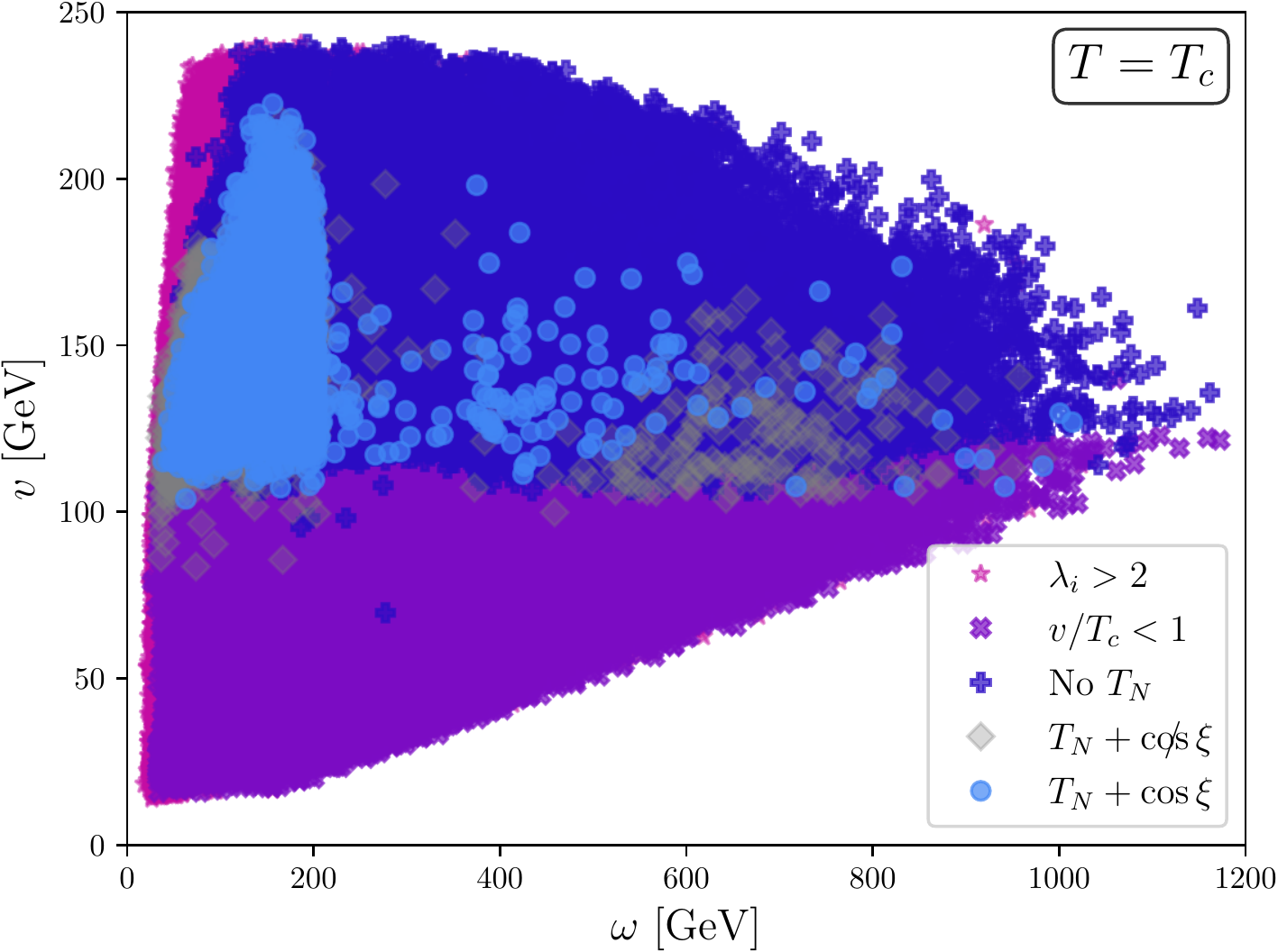}
    \includegraphics[width=0.49\textwidth]{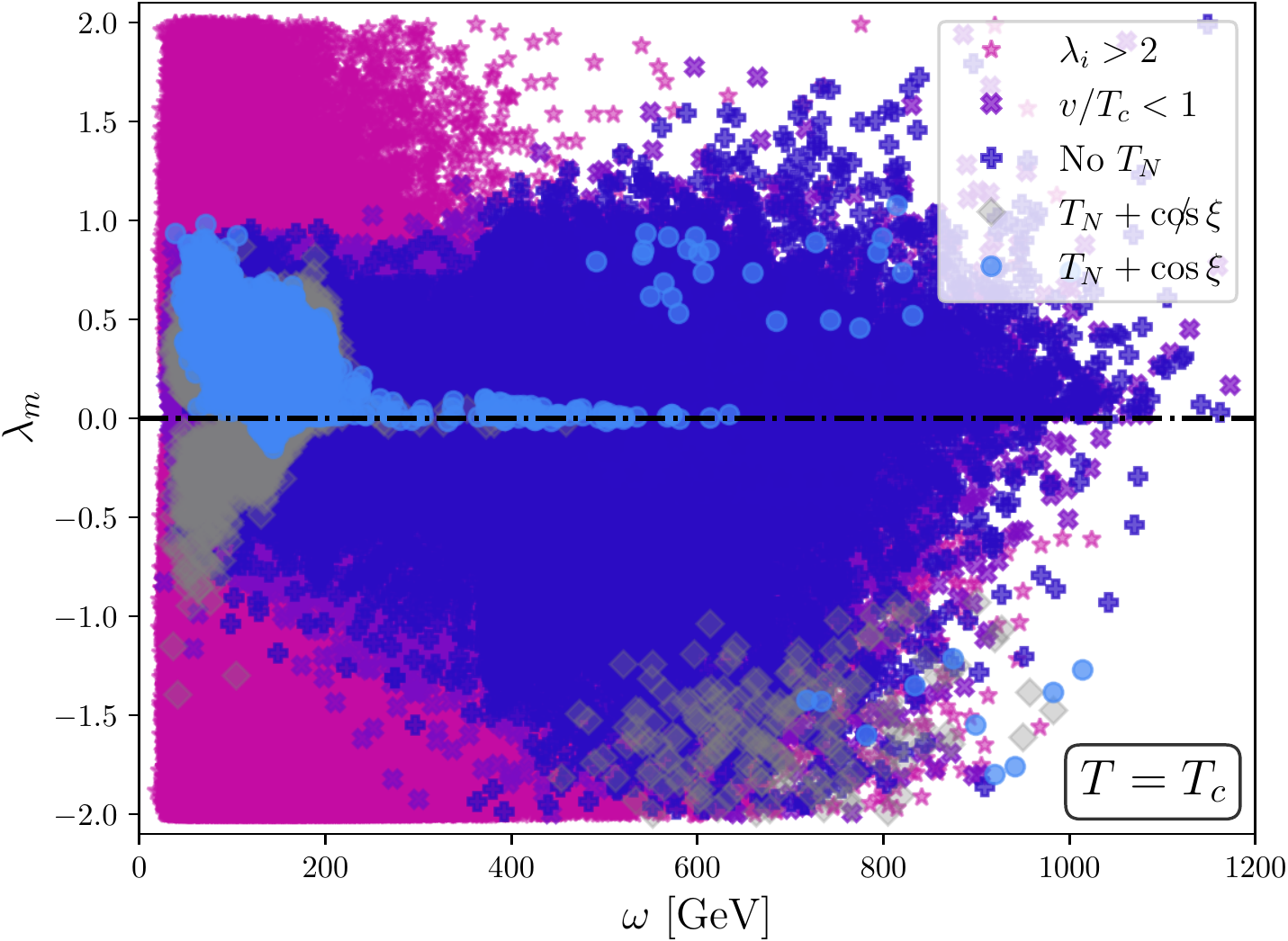}
    \includegraphics[width=0.49\textwidth]{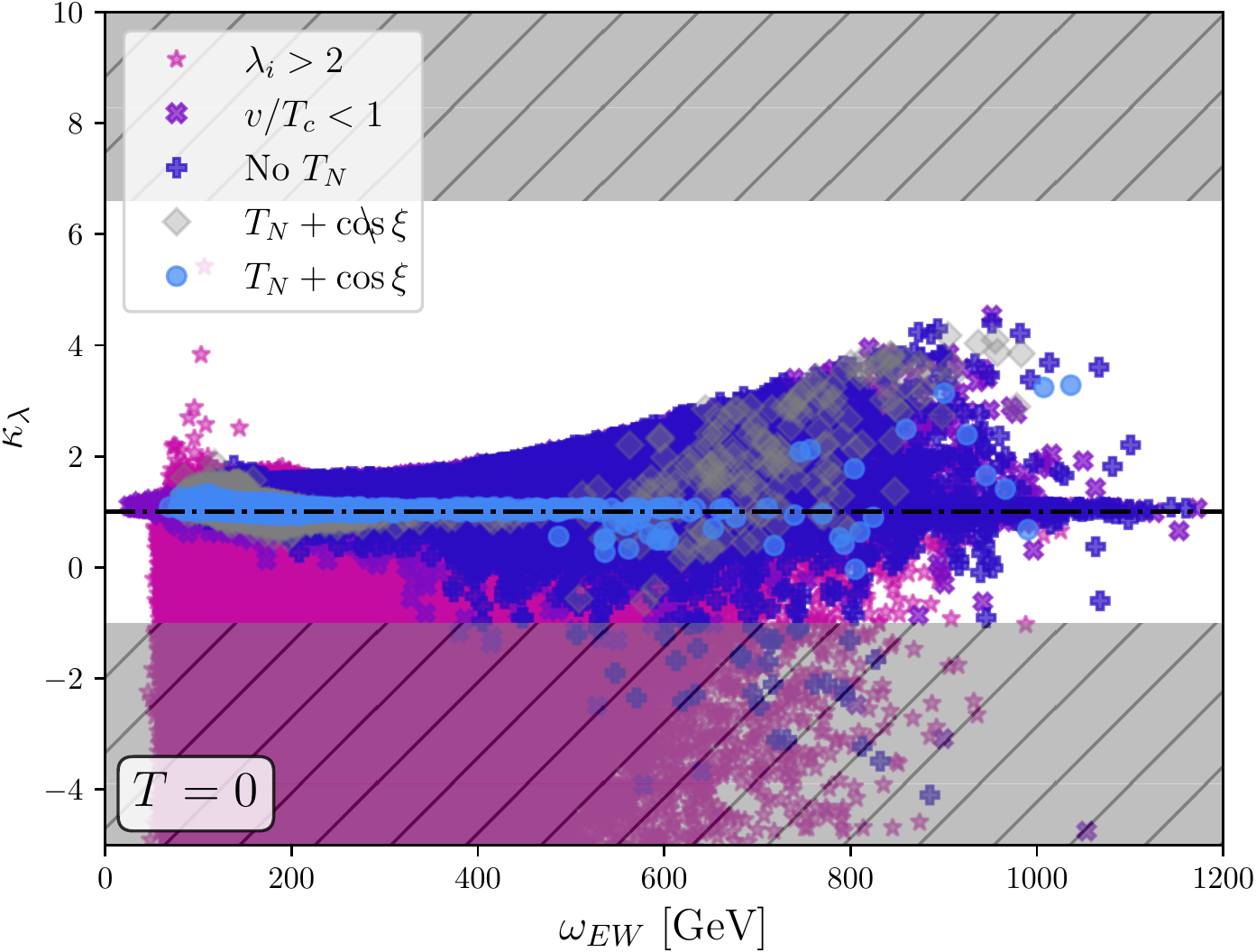}
    \caption{Correlations between different parameters in the scan with non-zero neutrino Yukawa coupling. Allowed regions are very similar and larger applicable for the scan without neutrino Yukawas. The color coding is the same as in Fig.~\ref{fig:Singlet_mass_vs_mixing} with the grey diamonds indicating the parameter space which can undergo successful nucleation but are excluded by their value of the scalar mixing $\xi$. In all panels the parameter represented ar at $T=T_c$ except for the bottom-right panel with the trilinear Higgs coupling for $T=0$.}
    \label{fig:Different_PS_slices}
\end{figure}

Through the comparison between the left and right panels of Fig.~\ref{fig:Singlet_mass_vs_mixing}, we can study the impact of the presence of heavy Dirac neutrinos. While we find new nucleating regions characterized by large $\omega$ and sizeable and negative $\lambda_m$ (as can be seen in Fig.~\ref{fig:Triangle_plot_Comparison}), these are largely ruled out by the constraints from Higgs signal strength measurements, as shown in  Fig.~\ref{fig:Triangle_plot}. Indeed, as shown in the upper-left panel of Fig.~\ref{fig:Different_PS_slices}, the points that pass the criteria for the case including the heavy neutrinos cluster at small values of $\mathcal{Y}_{N}$ or small $\omega$. The condition that leads to the (hyperbola-like) correlation shown in this panel is the requirement of the stability of the EW broken minimum imposed in Eq.~\eqref{eq:stable}, in particular that $m_s^2 >0$ at $T=0$. Since $\mathcal{Y}_{N}$ induces a negative evolution of $m_s^2$ from $T_c$ down to $T = 0$, values of $\mathcal{Y}_{N} > 1$ are constrained in the scan unless $T_c$ is small and/or $\lambda_m$ is negative to cancel its effect in Eq.~\eqref{eq:Running_consts}. This can be seen explicitly in the corresponding panels of Fig.~\ref{fig:Triangle_plot}. Such a cancellation is however prevented by the bounds on the scalar mixing, as indicated by the grey points in Fig.~\ref{fig:Triangle_plot}. Thus, from now on, we will only present the results of our scan with $\mathcal{Y}_{N} \neq 0$, i.e. in the presence of heavy Dirac neutrinos. However, the allowed regions should also be considered generally valid for the $\mathcal{Y}_{N}=0$ scenario without the heavy neutrinos, with the caveat that direct scalar searches at LEP and ATLAS further constrain the parameter space.
%Our finding that the presence of extra neutrinos with sizable $\mathcal{Y}_{N}$ hinders bubble nucleation is in contrast to the preliminary conclusion of Ref.~\cite{Cline:2009sn}, where the ratio $v/T_c$ was used to gauge successful SFOPT points. 
%We have verified that our parameter scan does agree with their results, which is that $\mathcal{Y}_{N}$ increases $v/T_c$ and makes the phase transition stronger, but it turns out to also hinder bubble nucleation by increasing the potential barrier between the two degenerate minima. \SR{Is this true? $\mathcal{Y}N$ enters into the $s^2$ term, basically increasing or decreasing the singlet vev (very roughly speaking no? It goes like $v^2\sim \mu^2/\lambda$). If $\mu_s^2$ turns out to be generally negative, larger $YN$ make it even larger in absolute value and then the singlet vev goes further and further away in configuration space so you're cooked. Probably the height of the barrier is also larger but I don't see this so clearly.}
%\TO{I see, but can we see this in the $\omega$-$\mathcal{Y}_{N}$ plot? ---hmm, larger $\mathcal{Y}_{N}$ seem to make $\omega$ smaller (both dark blue and light blue)...} \EFM{For me it is just a ``phenomenological'' observation. Because you said you find a smaller fraction of nucleating points when $Y_N \neq 0$ and because in the plots the light blue points cluster around small values of $Y_N$}

We present the distribution of scan points in various other interesting slices of the parameter space in Fig.~\ref{fig:Different_PS_slices}. 
For information on the distributions under the other parameters of the scan, we refer the reader to Fig.~\ref{fig:Triangle_plot}.
The color coding and symbols is the same as for Fig.~\ref{fig:Singlet_mass_vs_mixing}, but in addition to them, we now have the grey diamonds (labeled with ``$T_N+\Slash{\cos{\xi}}$'') indicating the parameter space points which can undergo successful nucleation but are excluded by their value of the scalar mixing $\sin{\xi}$. They correspond to the blue dots covered by the grey-shaded area in the right panel of Fig.~\ref{fig:Singlet_mass_vs_mixing}. In the upper right panel of Fig.~\ref{fig:Different_PS_slices} we show the correlations found in our scan between the two scalar vevs at $T=T_c$. As can be seen, the most significant constraint is the requirement of a sufficiently strong EW phase transition, $v/T_c > 1$. 
When imposing this together with $v_{EW}= v_{EW}^\mathrm{exp}$ and $M_H=M_H^\mathrm{exp}$, values of $v$ below $\sim 100$~GeV are ruled out. 
Besides this constraint, we find that large values of $\omega$, beyond $\sim 200$~GeV, are disfavored by the requirement of successful nucleation.
Indeed, we generally find that, if $\omega$ were too large, the field distance between the two minima would be too big to allow for bubble nucleation despite satisfying the rest of the requirements. Even though we find regions of parameter space successfully nucleating for singlet vevs as large as $\omega\sim 1000$~GeV at $T_c$, a detailed study of these regions shows that these transitions at $T_N$ occur from $(0,\omega-\Delta\omega)\rightarrow(0,\omega)$, with $\Delta\omega\ll \omega$, such as the distance travelled in field space is not qualitatively larger than for the region with $\omega\lesssim 200$~GeV. 

In the lower left panel of Fig.~\ref{fig:Different_PS_slices} we show the distribution of scan points in the $\omega-\lambda_m$ plane at $T=T_c$. We find that the light-blue dots, which pass all requirements and in particular the nucleation condition, display an anticorrelation between these two parameters.
Additionally, the bounds on $\sin{\xi}$ rule out most of the points with $\lambda_m<0$ unless $|\lambda_m|\ll 1$. 
These trends can be understood from the hyperbolic shape of the correlation between $\lambda_m$ and $\omega_p$, which is found in the corresponding panel of Fig.~\ref{fig:Triangle_plot}. 
Indeed, from Eq.~(\ref{eq:omegap}) this behavior is expected if $\mu_m$ is negative. 
Analyzing the accepted samples, we find that negative $\mu_m$ is preferred in order to satisfy our condition for successful nucleation. 
In fact, negative $\mu_m$ decreases the barrier between the two degenerate minima and thus we find no nucleating samples for positive $\mu_m$. Finally, the area with negative $\lambda_m<0$ and negative $\omega_p$ is ruled out by the constraints on the scalar singlet mixing since, as expected from Eq.~(\ref{eq:omegap}), $-m^2_{sh}$ would become too large.

\vspace{1mm}

\begin{figure}
    \centering
    \includegraphics[width=0.6\textwidth]{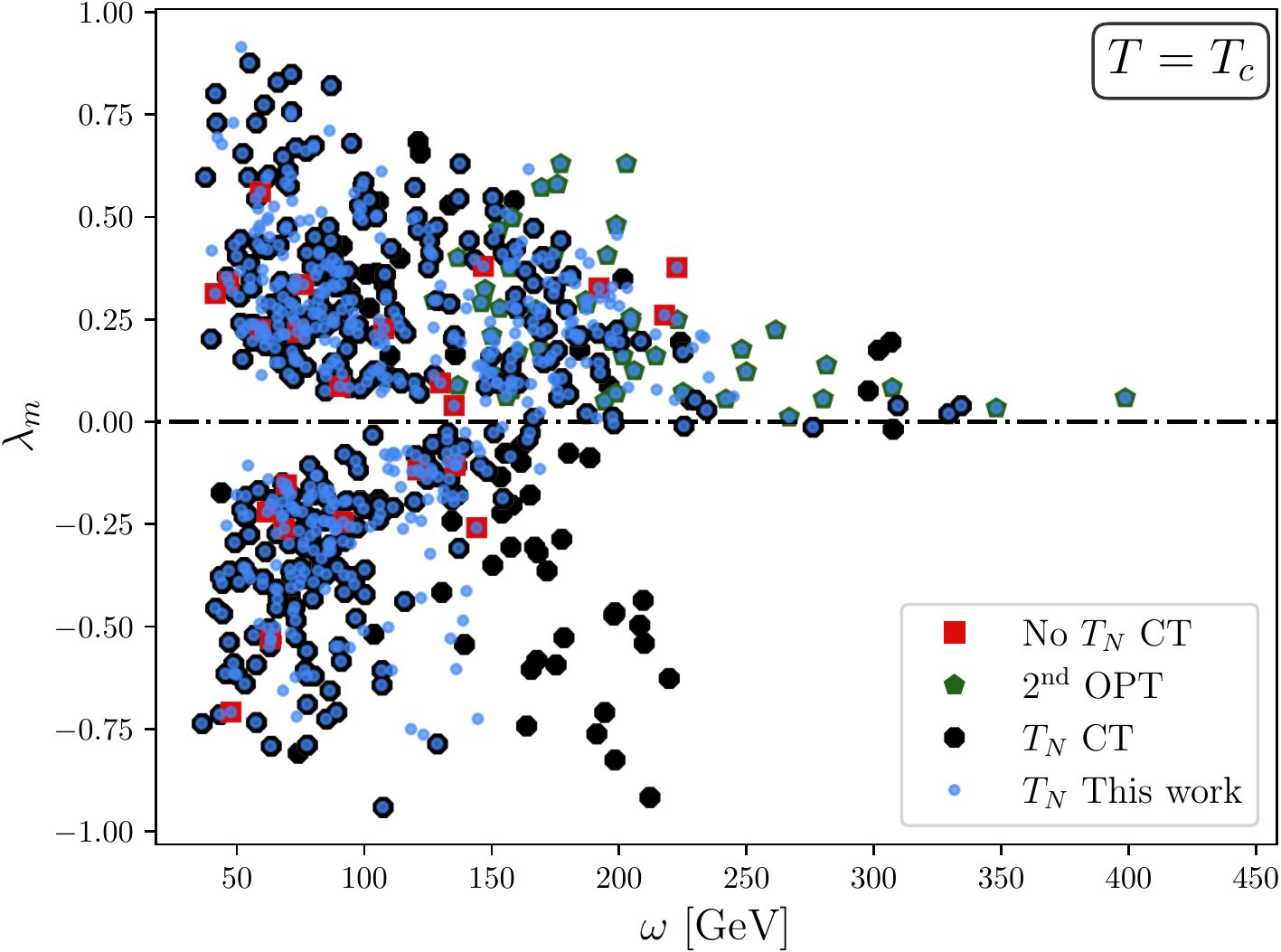}
    \caption{Sample comparison between the results obtained with our proxy (light blue circles) and {\tt CosmoTransitions} (CT) in the $\omega-\lambda_{m}$ plane for successfully nucleating points. According to CT, red squares correspond to non-nucleating points, green pentagons to those giving a $2^{\rm{nd}}$ order phase transition (OPT) and black octagons to successfully nucleating points.}
    \label{fig:Comparison_Proxy_CT}
\end{figure}

We have further analyzed the non-trivial correlation found between $\omega$ and $\lambda_m$ when imposing our criteria for nucleation, comparing these results with points that successfully nucleate according to the %output 
\textit{tunneling module} from {\tt CosmoTransitions}. As can be seen in Fig.~\ref{fig:Comparison_Proxy_CT}, for a subset of our sample featuring successful nucleation, the areas found by both our approximate estimate %for nucleation 
(light-blue points) and {\tt CosmoTransitions} (black octagons for a first-order EW phase transition) generally agree well. The two exceptions we identify are: {\textit (i)} the region with negative $\lambda_m$ and large $\omega$, where {\tt CosmoTransitions} finds successfully tunneling points 
%which feature a first-order transition that is 
which are not found by our approximation.\footnote{For positive $\lambda_m$ and large $\omega$, a few of the light-blue points yielding successful nucleation with our criteria are instead tagged as second-order phase transitions ($2^{\rm{nd}}$OPT, green pentagons) by {\tt CosmoTransitions}. While we have not explicitly discerned the order of the transition in our scan (which is beyond our present scope), we note that no qualitative new regions appear when considering such parameter points, as this region falls within the areas where a SFOPT is found by our method.}
This region %is found to 
corresponds to significantly more curved trajectories than those depicted in Fig.~\ref{fig:Contour_plotsTN}, not-well approximated by our straight-line assumption. Nevertheless, this whole area of the parameter space also leads to too large scalar singlet-doublet mixing (as described above) and is experimentally excluded. 
%For positive $\lambda_m$ and large $\omega$, some of the points found by our criteria for successful nucleation are instead tagged as second order phase transitions ($2^{\rm{nd}}$OPT) by {\tt CosmoTransitions} (green pentagons). Indeed, we did not explore the order of the transition in our scan and it is beyond the scope of this study. Nevertheless, no qualitative new regions appear when considering the possibility of a $2^{\rm{nd}}$OPT. Indeed, this region falls within the areas where a SFOPT is found.
{\textit (ii) The points for which {\tt CosmoTransitions} does not find an EW phase transition, whereas our nucleation proxy does, i.e.~the red squares and the blue points with no counterpart (neither red square, green pentagon or black octagon) in Fig.~\ref{fig:Comparison_Proxy_CT}. This should a priori never happen, since our criterium for nucleation is conservative. A careful investigation of such points reveals that the {\textit phase-tracking module} of {\tt CosmoTransitions} does not produce numerically reliable results}
%
%The red squares in Fig.~\ref{fig:Comparison_Proxy_CT} instead correspond to scalar potentials for which {\tt CosmoTransitions} does not find, a priori, an EW phase transition, whereas our nucleation proxy does. Even though very few points in the sample belong to this category, this should a priori never happen, since our criteria for nucleation is rather conservative. A more careful investigation of these points reveals that the result of {\tt CosmoTransitions} is not numerically reliable 
in such cases.\footnote{Even if by construction two phases are always present for our model parameter points, in these cases {\tt CosmoTransitions} fails to find one of them for the default numerical precision in the code. A significant increase in the numerical precision generally leads to {\tt CosmoTransitions} finding the second phase and identifying a first-order transition, in agreement with our estimate. Nevertheless, this increase in numerical precision makes the computation too slow to allow for an efficient scan of the parameter space.} 
%despite being selected by our proxy. Even though very few points in the sample belong to this category, this should a priori never happen, since our criteria for nucleation is rather conservative. We have therefore investigated more carefully these points and found that, even if by construction two phases were present, {\tt CosmoTransitions} failed to find one of them. Increasing the precision of the calculation in {\tt CosmoTransitions}, the second phase was eventually found for $\sim 50 \%$ of the points and the transitions tagged as first order, in agreement with our estimation. Nevertheless, this increase in precision makes the numerical computation too slow to allow for an efficient scan of the parameter space. 
%
%Finally, light-blue points with neither red square, green pentagon or black octagon counterpart correspond to cases in which {\tt CosmoTransitions} fails to produce any result for that point in the parameter space. 
%
We thus conclude that neither of these exceptions is meaningful, and our estimate for nucleation agrees well with the results from {\tt CosmoTransitions} for the values of the parameters for which {\tt CosmoTransitions} yields a reliable numerical result, thus representing an efficient and fast alternative for scans of the parameter space discriminating in a conservative way if nucleation could happen. 

% while {\tt CosmoTransitions} additionally finds a larger region for negative $\lambda_m$ and large $\omega$ with successful nucleation. This region corresponds to significantly more curved trajectories than those found in Fig.~\ref{fig:Contour_plotsTN}, not-well approximated by our straight line assumption, that avoided this increased barrier. Nevertheless, this whole area of the parameter space characterized by large $\omega$ also leads to too large scalar singlet mixing as described above and is finally excluded (yellow crosses and grey diamonds).
% \TO{I am not sure if we should step deep into the comparison between our proxy and CT here. Maybe here we say, "we will compare our approximated calculation of nucleation with the full numerical calculation with CT, in the distribution of the points in some parameter plane" and do it in Appendix?} \EFM{I think the comparison is interesting and would keep it here but updating to the new plot and changing the discussion accordingly.}

\vspace{2mm}

As discussed in Sec.~\ref{sec:pheno_singlet}, the inclusion of the singlet scalar causes a deviation of the Higgs trilinear coupling from its SM value, which can be parametrized as in Eq.~\eqref{eq:Kappa_lambda}. 
The distributions of $\kappa_{\lambda}$ as a function of $\omega_{EW}$ at $T=0$ is shown in the bottom-right panel of Fig.~\ref{fig:Different_PS_slices}, together with the current bounds from collider searches, which are given by Eq.~(\ref{eq:Kappa_lambda_bound}) and shown by the grey-shaded regions. 
%In the regions of parameter space satisfying all conditions from Ref.~\cite{Espinosa:2011ax} together with the correct Higgs mass and vev (all accepted points regardless of their color) , $\kappa_{\lambda}$ takes a value in the range of $-10 \lesssim \kappa_{\lambda} \lesssim 4$.
We find that the light-blue points satisfying all conditions tend to cluster in a narrow range around $\kappa_{\lambda}\sim 1$. 
%Although some of the allowed points predict values of $\kappa_{\lambda}$ deviating from the SM one, they are well beyond the reach of future collider precision. 
%
%Thus, a future discovery (e.g. at the HL-LHC, sensitive to $\kappa_\lambda \sim 2.2$~\cite{Cepeda:2019klc}) of significant deviations of the Higgs trilinear coupling from the SM value would \textbf{point towards new degrees of freedom other than the singlet scalar, even if the latter is responsible for the SFOPT.}
Thus, given the sensitivity to $\kappa_\lambda \sim 2.2$~\cite{Cepeda:2019klc} of future probes such as the HL-LHC, no deviations caused by a singlet scalar responsible for a SFOPT are to be expected in this observable.

\begin{figure}
    \centering
    \includegraphics[width=0.495\textwidth]{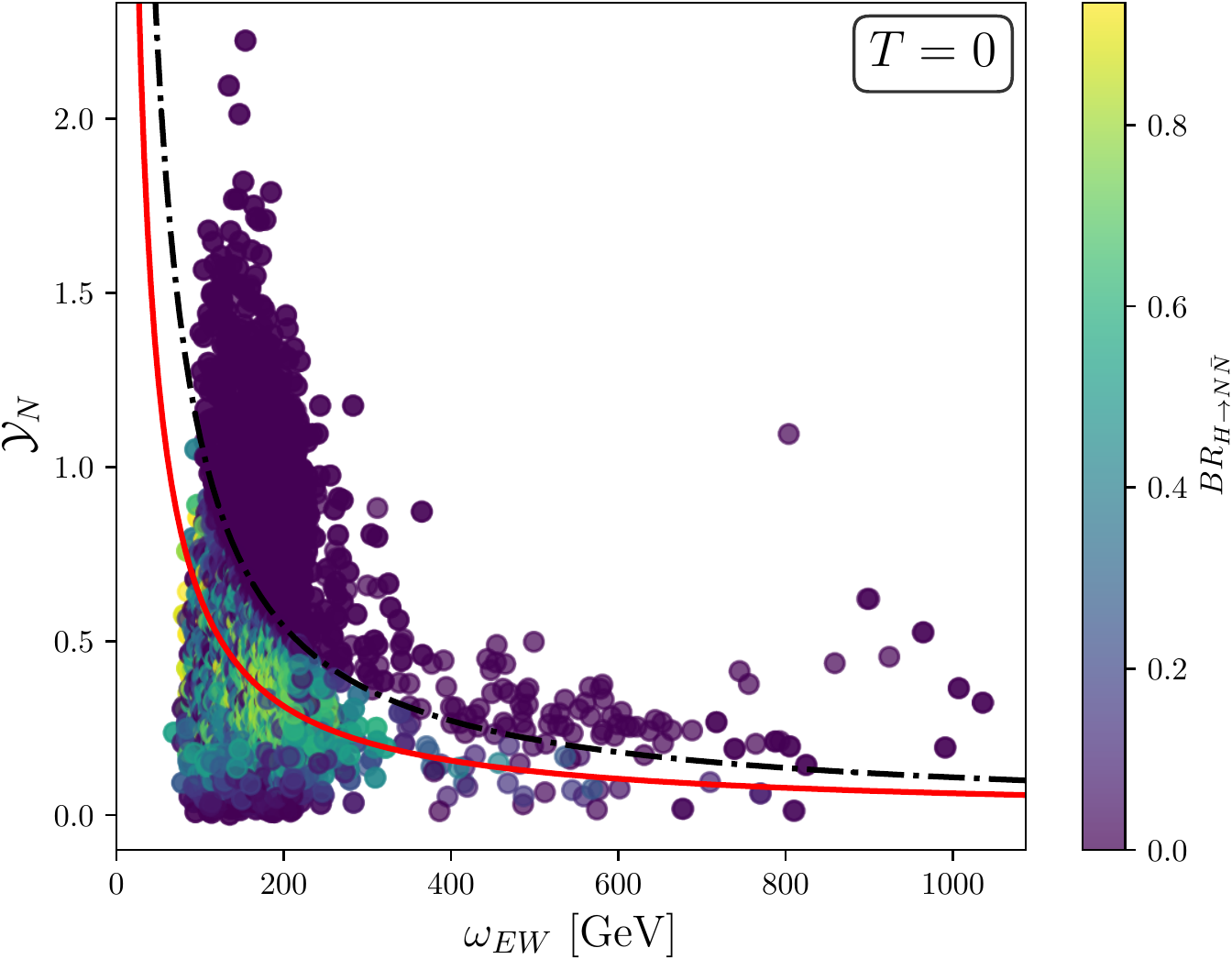}
    \includegraphics[width=0.495\textwidth]{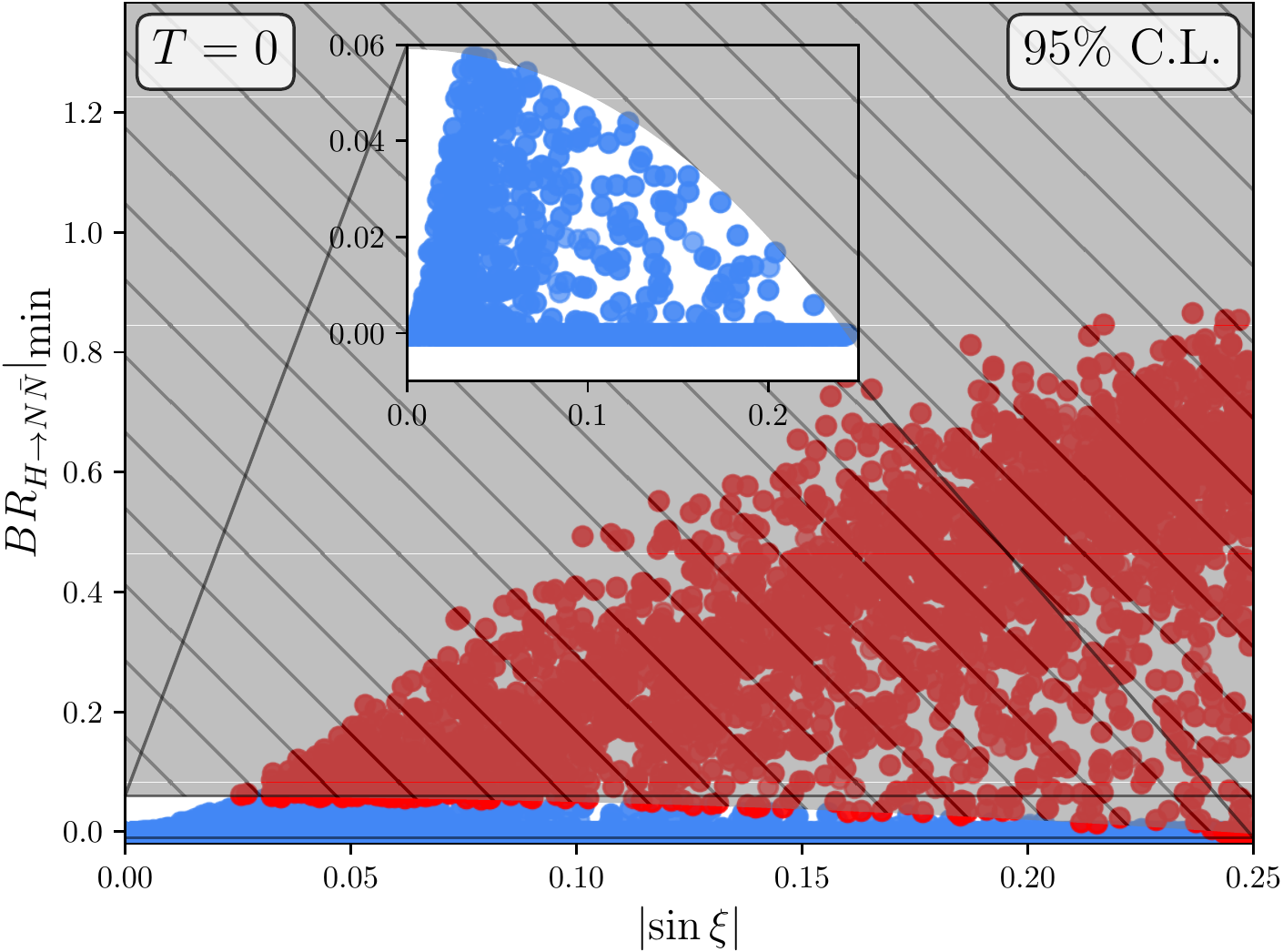}
    \caption{Bounds on the regions of parameter space giving rise to successful nucleation. The left panel correspond to the plane $\mathcal{Y}_N$ against $\omega_{EW}$ at $T=0$, with the color legend giving the size of ${\rm BR}_{H\rightarrow N\bar{N}}$ for $n=3$ degenerate heavy neutrinos. The solid red line corresponding to $\, \omega_{EW}\mathcal{Y}_N  = (M_H/2)$ separates the cases where the decay is always allowed (below) and where it depends strongly on the assumptions on the neutrino yukawas $Y_{N_i}$ (above). The dashed black line yields the boundary of kinematically allowed $H\rightarrow N_i\bar{N_i}$ decay for the $n= 3$ degenerate neutrino case. The right panel shows the excluded parameter space at $95\%$~C.L. from Higgs signal strength measurements, in which red points are excluded while blue ones comply with the bounds, %for points with $\omega_{EW}\mathcal{Y}_N<M_H/2$ 
    and is independent of $n$.}
    \label{fig:bounds_BR}
\end{figure}

Finally, in Fig.~\ref{fig:bounds_BR} we show the regions of the parameter space which are constrained by the possible new decay channel of the Higgs-like state $H$ into heavy neutrinos using Eqs.~(\ref{eq:Higgs_Signal_H_NN}) and (\ref{eq:total_DR_hN}) as described in Section~\ref{sec:Higgs_Signal_Strengths}, for points which pass all the constraints (i.e. light-blue in Fig.~\ref{fig:Different_PS_slices}) in our parameter scan. 
%undergo a SFOPT and are allowed considering the experimental constraints in the absence of the neutrino Yukawa coupling with the singlet. 
In the left panel we show the contribution to the corresponding branching ratio assuming a degenerate heavy neutrino spectrum ($Y_{N_i}^2=\mathcal{Y}_N^2/n$) with $n=3$, for which $\Gamma_{H \to N \bar{N}}$ can be comparable to or even exceed the SM Higgs boson total width $\Gamma_{\mathrm{SM}}$ in an important part of the parameter space. The solid red line separates ``Case 1'' and ``Case 2'' as discussed in Section~\ref{sec:Higgs_Signal_Strengths}.
Note that in the region above the solid red line (``Case 2''), a different combination of $Y_{N_i}$ Yukawa couplings (yielding the same value of $\mathcal{Y}_N^2$) could arbitrarily reduce the value of BR$_{H\rightarrow N\bar{N}}$ by making all neutrinos either too heavy for the Higgs to decay into or with negligible couplings. From this panel we can also infer that the heavy Dirac neutrinos are in general lighter than $\sim 300$~GeV.
In the right panel of Fig.~\ref{fig:bounds_BR} we instead show the minimum possible value of BR$_{H\rightarrow N\bar{N}}$ for each parameter point. 
Notice that for the points corresponding to ``Case 1'' (region below the solid red line in the left panel) the exclusion limits from Higgs signal strength measurements, shown in grey, are unavoidable and rule out a significant region of the parameter space, while for ``Case 2'' the BR$_{H\rightarrow N\bar{N}}$ can be made arbitrarily small and thus the bound can always be evaded. In Fig.~\ref{fig:Triangle_plot} we show in red the points excluded by $BR_{H\rightarrow N\bar{N}}$ in the different relevant planes in parameter space. Even though these constraints are important, as seen in Fig.~\ref{fig:bounds_BR}, they do not exclude particular regions of parameter space. 

%\JLP{Salva, in the right panel you consider the minimum in case A right?}
%\TO{As Jacobo said, in Case 1, $\Gamma_{\text{max}}$ Eq.~\eqref{eq:Gamma_max-Case1} cannot take $n \geq 3$,
%because $n=3$ leads $\mathcal{Y}_{N}^{2} \omega_{EW}^{2} = 3 M_{H}^{2}/10 > M_{H}^{2}/4$ and is contradict with the condition of Case 1. 
%Then, we take $n=2$ for max $\Gamma$?}

\section{Summary and Conclusions}
\label{sec:summary}

In this work we have explored the parameter space of the scalar singlet extension of the SM with the aim of identifying the regions in which a SFOPT, as required to explain the puzzle to the origin of the observed baryon asymmetry through the EWBG mechanism, can take place. The main goal of the study is to contribute to the predictability of the scenario by relating the areas where a SFOPT can happen with potentially testable observables or correlations among them. 

Previous studies~\cite{Espinosa:2011ax} showed the conditions that need to be met by the extended scalar potential in order to develop two degenerate minima at some critical temperature $T_c$. Together with the requirement of reproducing the correct mass and vacuum expectation value of the SM-like Higgs and of the required strength of the transition ($v/T_c > 1$), these set of constraints already impose stringent and non-trivial conditions of the allowed parameter space. 

Nevertheless, as advocated by~\cite{Carena:2022yvx,Biekotter:2021ysx,Baum:2020vfl,Kozaczuk:2019pet,Chen:2017qcz}, we find that the requirement that bubble nucleation may actually take place between the two minima is the most constraining requirement, reducing drastically the allowed parameter space. Furthermore, testing explicitly this condition is not possible in a fast and analytical way and relying on the numerical solvers available~\cite{ Wainwright:2011kj,Athron:2019nbd,Guada:2020xnz} necessarily limits the speed of the scan hindering the exploration of large parameter spaces. Moreover, given the complexity of the problem, for some points in the parameter space we find that some numerical solvers fail to find one of the phases, and hence the corresponding transition, or are unable to produce a result. 
For this reason we have adopted a fast and conservative approximation to the three-dimensional action of the bounce solution, $S_3^{\text{app}}$, that controls the transition rate between the two minima and requires $S_3^{\text{app}}/T_N\sim140$ at some nucleation temperature $T_N$. We find that for most of the sampled points in the parameter space that satisfy this criteria, {\tt CosmoTransitions} does indeed find a first order phase transition (with a small fraction of second order transitions, something we did not explicitly discriminate), thus validating our approach. We also point out that, for many of the points that passed our selection criteria, {\tt CosmoTransitions} did not provide an output. Thus, larger regions of the parameter space may be explored in a fast and efficient way through the approximation adopted, although it should be taken as a conservative estimate and not as an exact result. 

In our scan of the parameter space we find that the regions with the correct mass and vev for the Higgs and successful nucleation are mainly characterized by values of the singlet vev $\omega_{EW} \lesssim 300$ GeV. Indeed, if $\omega_{EW}$ is too large, the two minima tend to be too far apart in field space\footnote{Without loss of generality, exploiting the shift symmetry of the potential, we choose the value of the singlet vev at the EW symmetric minimum to also vanish at the critical temperature.} and nucleation may not happen. This in turn translates into values of the scalar singlet mass that cluster around $M_S\lesssim 300$~GeV. The exception to this rule is a clustering of allowed points with large $\omega$ and values of $M_S$ in the 500-1000~GeV range, which in any case do not produce an EW phase transition given that $v(T)$ smoothly goes from 0 to $v_{EW}$ as the Universe expands. We have verified that for these points the actual jump in $\omega$ during the phase transition is also small. Regarding the most constraining observables, we find that the bounds on the singlet-doublet mixing from Higgs signal strength measurements by ATLAS and CMS are already ruling out important regions of the parameter space. Direct searches for the singlet scalar when its decays are SM Higgs-like both at LEP and at LHC are also relevant. 

We have also investigated how this picture is affected when the scalar singlet is not alone, but part of larger dark sector it may interact with. As a particularly motivated scenario, we considered as case study the addition of extra sterile neutrino singlets of both chiralities. These new states will have Yukawa couplings $Y_{N}$ to the scalar singlet, which would induce Dirac masses around the EW scale for these heavy neutral leptons. Furthermore, a Yukawa coupling $Y_{\nu}$ among the SM Higgs doublet, the SM neutrinos and the heavy neutrinos would generally also be allowed. The simultaneous presence of $Y_N$ and $Y_\nu$ implies a new source of CP-violation that may be enough to induce the BAU via EWBG~\cite{Hernandez:1996bu, Fernandez-Martinez:2020szk} (the so-called $\nu$-EWBG scenario~\cite{Fernandez-Martinez:2020szk}). Furthermore, if a small source of lepton-number violation is introduced, the presence of $Y_\nu$ would induce small neutrino masses able to explain the neutrino oscillation phenomenon in the manner of the low-scale symmetry-protected seesaws like the inverse or linear seesaw variants.

\begin{figure}
    \centering
    \unitlength=1cm
    \begin{picture}(18,19)
    \put(-2,0){\includegraphics[width=19cm]{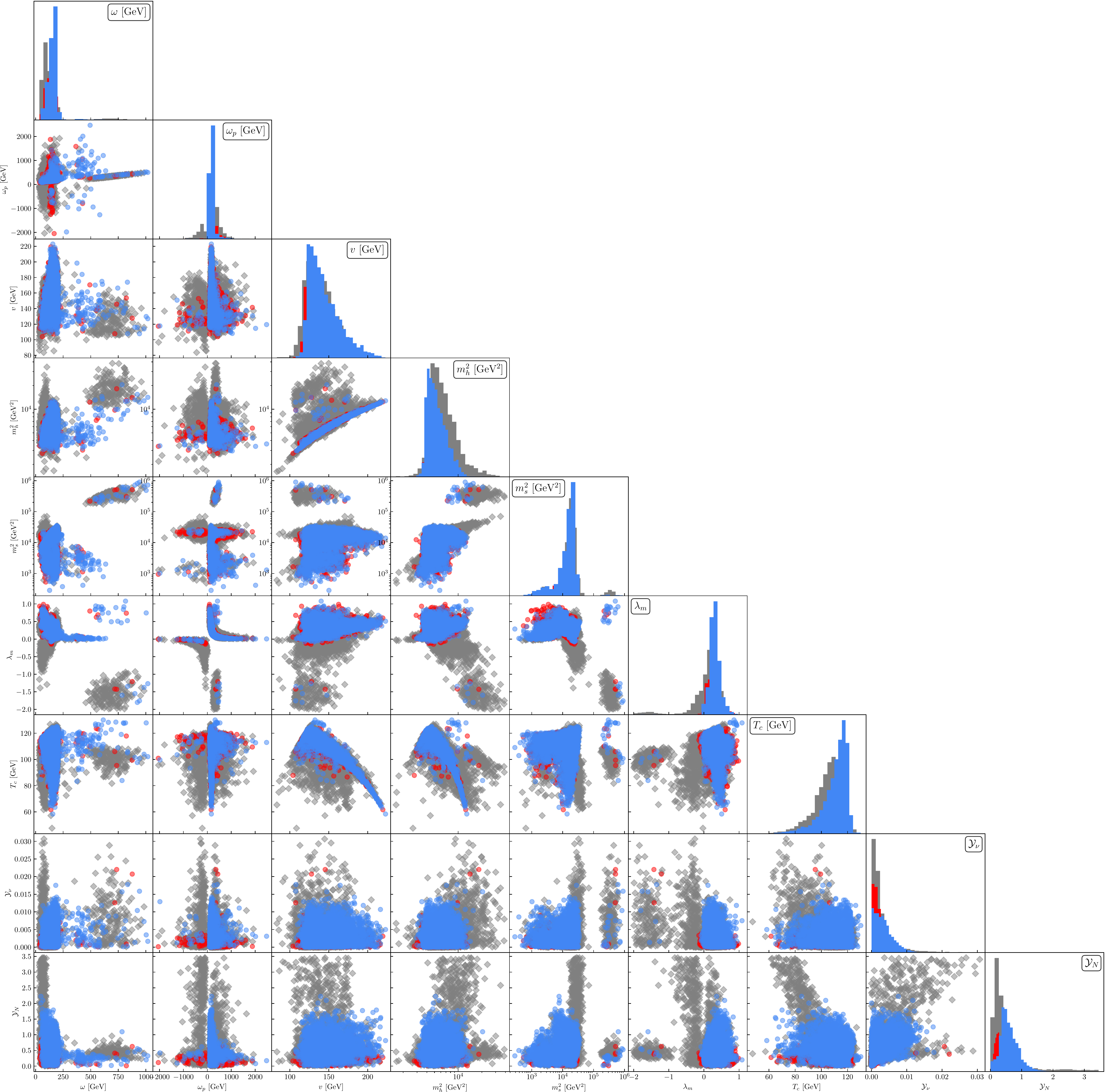}}
    %\put(0,19){\circle*{0.5}}
    \end{picture}
    \caption{Triangle plot for non-zero Yukawas, only for points successfully nucleating. The grey diamonds correspond to regions of parameter space excluded by constraints on the scalar mixing while red dots are excluded by the constraint from $H \rightarrow N\bar{N}$. The blue points satisfy all phenomenological bounds. We note the very strong correlation between the singlet vev $\omega$ and $\omega_p$ at $T_c$ for blue points, as well as between $\lambda_m$ and $\omega_p$.}
    \label{fig:Triangle_plot}
\end{figure}

\begin{figure}
    \centering
    \unitlength=1cm
    \begin{picture}(18,19)
    \put(-2,0){\includegraphics[width=19cm]{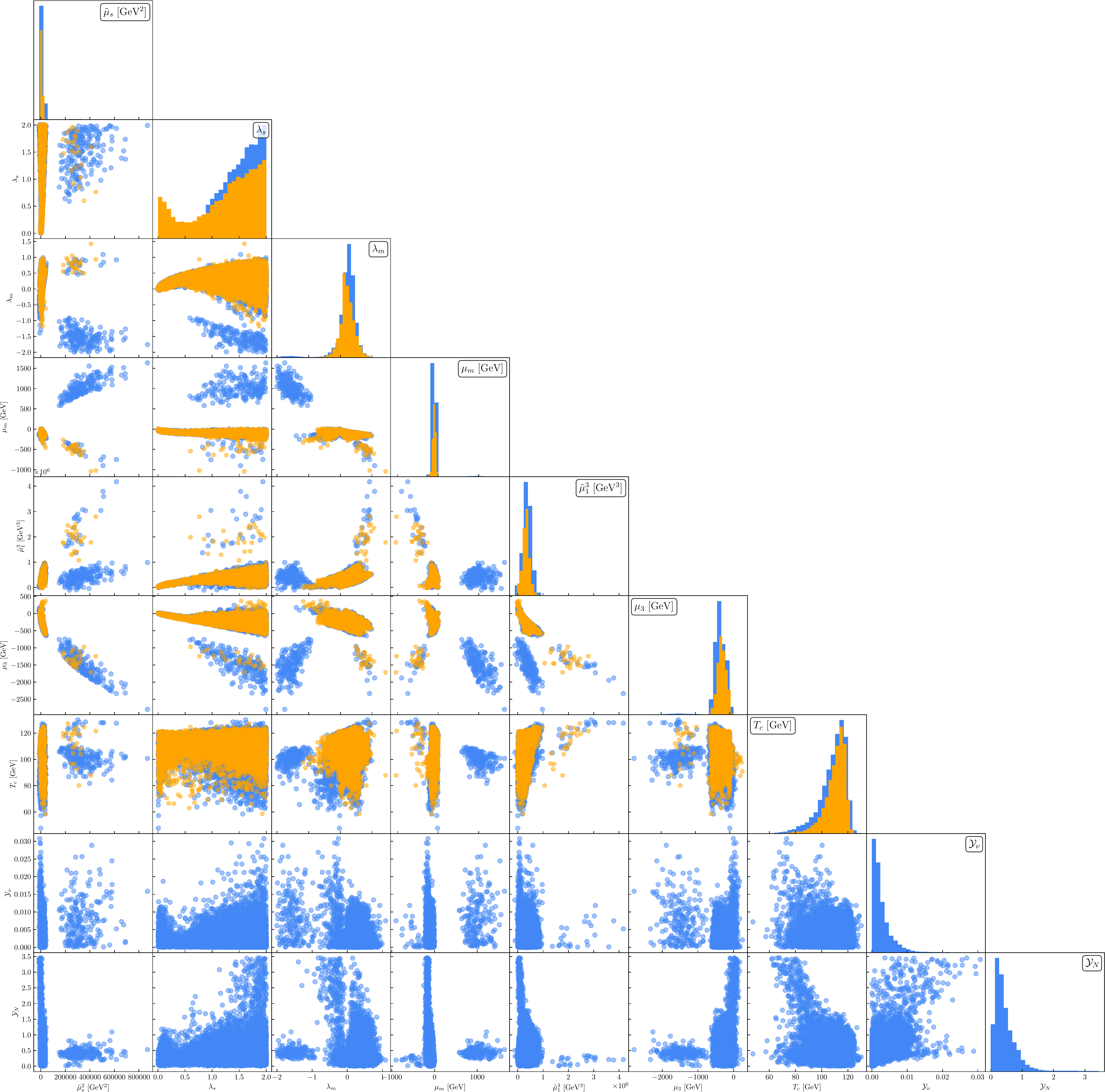}}
    %\put(0,19){\circle*{0.5}}
    \end{picture}
    \caption{Triangle plot comparing the successfully nucleating points for the singlet scalar alone (orange pentagons) and including the neutral fermions (blue dots), without imposing any phenomenological bound, at $T=0$ for the parameters defined in Eq.~(\ref{eq:ScalarPot1}). In the singlet-scalar-only case we have $\mathcal{Y}_{\nu(N)}=0$ and it is not shown in the plot.}
    \label{fig:Triangle_plot_Comparison}
\end{figure}

Previous studies~\cite{Cline:2009sn} showed that the presence of the heavy neutrinos increases the strength of the transition by enhancing $v/T_c$. We reproduce this result, but find that sizable $Y_N$, unless accompanied by small $\omega$,
can also destabilize the broken minimum. Thus, when our criteria for nucleation and stability are imposed, for the allowed values of the Yukawa couplings the regions of the scalar potential parameter space are comparable to the scenario without heavy neutrinos. Hence the early universe phenomenology regarding the possibility of a SFOPT of both scenarios is very similar, as summarized in Figs.~\ref{fig:Triangle_plot} and \ref{fig:Triangle_plot_Comparison}. Even if new areas appear when including the neutrinos, we observe in Fig.~\ref{fig:Triangle_plot} that they are excluded by Higgs signal strength measurements. The small values of the Yukawa couplings and the scalar singlet vev $\omega$ preferred, seem to make the generation of the BAU via EWBG difficult according to the findings of~\cite{Fernandez-Martinez:2020szk}, but a dedicated analysis would be required to confirm or rule out its viability. 

Conversely, the presence of the heavy sterile neutrinos may significantly affect the collider phenomenology of the scalar singlet extension. Indeed, while very large values of $Y_N$ could hinder vacuum stability%bubble nucleation
, values around $Y_N \sim \mathcal{O}(1)$ are perfectly allowed. Such a sizable coupling would on the one hand imply that the scalar singlet decays would be overwhelmingly dominated to the heavy sterile neutrino channel, given that the singlet-doublet mixing is more strongly constrained and the smaller SM Yukawa couplings. This would in turn invalidate the bounds on the scalar mixing derived from direct searches of the singlet with SM-like decays at LEP and LHC. On the other hand, dedicated searches for this new decay channel should be considered. 

Furthermore, if allowed by phase space, the decay to heavy sterile neutrinos could also be sizable for the SM-like Higgs scalar via its mixing with the singlet. We have found that this in fact strengthens the Higgs signal strength constraints in significant portions of the parameter space, corresponding to the red points in Fig.~\ref{fig:bounds_BR}. Interestingly, the possibility that this is the dominant channel to produce and test for the heavy neutrinos at collider searches also remains open in parts of the parameter space. Indeed, the mixing of the heavy neutrinos with their active counterparts induced by $Y_\nu$ is more strongly constrained from flavour and electroweak precision observables, as well as collider searches via Drell-Yan production. Thus, if $Y_N$ is more sizable than $Y_\nu$, the heavy neutrinos would be more easily produced via Higgs or singlet decays. For small enough values of $Y_\nu$, the decays of the heavy neutrinos themselves would not be prompt and may induce interesting signatures with displaced vertexes. We thus find that the viable parameter space allows for very striking and non-standard collider phenomenology which will be interesting to pursue in future dedicated studies.

\section*{Acknowledgments}
We acknowledge very helpful discussions with Josu Hernandez-Garcia and Daniel Naredo-Tuero. We also thank Jose Ram\'on Espinosa for invaluable discussions regarding tunneling.
The work of J.M.N. was supported by the Ram\'on y Cajal Fellowship contract RYC-2017-22986, and by grant PGC2018-096646-A-I00 from the Spanish Proyectos de I+D de Generaci\'on de Conocimiento. E.F.M., J.M.N and T.O. acknowledge partial financial support by the Spanish Research Agency (Agencia Estatal de Investigaci\'on) through the grant IFT Centro de Excelencia Severo Ochoa No CEX2020-001007-S and (E.F.M and T.O.) by the grant PID2019-108892RB-I00 funded by MCIN/AEI/ 10.13039/501100011033. The authors also acknowledge support through the European Union's Horizon 2020 research and innovation programme under the Marie Sklodowska-Curie grant agreements No 860881-HIDDeN and No 101086085-ASYMMETRY. JLP also acknowledges support from Generalitat Valenciana through the plan GenT program (CIDEGENT/2018/019) and the Spanish Ministerio de Ciencia e Innovacion project PID2020-113644GB-I00.

%%%%%%%%%%%%%%%%%%%%%%%%%%%%%%%%%%%%%%%%%%%%%
%\appendix
%\section{Alternative parametrization of the potential}\label{app:pot_param}

%The potential Eq.~\eqref{eq:V-2degmin-Espinosa} 
%can be specified with the following parameters:
%$\{\omega,\omega_{p},\omega_{0},v, m_{h}^{2}, m_{s}^{2}, \lambda_{m}\}$,
%and we provide those parameters at $T_{c}$ that is also 
%a parameter we have to generate in the scan.
%
%$\omega_{0}$ can be set to be zero,
%which corresponds to a shift of $\mu_{1}$, 
%without loss of generality.
%

%To bring the potential down to $T<T_{c}$ to compute 
%the action for the nucleation condition and the electroweak phenomenology,
%we need to include the thermal corrections which 
%contains two more free parameters,
%$\mathcal{Y}_{\nu}$ and $\mathcal{Y}_{N}$.
%
%The parameters we scan are 
%$\{\omega,\omega_{p},v, m_{h}^{2}, m_{s}^{2}, \lambda_{m}, 
%T_{c}, \mathcal{Y}_{\nu}, \mathcal{Y}_{N} \}$.
%
%The coefficients in the ponteital Eq.~\eqref{eq:ScalarPot1}
%at $T_{c}$ are expressed with those input parameters as

%They are related to the coefficients at any temperature lower than $T_{c}$ through Eqs.~\eqref{eq:ScalarPot2} and \eqref{eq:Running_consts},
%with which we can compute the actions for the nucleation condition
%and various phenomenological observables.

\appendix

\section{Parameter scan with weight function}\label{appendix}

We describe in this section the ad-hoc weight function used for the parameter scan 
and also show the distribution of the parameter points that satisfy the nucleation condition in the full parameter space at Fig.~\ref{fig:Triangle_plot}.

A set of necessary conditions for a successful SFOPT were identified, analytically, in Ref.~\cite{Espinosa:2011ax}, and can be found in Table~1 therein. 
They fall into the following three categories:
\begin{itemize}
    \item
    Conditions to have degenerate minima at $T=T_c$: 
    
    In order for the two minima to be stable we impose
    \begin{equation}
    \begin{gathered}
        \det(\mathcal{M}_s)
        -
        \frac{v^2}{\omega^2}(m_h^2|_0)m_h^2>0,
        \\
        m_h^2|_0, \;  m_h^2, \; m_s^2 >0,
    \end{gathered}
    \end{equation}
    with $m_h^2|_0\equiv(\omega/2)[\lambda_m-m_h^2/\omega^2-2m_{sh}^2/(v\omega)]$. 
    These expressions are valid under the assumption that the symmetric minimum sits at $(0,0)$, which we can assume without loss of generality thanks to the shift symmetry present in the potential.

    \item Matching conditions at $T_c$:
    
    Once a viable degenerate potential is found at $T_c$ with the reduced set of parameters $\lbrace\omega,\omega_p,v,m_h^2,m_s^2,\lambda_m\rbrace$, we need to set the parameters $\lambda^2$, $m_{*}$ (given by Eq.~(\ref{eq:lambda_m*})) and $m_{sh}^2$ to particular values so as to have the general potential from Eq.~(\ref{eq:ScalarPot_wv}).
    In this step we ensure that the running of the potential according to temperature change makes the broken minimum the deepest one by imposing
    \begin{equation}
        \frac{d(V_{T,b}-V_{T,s})}{dT^2}\Bigg|_{T_c}=c_h v^2+\omega(c_s \omega+2m_3)>0,
    \end{equation}
    where $V_{T,b(s)}$ correspond to the potential in the broken (symmetric) minimum. Notice that in this step the neutrino Yukawa couplings $\mathcal{Y}_{\nu(N)}$ may play an important role as they enter into the $c_h$ and $c_s$ constants which set the running of the potential with the temperature.

    \item Conditions for potential at $T\leq T_c$:
    
    We require that 
    the potential is bounded from below, 
    and the broken minimum is the global minimum and stable, 
    which are translated into the following conditions on the parameters at $T=0$.
    \begin{equation}
        \begin{gathered}
        \begin{cases}
          \lambda^{2} > 0, 
          &\quad \text{for}\quad 
          \lambda_{m} \leq 0,
          \\
          \lambda^{2} + \frac{1}{4}\lambda_{m}^{2} > 0,
          &\quad \text{for}\quad
          \lambda_{m} > 0,
        \end{cases}
        \\
            \det(\mathcal{M}_{s}^{2}),
            \;
            m_h^2,
            \;
            m_s^2>0,
            \\
            \lambda^2
            -
            \frac{4 m_{*}^{2} v^{2}}
            {\det(\mathcal{M}_{s}^{2})}
            \geq 0.
        \end{gathered}
        \label{eq:stable}
    \end{equation}
    Notice that, as seems to be the case in the SM, the electroweak minimum could be metastable at $T=0$ with a lifetime longer than the age of the Universe. Thus, this condition is more restrictive than strictly necessary. Nevertheless, it is convenient since it allows to efficiently scan the potential without the need of computing the lifetime of the vacuum.
    %which the precise computation of the lifetime of the vacua would forbid.

\end{itemize}
All the conditions listed above take the form ``$C_X >0$'' with $C_X$ a given function of the parameters in the potential. In order to guide our scan towards the areas where these conditions are met we construct the following weight function
\begin{equation}
    \mathcal{W}=\sum_X W_{X},
    \label{eq:chi2}
\end{equation}
with $W_X$ defined as
\begin{equation}
W_X = 
     \begin{cases}
       (10^{6} C_{X})^{2}, 
       &\quad\text{if}\quad C_X\leq0
       \\
       \text{0,} &\quad\text{if}\quad C_X>0. \\
     \end{cases}
\end{equation}
The factor $10^6$ is a penalty to the points that do not satisfy one of the conditions,
with which we can make sure any point accepted in the MCMC satisfies all the necessary conditions for the SFOPT. 
%ven though we set $m_h^2$ at $T_c$ such that the Higgs mass and vev at $T=0$ agree with the experimental values, 
We also add a Gaussian $\chi^2$ term for the Higgs mass as well as for the constraint on $\mathcal{Y}_\nu$ from Eq.~(\ref{eq:bound_mixing}) as Gaussian priors to $\mathcal{W}$
\begin{equation}
    \mathcal{W}
    =
    \sum_X W_{X}
    +\left(\frac{M_H-M_{H}^\text{exp}}
    { \sigma_{M_{H}^\text{exp}}}\right)^2
    +\left(\frac{\frac{n^2v_{EW}^2}{2\omega_{EW}^2\mathcal{Y}_N}\mathcal{Y}^2_\nu-\mathrm{tr}\left(\theta\theta^{\dagger}\right)_\text{exp}}
    { \sigma_{\mathrm{tr}\left(\theta\theta^{\dagger}\right)}}\right)^2,
\end{equation}
where $M_{H}^\text{exp}=125.10$~GeV and $\sigma_{M_{H}^\text{exp}}=0.17$~GeV is its uncertainty~\cite{Workman:2022ynf} and $\mathrm{tr}\left(\theta\theta^{\dagger}\right)_\text{exp}=0.0014$ with $\sigma_{\mathrm{tr}\left(\theta\theta^{\dagger}\right)}=0.0014$ \cite{Fernandez-Martinez:2016lgt,Dani}, and we take $n=3$.

%%%%%%%%%%%%%%%%%%%%%%%%%%%%%%%%%%%%%%%%%%%%%%%%%%
\bibliographystyle{JHEP}
\bibliography{Baryogenesis_bib}

\providecommand{\href}[2]{#2}\begingroup\begin{thebibliography}{100}

\bibitem{Sakharov:1967dj}
A.~D. Sakharov, \emph{{Violation of CP Invariance, C asymmetry, and baryon
  asymmetry of the universe}},
  \href{http://dx.doi.org/10.1070/PU1991v034n05ABEH002497}{\emph{Pisma Zh.
  Eksp. Teor. Fiz.} {\bfseries 5} (1967) 32--35}.

\bibitem{Shaposhnikov:1986jp}
M.~E. Shaposhnikov, \emph{{Possible Appearance of the Baryon Asymmetry of the
  Universe in an Electroweak Theory}}, {\emph{JETP Lett.} {\bfseries 44} (1986)
  465--468}.

\bibitem{Shaposhnikov:1987tw}
M.~E. Shaposhnikov, \emph{{Baryon Asymmetry of the Universe in Standard
  Electroweak Theory}},
  \href{http://dx.doi.org/10.1016/0550-3213(87)90127-1}{\emph{Nucl. Phys.}
  {\bfseries B287} (1987) 757--775}.

\bibitem{Cohen:1987vi}
A.~G. Cohen and D.~B. Kaplan, \emph{{Thermodynamic Generation of the Baryon
  Asymmetry}},
  \href{http://dx.doi.org/10.1016/0370-2693(87)91369-4}{\emph{Phys. Lett.}
  {\bfseries B199} (1987) 251--258}.

\bibitem{Cohen:1990it}
A.~G. Cohen, D.~B. Kaplan and A.~E. Nelson, \emph{{Baryogenesis at the weak
  phase transition}},
  \href{http://dx.doi.org/10.1016/0550-3213(91)90395-E}{\emph{Nucl. Phys.}
  {\bfseries B349} (1991) 727--742}.

\bibitem{Cohen:1990py}
A.~G. Cohen, D.~B. Kaplan and A.~E. Nelson, \emph{{WEAK SCALE BARYOGENESIS}},
  \href{http://dx.doi.org/10.1016/0370-2693(90)90690-8}{\emph{Phys. Lett.}
  {\bfseries B245} (1990) 561--564}.

\bibitem{Nelson:1991ab}
A.~E. Nelson, D.~B. Kaplan and A.~G. Cohen, \emph{{Why there is something
  rather than nothing: Matter from weak interactions}},
  \href{http://dx.doi.org/10.1016/0550-3213(92)90440-M}{\emph{Nucl. Phys.}
  {\bfseries B373} (1992) 453--478}.

\bibitem{Gavela:1993ts}
M.~B. Gavela, P.~Hernandez, J.~Orloff and O.~Pene, \emph{{Standard model CP
  violation and baryon asymmetry}},
  \href{http://dx.doi.org/10.1142/S0217732394000629}{\emph{Mod. Phys. Lett.}
  {\bfseries A9} (1994) 795--810}
  [\href{https://arxiv.org/abs/hep-ph/9312215}{{\ttfamily
  arXiv:hep-ph/9312215}}].

\bibitem{Gavela:1994ds}
M.~B. Gavela, M.~Lozano, J.~Orloff and O.~Pene, \emph{{Standard model CP
  violation and baryon asymmetry. Part 1: Zero temperature}},
  \href{http://dx.doi.org/10.1016/0550-3213(94)00409-9}{\emph{Nucl. Phys.}
  {\bfseries B430} (1994) 345--381}
  [\href{https://arxiv.org/abs/hep-ph/9406288}{{\ttfamily
  arXiv:hep-ph/9406288}}].

\bibitem{Gavela:1994dt}
M.~B. Gavela, P.~Hernandez, J.~Orloff, O.~Pene and C.~Quimbay, \emph{{Standard
  model CP violation and baryon asymmetry. Part 2: Finite temperature}},
  \href{http://dx.doi.org/10.1016/0550-3213(94)00410-2}{\emph{Nucl. Phys.}
  {\bfseries B430} (1994) 382--426}
  [\href{https://arxiv.org/abs/hep-ph/9406289}{{\ttfamily
  arXiv:hep-ph/9406289}}].

\bibitem{Kajantie:1996mn}
K.~Kajantie, M.~Laine, K.~Rummukainen and M.~E. Shaposhnikov, \emph{{Is there a
  hot electroweak phase transition at m(H) larger or equal to m(W)?}},
  \href{http://dx.doi.org/10.1103/PhysRevLett.77.2887}{\emph{Phys. Rev. Lett.}
  {\bfseries 77} (1996) 2887--2890}
  [\href{https://arxiv.org/abs/hep-ph/9605288}{{\ttfamily
  arXiv:hep-ph/9605288}}].

\bibitem{Degrassi:2012ry}
G.~Degrassi, S.~Di~Vita, J.~Elias-Miro, J.~R. Espinosa, G.~F. Giudice,
  G.~Isidori et~al., \emph{{Higgs mass and vacuum stability in the Standard
  Model at NNLO}}, \href{http://dx.doi.org/10.1007/JHEP08(2012)098}{\emph{JHEP}
  {\bfseries 08} (2012) 098} [\href{https://arxiv.org/abs/1205.6497}{{\ttfamily
  arXiv:1205.6497}}].

\bibitem{Espinosa:1993bs}
J.~R. Espinosa and M.~Quiros, \emph{{The Electroweak phase transition with a
  singlet}}, \href{http://dx.doi.org/10.1016/0370-2693(93)91111-Y}{\emph{Phys.
  Lett.} {\bfseries B305} (1993) 98--105}
  [\href{https://arxiv.org/abs/hep-ph/9301285}{{\ttfamily
  arXiv:hep-ph/9301285}}].

\bibitem{Profumo:2007wc}
S.~Profumo, M.~J. Ramsey-Musolf and G.~Shaughnessy, \emph{{Singlet Higgs
  phenomenology and the electroweak phase transition}},
  \href{http://dx.doi.org/10.1088/1126-6708/2007/08/010}{\emph{JHEP} {\bfseries
  08} (2007) 010} [\href{https://arxiv.org/abs/0705.2425}{{\ttfamily
  arXiv:0705.2425}}].

\bibitem{Espinosa:2011ax}
J.~R. Espinosa, T.~Konstandin and F.~Riva, \emph{{Strong Electroweak Phase
  Transitions in the Standard Model with a Singlet}},
  \href{http://dx.doi.org/10.1016/j.nuclphysb.2011.09.010}{\emph{Nucl. Phys.}
  {\bfseries B854} (2012) 592--630}
  [\href{https://arxiv.org/abs/1107.5441}{{\ttfamily arXiv:1107.5441}}].

\bibitem{Profumo:2014opa}
S.~Profumo, M.~J. Ramsey-Musolf, C.~L. Wainwright and P.~Winslow,
  \emph{{Singlet-catalyzed electroweak phase transitions and precision Higgs
  boson studies}},
  \href{http://dx.doi.org/10.1103/PhysRevD.91.035018}{\emph{Phys. Rev.}
  {\bfseries D91} (2015) 035018}
  [\href{https://arxiv.org/abs/1407.5342}{{\ttfamily arXiv:1407.5342}}].

\bibitem{Chen:2017qcz}
C.-Y. Chen, J.~Kozaczuk and I.~M. Lewis, \emph{{Non-resonant Collider
  Signatures of a Singlet-Driven Electroweak Phase Transition}},
  \href{http://dx.doi.org/10.1007/JHEP08(2017)096}{\emph{JHEP} {\bfseries 08}
  (2017) 096} [\href{https://arxiv.org/abs/1704.05844}{{\ttfamily
  arXiv:1704.05844}}].

\bibitem{Hernandez:1996bu}
P.~Hernandez and N.~Rius, \emph{{Neutral heavy leptons and electroweak
  baryogenesis}},
  \href{http://dx.doi.org/10.1016/S0550-3213(97)00193-4}{\emph{Nucl. Phys.}
  {\bfseries B495} (1997) 57--80}
  [\href{https://arxiv.org/abs/hep-ph/9611227}{{\ttfamily
  arXiv:hep-ph/9611227}}].

\bibitem{Fernandez-Martinez:2020szk}
E.~Fern\'andez-Mart\'\i{}nez, J.~L\'opez-Pav\'on, T.~Ota and
  S.~Rosauro-Alcaraz, \emph{{$\nu$ electroweak baryogenesis}},
  \href{http://dx.doi.org/10.1007/JHEP10(2020)063}{\emph{JHEP} {\bfseries 10}
  (2020) 063} [\href{https://arxiv.org/abs/2007.11008}{{\ttfamily
  arXiv:2007.11008}}].

\bibitem{Carena:2019une}
M.~Carena, Z.~Liu and Y.~Wang, \emph{{Electroweak phase transition with
  spontaneous Z$_{2}$-breaking}},
  \href{http://dx.doi.org/10.1007/JHEP08(2020)107}{\emph{JHEP} {\bfseries 08}
  (2020) 107} [\href{https://arxiv.org/abs/1911.10206}{{\ttfamily
  arXiv:1911.10206}}].

\bibitem{Cline:2009sn}
J.~M. Cline, G.~Laporte, H.~Yamashita and S.~Kraml, \emph{{Electroweak Phase
  Transition and LHC Signatures in the Singlet Majoron Model}},
  \href{http://dx.doi.org/10.1088/1126-6708/2009/07/040}{\emph{JHEP} {\bfseries
  07} (2009) 040} [\href{https://arxiv.org/abs/0905.2559}{{\ttfamily
  arXiv:0905.2559}}].

\bibitem{Hernandez:2016kel}
P.~Hernández, M.~Kekic, J.~López-Pavón, J.~Racker and J.~Salvado,
  \emph{{Testable Baryogenesis in Seesaw Models}},
  \href{http://dx.doi.org/10.1007/JHEP08(2016)157}{\emph{JHEP} {\bfseries 08}
  (2016) 157} [\href{https://arxiv.org/abs/1606.06719}{{\ttfamily
  arXiv:1606.06719}}].

\bibitem{Arcadi:2019lka}
G.~Arcadi, A.~Djouadi and M.~Raidal, \emph{{Dark Matter through the Higgs
  portal}}, \href{http://dx.doi.org/10.1016/j.physrep.2019.11.003}{\emph{Phys.
  Rept.} {\bfseries 842} (2020) 1--180}
  [\href{https://arxiv.org/abs/1903.03616}{{\ttfamily arXiv:1903.03616}}].

\bibitem{Falkowski:2009yz}
A.~Falkowski, J.~Juknevich and J.~Shelton, \emph{{Dark Matter Through the
  Neutrino Portal}},  \href{https://arxiv.org/abs/0908.1790}{{\ttfamily
  arXiv:0908.1790}}.

\bibitem{Lindner:2010rr}
M.~Lindner, A.~Merle and V.~Niro, \emph{{Enhancing Dark Matter Annihilation
  into Neutrinos}},
  \href{http://dx.doi.org/10.1103/PhysRevD.82.123529}{\emph{Phys. Rev. D}
  {\bfseries 82} (2010) 123529}
  [\href{https://arxiv.org/abs/1005.3116}{{\ttfamily arXiv:1005.3116}}].

\bibitem{Bertoni:2014mva}
B.~Bertoni, S.~Ipek, D.~McKeen and A.~E. Nelson, \emph{{Constraints and
  consequences of reducing small scale structure via large dark matter-neutrino
  interactions}}, \href{http://dx.doi.org/10.1007/JHEP04(2015)170}{\emph{JHEP}
  {\bfseries 04} (2015) 170} [\href{https://arxiv.org/abs/1412.3113}{{\ttfamily
  arXiv:1412.3113}}].

\bibitem{GonzalezMacias:2015rxl}
V.~Gonzalez~Macias and J.~Wudka, \emph{{Effective theories for Dark Matter
  interactions and the neutrino portal paradigm}},
  \href{http://dx.doi.org/10.1007/JHEP07(2015)161}{\emph{JHEP} {\bfseries 07}
  (2015) 161} [\href{https://arxiv.org/abs/1506.03825}{{\ttfamily
  arXiv:1506.03825}}].

\bibitem{Batell:2017cmf}
B.~Batell, T.~Han, D.~McKeen and B.~Shams Es~Haghi, \emph{{Thermal Dark Matter
  Through the Dirac Neutrino Portal}},
  \href{http://dx.doi.org/10.1103/PhysRevD.97.075016}{\emph{Phys. Rev. D}
  {\bfseries 97} (2018) 075016}
  [\href{https://arxiv.org/abs/1709.07001}{{\ttfamily arXiv:1709.07001}}].

\bibitem{Blennow:2019fhy}
M.~Blennow, E.~Fernandez-Martinez, A.~Olivares-Del~Campo, S.~Pascoli,
  S.~Rosauro-Alcaraz and A.~V. Titov, \emph{{Neutrino Portals to Dark Matter}},
  \href{http://dx.doi.org/10.1140/epjc/s10052-019-7060-5}{\emph{Eur. Phys. J.
  C} {\bfseries 79} (2019) 555}
  [\href{https://arxiv.org/abs/1903.00006}{{\ttfamily arXiv:1903.00006}}].

\bibitem{Baker:2019ndr}
M.~J. Baker, J.~Kopp and A.~J. Long, \emph{{Filtered Dark Matter at a First
  Order Phase Transition}},
  \href{http://dx.doi.org/10.1103/PhysRevLett.125.151102}{\emph{Phys. Rev.
  Lett.} {\bfseries 125} (2020) 151102}
  [\href{https://arxiv.org/abs/1912.02830}{{\ttfamily arXiv:1912.02830}}].

\bibitem{Baker:2021zsf}
M.~J. Baker, M.~Breitbach, J.~Kopp, L.~Mittnacht and Y.~Soreq, \emph{{Filtered
  Baryogenesis}},  \href{https://arxiv.org/abs/2112.08987}{{\ttfamily
  arXiv:2112.08987}}.

\bibitem{Minkowski:1977sc}
P.~Minkowski, \emph{{$\mu \to e\gamma$ at a Rate of One Out of $10^{9}$ Muon
  Decays?}}, \href{http://dx.doi.org/10.1016/0370-2693(77)90435-X}{\emph{Phys.
  Lett.} {\bfseries 67B} (1977) 421--428}.

\bibitem{Mohapatra:1979ia}
R.~N. Mohapatra and G.~Senjanovic, \emph{{Neutrino Mass and Spontaneous Parity
  Nonconservation}},
  \href{http://dx.doi.org/10.1103/PhysRevLett.44.912}{\emph{Phys. Rev. Lett.}
  {\bfseries 44} (1980) 912}.

\bibitem{Yanagida:1979as}
T.~Yanagida, \emph{{Horizontal gauge symmetry and masses of neutrinos}},
  {\emph{Conf. Proc.} {\bfseries C7902131} (1979) 95--99}.

\bibitem{Gell-Mann:1979vob}
M.~Gell-Mann, P.~Ramond and R.~Slansky, \emph{{Complex Spinors and Unified
  Theories}}, {\emph{Conf. Proc. C} {\bfseries 790927} (1979) 315--321}
  [\href{https://arxiv.org/abs/1306.4669}{{\ttfamily arXiv:1306.4669}}].

\bibitem{Branco:1988ex}
G.~C. Branco, W.~Grimus and L.~Lavoura, \emph{{THE SEESAW MECHANISM IN THE
  PRESENCE OF A CONSERVED LEPTON NUMBER}},
  \href{http://dx.doi.org/10.1016/0550-3213(89)90304-0}{\emph{Nucl. Phys.}
  {\bfseries B312} (1989) 492}.

\bibitem{Kersten:2007vk}
J.~Kersten and A.~{\relax Yu}. Smirnov, \emph{{Right-Handed Neutrinos at CERN
  LHC and the Mechanism of Neutrino Mass Generation}},
  \href{http://dx.doi.org/10.1103/PhysRevD.76.073005}{\emph{Phys. Rev.}
  {\bfseries D76} (2007) 073005}
  [\href{https://arxiv.org/abs/0705.3221}{{\ttfamily arXiv:0705.3221}}].

\bibitem{Abada:2007ux}
A.~Abada, C.~Biggio, F.~Bonnet, M.~B. Gavela and T.~Hambye, \emph{{Low energy
  effects of neutrino masses}},
  \href{http://dx.doi.org/10.1088/1126-6708/2007/12/061}{\emph{JHEP} {\bfseries
  12} (2007) 061} [\href{https://arxiv.org/abs/0707.4058}{{\ttfamily
  arXiv:0707.4058}}].

\bibitem{Mohapatra:1986aw}
R.~N. Mohapatra, \emph{{Mechanism for Understanding Small Neutrino Mass in
  Superstring Theories}},
  \href{http://dx.doi.org/10.1103/PhysRevLett.56.561}{\emph{Phys. Rev. Lett.}
  {\bfseries 56} (1986) 561--563}.

\bibitem{Mohapatra:1986bd}
R.~N. Mohapatra and J.~W.~F. Valle, \emph{{Neutrino Mass and Baryon Number
  Nonconservation in Superstring Models}},
  \href{http://dx.doi.org/10.1103/PhysRevD.34.1642}{\emph{Phys. Rev.}
  {\bfseries D34} (1986) 1642}.

\bibitem{Akhmedov:1995ip}
E.~K. Akhmedov, M.~Lindner, E.~Schnapka and J.~W.~F. Valle, \emph{{Left-right
  symmetry breaking in NJL approach}},
  \href{http://dx.doi.org/10.1016/0370-2693(95)01504-3}{\emph{Phys. Lett. B}
  {\bfseries 368} (1996) 270--280}
  [\href{https://arxiv.org/abs/hep-ph/9507275}{{\ttfamily
  arXiv:hep-ph/9507275}}].

\bibitem{Malinsky:2005bi}
M.~Malinsky, J.~C. Romao and J.~W.~F. Valle, \emph{{Novel supersymmetric SO(10)
  seesaw mechanism}},
  \href{http://dx.doi.org/10.1103/PhysRevLett.95.161801}{\emph{Phys. Rev.
  Lett.} {\bfseries 95} (2005) 161801}
  [\href{https://arxiv.org/abs/hep-ph/0506296}{{\ttfamily
  arXiv:hep-ph/0506296}}].

\bibitem{Fernandez-Martinez:2022gsu}
E.~Fern\'andez-Mart\'\i{}nez, X.~Marcano and D.~Naredo-Tuero, \emph{{HNL mass
  degeneracy: implications for low-scale seesaws, LNV at colliders and
  leptogenesis}},  \href{https://arxiv.org/abs/2209.04461}{{\ttfamily
  arXiv:2209.04461}}.

\bibitem{Coleman:1973jx}
S.~R. Coleman and E.~J. Weinberg, \emph{{Radiative Corrections as the Origin of
  Spontaneous Symmetry Breaking}},
  \href{http://dx.doi.org/10.1103/PhysRevD.7.1888}{\emph{Phys. Rev. D}
  {\bfseries 7} (1973) 1888--1910}.

\bibitem{Dolan:1973qd}
L.~Dolan and R.~Jackiw, \emph{{Symmetry Behavior at Finite Temperature}},
  \href{http://dx.doi.org/10.1103/PhysRevD.9.3320}{\emph{Phys. Rev. D}
  {\bfseries 9} (1974) 3320--3341}.

\bibitem{Weinberg:1974hy}
S.~Weinberg, \emph{{Gauge and Global Symmetries at High Temperature}},
  \href{http://dx.doi.org/10.1103/PhysRevD.9.3357}{\emph{Phys. Rev. D}
  {\bfseries 9} (1974) 3357--3378}.

\bibitem{Patel:2011th}
H.~H. Patel and M.~J. Ramsey-Musolf, \emph{{Baryon Washout, Electroweak Phase
  Transition, and Perturbation Theory}},
  \href{http://dx.doi.org/10.1007/JHEP07(2011)029}{\emph{JHEP} {\bfseries 07}
  (2011) 029} [\href{https://arxiv.org/abs/1101.4665}{{\ttfamily
  arXiv:1101.4665}}].

\bibitem{Croon:2020cgk}
D.~Croon, O.~Gould, P.~Schicho, T.~V.~I. Tenkanen and G.~White,
  \emph{{Theoretical uncertainties for cosmological first-order phase
  transitions}}, \href{http://dx.doi.org/10.1007/JHEP04(2021)055}{\emph{JHEP}
  {\bfseries 04} (2021) 055}
  [\href{https://arxiv.org/abs/2009.10080}{{\ttfamily arXiv:2009.10080}}].

\bibitem{Papaefstathiou:2020iag}
A.~Papaefstathiou and G.~White, \emph{{The electro-weak phase transition at
  colliders: confronting theoretical uncertainties and complementary
  channels}}, \href{http://dx.doi.org/10.1007/JHEP05(2021)099}{\emph{JHEP}
  {\bfseries 05} (2021) 099}
  [\href{https://arxiv.org/abs/2010.00597}{{\ttfamily arXiv:2010.00597}}].

\bibitem{Kajantie:1995dw}
K.~Kajantie, M.~Laine, K.~Rummukainen and M.~E. Shaposhnikov, \emph{{Generic
  rules for high temperature dimensional reduction and their application to the
  standard model}},
  \href{http://dx.doi.org/10.1016/0550-3213(95)00549-8}{\emph{Nucl. Phys. B}
  {\bfseries 458} (1996) 90--136}
  [\href{https://arxiv.org/abs/hep-ph/9508379}{{\ttfamily
  arXiv:hep-ph/9508379}}].

\bibitem{Brauner:2016fla}
T.~Brauner, T.~V.~I. Tenkanen, A.~Tranberg, A.~Vuorinen and D.~J. Weir,
  \emph{{Dimensional reduction of the Standard Model coupled to a new singlet
  scalar field}}, \href{http://dx.doi.org/10.1007/JHEP03(2017)007}{\emph{JHEP}
  {\bfseries 03} (2017) 007}
  [\href{https://arxiv.org/abs/1609.06230}{{\ttfamily arXiv:1609.06230}}].

\bibitem{Schicho:2021gca}
P.~M. Schicho, T.~V.~I. Tenkanen and J.~\"Osterman, \emph{{Robust approach to
  thermal resummation: Standard Model meets a singlet}},
  \href{http://dx.doi.org/10.1007/JHEP06(2021)130}{\emph{JHEP} {\bfseries 06}
  (2021) 130} [\href{https://arxiv.org/abs/2102.11145}{{\ttfamily
  arXiv:2102.11145}}].

\bibitem{Niemi:2021qvp}
L.~Niemi, P.~Schicho and T.~V.~I. Tenkanen, \emph{{Singlet-assisted electroweak
  phase transition at two loops}},
  \href{http://dx.doi.org/10.1103/PhysRevD.103.115035}{\emph{Phys. Rev. D}
  {\bfseries 103} (2021) 115035}
  [\href{https://arxiv.org/abs/2103.07467}{{\ttfamily arXiv:2103.07467}}].

\bibitem{Schicho:2022wty}
P.~Schicho, T.~V.~I. Tenkanen and G.~White, \emph{{Combining thermal
  resummation and gauge invariance for electroweak phase transition}},
  \href{https://arxiv.org/abs/2203.04284}{{\ttfamily arXiv:2203.04284}}.

\bibitem{Fuyuto:2014yia}
K.~Fuyuto and E.~Senaha, \emph{{Improved sphaleron decoupling condition and the
  Higgs coupling constants in the real singlet-extended standard model}},
  \href{http://dx.doi.org/10.1103/PhysRevD.90.015015}{\emph{Phys. Rev. D}
  {\bfseries 90} (2014) 015015}
  [\href{https://arxiv.org/abs/1406.0433}{{\ttfamily arXiv:1406.0433}}].

\bibitem{Kotwal:2016tex}
A.~V. Kotwal, M.~J. Ramsey-Musolf, J.~M. No and P.~Winslow,
  \emph{{Singlet-catalyzed electroweak phase transitions in the 100 TeV
  frontier}}, \href{http://dx.doi.org/10.1103/PhysRevD.94.035022}{\emph{Phys.
  Rev. D} {\bfseries 94} (2016) 035022}
  [\href{https://arxiv.org/abs/1605.06123}{{\ttfamily arXiv:1605.06123}}].

\bibitem{Hashino:2016xoj}
K.~Hashino, M.~Kakizaki, S.~Kanemura, P.~Ko and T.~Matsui, \emph{{Gravitational
  waves and Higgs boson couplings for exploring first order phase transition in
  the model with a singlet scalar field}},
  \href{http://dx.doi.org/10.1016/j.physletb.2016.12.052}{\emph{Phys. Lett. B}
  {\bfseries 766} (2017) 49--54}
  [\href{https://arxiv.org/abs/1609.00297}{{\ttfamily arXiv:1609.00297}}].

\bibitem{Kurup:2017dzf}
G.~Kurup and M.~Perelstein, \emph{{Dynamics of Electroweak Phase Transition In
  Singlet-Scalar Extension of the Standard Model}},
  \href{http://dx.doi.org/10.1103/PhysRevD.96.015036}{\emph{Phys. Rev. D}
  {\bfseries 96} (2017) 015036}
  [\href{https://arxiv.org/abs/1704.03381}{{\ttfamily arXiv:1704.03381}}].

\bibitem{Chiang:2018gsn}
C.-W. Chiang, Y.-T. Li and E.~Senaha, \emph{{Revisiting electroweak phase
  transition in the standard model with a real singlet scalar}},
  \href{http://dx.doi.org/10.1016/j.physletb.2018.12.017}{\emph{Phys. Lett. B}
  {\bfseries 789} (2019) 154--159}
  [\href{https://arxiv.org/abs/1808.01098}{{\ttfamily arXiv:1808.01098}}].

\bibitem{Kozaczuk:2019pet}
J.~Kozaczuk, M.~J. Ramsey-Musolf and J.~Shelton, \emph{{Exotic Higgs boson
  decays and the electroweak phase transition}},
  \href{http://dx.doi.org/10.1103/PhysRevD.101.115035}{\emph{Phys. Rev. D}
  {\bfseries 101} (2020) 115035}
  [\href{https://arxiv.org/abs/1911.10210}{{\ttfamily arXiv:1911.10210}}].

\bibitem{Liu:2021jyc}
W.~Liu and K.-P. Xie, \emph{{Probing electroweak phase transition with
  multi-TeV muon colliders and gravitational waves}},
  \href{http://dx.doi.org/10.1007/JHEP04(2021)015}{\emph{JHEP} {\bfseries 04}
  (2021) 015} [\href{https://arxiv.org/abs/2101.10469}{{\ttfamily
  arXiv:2101.10469}}].

\bibitem{Carena:2022yvx}
M.~Carena, J.~Kozaczuk, Z.~Liu, T.~Ou, M.~J. Ramsey-Musolf, J.~Shelton et~al.,
  \emph{{Probing the Electroweak Phase Transition with Exotic Higgs Decays}},
  in \emph{{2022 Snowmass Summer Study}}, 3, 2022
  [\href{https://arxiv.org/abs/2203.08206}{{\ttfamily arXiv:2203.08206}}].

\bibitem{Azatov:2022tii}
A.~Azatov, G.~Barni, S.~Chakraborty, M.~Vanvlasselaer and W.~Yin,
  \emph{{Ultra-relativistic bubbles from the simplest Higgs portal and their
  cosmological consequences}},
  \href{http://dx.doi.org/10.1007/JHEP10(2022)017}{\emph{JHEP} {\bfseries 10}
  (2022) 017} [\href{https://arxiv.org/abs/2207.02230}{{\ttfamily
  arXiv:2207.02230}}].

\bibitem{Workman:2022ynf}
{\scshape Particle Data Group} Collaboration, R.~L. Workman, \emph{{Review of
  Particle Physics}}, {\emph{PTEP} {\bfseries 2022} (2022) 083C01}.

\bibitem{CDF:2022hxs}
{\scshape CDF} Collaboration, T.~Aaltonen et~al., \emph{{High-precision
  measurement of the W boson mass with the CDF II detector}},
  \href{http://dx.doi.org/10.1126/science.abk1781}{\emph{Science} {\bfseries
  376} (2022) 170--176}.

\bibitem{Blennow:2022yfm}
M.~Blennow, P.~Coloma, E.~Fern\'andez-Mart\'\i{}nez and M.~Gonz\'alez-L\'opez,
  \emph{{Right-handed neutrinos and the CDF II anomaly}},
  \href{https://arxiv.org/abs/2204.04559}{{\ttfamily arXiv:2204.04559}}.

\bibitem{Fernandez-Martinez:2016lgt}
E.~Fernandez-Martinez, J.~Hernandez-Garcia and J.~Lopez-Pavon, \emph{{Global
  constraints on heavy neutrino mixing}},
  \href{http://dx.doi.org/10.1007/JHEP08(2016)033}{\emph{JHEP} {\bfseries 08}
  (2016) 033} [\href{https://arxiv.org/abs/1605.08774}{{\ttfamily
  arXiv:1605.08774}}].

\bibitem{Dani}
D.~Naredo-Tuero. Private communication for updated results. Work in progress.

\bibitem{Bolton:2019pcu}
P.~D. Bolton, F.~F. Deppisch and P.~S. Bhupal~Dev, \emph{{Neutrinoless double
  beta decay versus other probes of heavy sterile neutrinos}},
  \href{http://dx.doi.org/10.1007/JHEP03(2020)170}{\emph{JHEP} {\bfseries 03}
  (2020) 170} [\href{https://arxiv.org/abs/1912.03058}{{\ttfamily
  arXiv:1912.03058}}].

\bibitem{MatheusRepository}
M.~Hostert. {\it Heavy Neutrino Limits} GitHub repository,
  \url{https://github.com/mhostert/Heavy-Neutrino-Limits}.

\bibitem{ATLAS:2019nkf}
{\scshape ATLAS} Collaboration, G.~Aad et~al., \emph{{Combined measurements of
  Higgs boson production and decay using up to $80$ fb$^{-1}$ of proton-proton
  collision data at $\sqrt{s}=$ 13 TeV collected with the ATLAS experiment}},
  \href{http://dx.doi.org/10.1103/PhysRevD.101.012002}{\emph{Phys. Rev. D}
  {\bfseries 101} (2020) 012002}
  [\href{https://arxiv.org/abs/1909.02845}{{\ttfamily arXiv:1909.02845}}].

\bibitem{CMS:2020gsy}
{\scshape CMS} Collaboration, \emph{{Combined Higgs boson production and decay
  measurements with up to 137 fb$^{-1}$ of proton-proton collision data at
  $\sqrt s$ = 13 TeV}}, .

\bibitem{Robens:2015gla}
T.~Robens and T.~Stefaniak, \emph{{Status of the Higgs Singlet Extension of the
  Standard Model after LHC Run 1}},
  \href{http://dx.doi.org/10.1140/epjc/s10052-015-3323-y}{\emph{Eur. Phys. J.
  C} {\bfseries 75} (2015) 104}
  [\href{https://arxiv.org/abs/1501.02234}{{\ttfamily arXiv:1501.02234}}].

\bibitem{Buttazzo:2015bka}
D.~Buttazzo, F.~Sala and A.~Tesi, \emph{{Singlet-like Higgs bosons at present
  and future colliders}},
  \href{http://dx.doi.org/10.1007/JHEP11(2015)158}{\emph{JHEP} {\bfseries 11}
  (2015) 158} [\href{https://arxiv.org/abs/1505.05488}{{\ttfamily
  arXiv:1505.05488}}].

\bibitem{Fuchs:2020cmm}
E.~Fuchs, O.~Matsedonskyi, I.~Savoray and M.~Schlaffer, \emph{{Collider
  searches for scalar singlets across lifetimes}},
  \href{http://dx.doi.org/10.1007/JHEP04(2021)019}{\emph{JHEP} {\bfseries 04}
  (2021) 019} [\href{https://arxiv.org/abs/2008.12773}{{\ttfamily
  arXiv:2008.12773}}].

\bibitem{Dawson:2021jcl}
S.~Dawson, P.~P. Giardino and S.~Homiller, \emph{{Uncovering the High Scale
  Higgs Singlet Model}},
  \href{http://dx.doi.org/10.1103/PhysRevD.103.075016}{\emph{Phys. Rev. D}
  {\bfseries 103} (2021) 075016}
  [\href{https://arxiv.org/abs/2102.02823}{{\ttfamily arXiv:2102.02823}}].

\bibitem{ATLAS:2022vkf}
{\scshape ATLAS} Collaboration, \emph{{A detailed map of Higgs boson
  interactions by the ATLAS experiment ten years after the discovery}},
  \href{http://dx.doi.org/10.1038/s41586-022-04893-w}{\emph{Nature} {\bfseries
  607} (2022) 52--59} [\href{https://arxiv.org/abs/2207.00092}{{\ttfamily
  arXiv:2207.00092}}].

\bibitem{CMS:2022dwd}
{\scshape CMS} Collaboration, \emph{{A portrait of the Higgs boson by the CMS
  experiment ten years after the discovery}},
  \href{http://dx.doi.org/10.1038/s41586-022-04892-x}{\emph{Nature} {\bfseries
  607} (2022) 60--68} [\href{https://arxiv.org/abs/2207.00043}{{\ttfamily
  arXiv:2207.00043}}].

\bibitem{Feldman:1997qc}
G.~J. Feldman and R.~D. Cousins, \emph{{A Unified approach to the classical
  statistical analysis of small signals}},
  \href{http://dx.doi.org/10.1103/PhysRevD.57.3873}{\emph{Phys. Rev. D}
  {\bfseries 57} (1998) 3873--3889}
  [\href{https://arxiv.org/abs/physics/9711021}{{\ttfamily
  arXiv:physics/9711021}}].

\bibitem{Das:2017rsu}
A.~Das, Y.~Gao and T.~Kamon, \emph{{Heavy neutrino search via semileptonic
  Higgs decay at the LHC}},
  \href{http://dx.doi.org/10.1140/epjc/s10052-019-6937-7}{\emph{Eur. Phys. J.
  C} {\bfseries 79} (2019) 424}
  [\href{https://arxiv.org/abs/1704.00881}{{\ttfamily arXiv:1704.00881}}].

\bibitem{Das:2017zjc}
A.~Das, P.~S.~B. Dev and C.~S. Kim, \emph{{Constraining Sterile Neutrinos from
  Precision Higgs Data}},
  \href{http://dx.doi.org/10.1103/PhysRevD.95.115013}{\emph{Phys. Rev. D}
  {\bfseries 95} (2017) 115013}
  [\href{https://arxiv.org/abs/1704.00880}{{\ttfamily arXiv:1704.00880}}].

\bibitem{LHC_Higgs_XS_WG}
\emph{{LHC Higgs Working Group}},
  https://twiki.cern.ch/twiki/bin/view/LHCPhysics/LHCHWG.

\bibitem{ATLAS:2020tlo}
{\scshape ATLAS} Collaboration, G.~Aad et~al., \emph{{Search for heavy
  resonances decaying into a pair of Z bosons in the $\ell ^+\ell ^-\ell
  '^+\ell '^-$ and $\ell ^+\ell ^-\nu {{\bar{\nu }}}$ final states using 139
  $\mathrm {fb}^{-1}$ of proton\textendash{}proton collisions at $\sqrt{s} =
  13\,$TeV with the ATLAS detector}},
  \href{http://dx.doi.org/10.1140/epjc/s10052-021-09013-y}{\emph{Eur. Phys. J.
  C} {\bfseries 81} (2021) 332}
  [\href{https://arxiv.org/abs/2009.14791}{{\ttfamily arXiv:2009.14791}}].

\bibitem{Graesser:2007pc}
M.~L. Graesser, \emph{{Experimental Constraints on Higgs Boson Decays to
  TeV-scale Right-Handed Neutrinos}},
  \href{https://arxiv.org/abs/0705.2190}{{\ttfamily arXiv:0705.2190}}.

\bibitem{Caputo:2017pit}
A.~Caputo, P.~Hernandez, J.~Lopez-Pavon and J.~Salvado, \emph{{The seesaw
  portal in testable models of neutrino masses}},
  \href{http://dx.doi.org/10.1007/JHEP06(2017)112}{\emph{JHEP} {\bfseries 06}
  (2017) 112} [\href{https://arxiv.org/abs/1704.08721}{{\ttfamily
  arXiv:1704.08721}}].

\bibitem{Gorbahn:2015gxa}
M.~Gorbahn, J.~M. No and V.~Sanz, \emph{{Benchmarks for Higgs Effective Theory:
  Extended Higgs Sectors}},
  \href{http://dx.doi.org/10.1007/JHEP10(2015)036}{\emph{JHEP} {\bfseries 10}
  (2015) 036} [\href{https://arxiv.org/abs/1502.07352}{{\ttfamily
  arXiv:1502.07352}}].

\bibitem{Beniwal:2018hyi}
A.~Beniwal, M.~Lewicki, M.~White and A.~G. Williams, \emph{{Gravitational waves
  and electroweak baryogenesis in a global study of the extended scalar singlet
  model}}, \href{http://dx.doi.org/10.1007/JHEP02(2019)183}{\emph{JHEP}
  {\bfseries 02} (2019) 183}
  [\href{https://arxiv.org/abs/1810.02380}{{\ttfamily arXiv:1810.02380}}].

\bibitem{Peskin:1991sw}
M.~E. Peskin and T.~Takeuchi, \emph{{Estimation of oblique electroweak
  corrections}}, \href{http://dx.doi.org/10.1103/PhysRevD.46.381}{\emph{Phys.
  Rev. D} {\bfseries 46} (1992) 381--409}.

\bibitem{Haller:2018nnx}
J.~Haller, A.~Hoecker, R.~Kogler, K.~M\"onig, T.~Peiffer and J.~Stelzer,
  \emph{{Update of the global electroweak fit and constraints on
  two-Higgs-doublet models}},
  \href{http://dx.doi.org/10.1140/epjc/s10052-018-6131-3}{\emph{Eur. Phys. J.
  C} {\bfseries 78} (2018) 675}
  [\href{https://arxiv.org/abs/1803.01853}{{\ttfamily arXiv:1803.01853}}].

\bibitem{Loinaz:2002ep}
W.~Loinaz, N.~Okamura, T.~Takeuchi and L.~C.~R. Wijewardhana, \emph{{The NuTeV
  anomaly, neutrino mixing, and a heavy Higgs boson}},
  \href{http://dx.doi.org/10.1103/PhysRevD.67.073012}{\emph{Phys. Rev. D}
  {\bfseries 67} (2003) 073012}
  [\href{https://arxiv.org/abs/hep-ph/0210193}{{\ttfamily
  arXiv:hep-ph/0210193}}].

\bibitem{Loinaz:2004qc}
W.~Loinaz, N.~Okamura, S.~Rayyan, T.~Takeuchi and L.~C.~R. Wijewardhana,
  \emph{{The NuTeV anomaly, lepton universality, and nonuniversal neutrino
  gauge couplings}},
  \href{http://dx.doi.org/10.1103/PhysRevD.70.113004}{\emph{Phys. Rev. D}
  {\bfseries 70} (2004) 113004}
  [\href{https://arxiv.org/abs/hep-ph/0403306}{{\ttfamily
  arXiv:hep-ph/0403306}}].

\bibitem{Akhmedov:2013hec}
E.~Akhmedov, A.~Kartavtsev, M.~Lindner, L.~Michaels and J.~Smirnov,
  \emph{{Improving Electro-Weak Fits with TeV-scale Sterile Neutrinos}},
  \href{http://dx.doi.org/10.1007/JHEP05(2013)081}{\emph{JHEP} {\bfseries 05}
  (2013) 081} [\href{https://arxiv.org/abs/1302.1872}{{\ttfamily
  arXiv:1302.1872}}].

\bibitem{Fernandez-Martinez:2015hxa}
E.~Fernandez-Martinez, J.~Hernandez-Garcia, J.~Lopez-Pavon and M.~Lucente,
  \emph{{Loop level constraints on Seesaw neutrino mixing}},
  \href{http://dx.doi.org/10.1007/JHEP10(2015)130}{\emph{JHEP} {\bfseries 10}
  (2015) 130} [\href{https://arxiv.org/abs/1508.03051}{{\ttfamily
  arXiv:1508.03051}}].

\bibitem{CMS:2013zmy}
{\scshape CMS} Collaboration, S.~Chatrchyan et~al., \emph{{Measurement of Higgs
  Boson Production and Properties in the WW Decay Channel with Leptonic Final
  States}}, \href{http://dx.doi.org/10.1007/JHEP01(2014)096}{\emph{JHEP}
  {\bfseries 01} (2014) 096} [\href{https://arxiv.org/abs/1312.1129}{{\ttfamily
  arXiv:1312.1129}}].

\bibitem{ATLAS:2015pre}
{\scshape ATLAS} Collaboration, G.~Aad et~al., \emph{{Search for an additional,
  heavy Higgs boson in the $H\rightarrow ZZ$ decay channel at $\sqrt{s} =
  8\;\text{ TeV }$ in $pp$ collision data with the ATLAS detector}},
  \href{http://dx.doi.org/10.1140/epjc/s10052-015-3820-z}{\emph{Eur. Phys. J.
  C} {\bfseries 76} (2016) 45}
  [\href{https://arxiv.org/abs/1507.05930}{{\ttfamily arXiv:1507.05930}}].

\bibitem{ALEPH:2001roc}
{\scshape ALEPH} Collaboration, A.~Heister et~al., \emph{{Final results of the
  searches for neutral Higgs bosons in e+ e- collisions at s**(1/2) up to
  209-GeV}}, \href{http://dx.doi.org/10.1016/S0370-2693(01)01487-3}{\emph{Phys.
  Lett. B} {\bfseries 526} (2002) 191--205}
  [\href{https://arxiv.org/abs/hep-ex/0201014}{{\ttfamily
  arXiv:hep-ex/0201014}}].

\bibitem{DELPHI:2003azm}
{\scshape DELPHI} Collaboration, J.~Abdallah et~al., \emph{{Searches for
  invisibly decaying Higgs bosons with the DELPHI detector at LEP}},
  \href{http://dx.doi.org/10.1140/epjc/s2003-01469-8}{\emph{Eur. Phys. J. C}
  {\bfseries 32} (2004) 475--492}
  [\href{https://arxiv.org/abs/hep-ex/0401022}{{\ttfamily
  arXiv:hep-ex/0401022}}].

\bibitem{L3:2004svb}
{\scshape L3} Collaboration, P.~Achard et~al., \emph{{Search for an
  invisibly-decaying Higgs boson at LEP}},
  \href{http://dx.doi.org/10.1016/j.physletb.2005.01.030}{\emph{Phys. Lett. B}
  {\bfseries 609} (2005) 35--48}
  [\href{https://arxiv.org/abs/hep-ex/0501033}{{\ttfamily
  arXiv:hep-ex/0501033}}].

\bibitem{OPAL:2007qwz}
{\scshape OPAL} Collaboration, G.~Abbiendi et~al., \emph{{Search for invisibly
  decaying Higgs bosons in e+ e- ---\ensuremath{>} Z0 h0 production at s**(1/2)
  = 183-GeV - 209-GeV}},
  \href{http://dx.doi.org/10.1016/j.physletb.2009.09.010}{\emph{Phys. Lett. B}
  {\bfseries 682} (2010) 381--390}
  [\href{https://arxiv.org/abs/0707.0373}{{\ttfamily arXiv:0707.0373}}].

\bibitem{ATLAS:2021ifb}
{\scshape ATLAS} Collaboration, G.~Aad et~al., \emph{{Search for Higgs boson
  pair production in the two bottom quarks plus two photons final state in $pp$
  collisions at $\sqrt{s}=13$ TeV with the ATLAS detector}},
  \href{http://dx.doi.org/10.1103/PhysRevD.106.052001}{\emph{Phys. Rev. D}
  {\bfseries 106} (2022) 052001}
  [\href{https://arxiv.org/abs/2112.11876}{{\ttfamily arXiv:2112.11876}}].

\bibitem{ATLAS:2022xzm}
{\scshape ATLAS} Collaboration, \emph{{Search for resonant and non-resonant
  Higgs boson pair production in the $b\bar b\tau^+\tau^-$ decay channel using
  13 TeV $pp$ collision data from the ATLAS detector}},
  \href{https://arxiv.org/abs/2209.10910}{{\ttfamily arXiv:2209.10910}}.

\bibitem{Lee:1977eg}
B.~W. Lee, C.~Quigg and H.~B. Thacker, \emph{{Weak Interactions at Very
  High-Energies: The Role of the Higgs Boson Mass}},
  \href{http://dx.doi.org/10.1103/PhysRevD.16.1519}{\emph{Phys. Rev. D}
  {\bfseries 16} (1977) 1519}.

\bibitem{Mitra:2016kov}
M.~Mitra, R.~Ruiz, D.~J. Scott and M.~Spannowsky, \emph{{Neutrino Jets from
  High-Mass $W_R$ Gauge Bosons in TeV-Scale Left-Right Symmetric Models}},
  \href{http://dx.doi.org/10.1103/PhysRevD.94.095016}{\emph{Phys. Rev. D}
  {\bfseries 94} (2016) 095016}
  [\href{https://arxiv.org/abs/1607.03504}{{\ttfamily arXiv:1607.03504}}].

\bibitem{Mattelaer:2016ynf}
O.~Mattelaer, M.~Mitra and R.~Ruiz, \emph{{Automated Neutrino Jet and Top Jet
  Predictions at Next-to-Leading-Order with Parton Shower Matching in Effective
  Left-Right Symmetric Models}},
  \href{https://arxiv.org/abs/1610.08985}{{\ttfamily arXiv:1610.08985}}.

\bibitem{Degrande:2016aje}
C.~Degrande, O.~Mattelaer, R.~Ruiz and J.~Turner, \emph{{Fully-Automated
  Precision Predictions for Heavy Neutrino Production Mechanisms at Hadron
  Colliders}}, \href{http://dx.doi.org/10.1103/PhysRevD.94.053002}{\emph{Phys.
  Rev. D} {\bfseries 94} (2016) 053002}
  [\href{https://arxiv.org/abs/1602.06957}{{\ttfamily arXiv:1602.06957}}].

\bibitem{CMS:2022irq}
{\scshape CMS} Collaboration, \emph{{Search for ${\mathrm{Z}}^{\prime}$ bosons
  decaying to pairs of heavy Majorana neutrinos in proton-proton collisions at
  $\sqrt{s} = 13 ~\mathrm{TeV}$}}, .

\bibitem{Cai:2017mow}
Y.~Cai, T.~Han, T.~Li and R.~Ruiz, \emph{{Lepton Number Violation: Seesaw
  Models and Their Collider Tests}},
  \href{http://dx.doi.org/10.3389/fphy.2018.00040}{\emph{Front. in Phys.}
  {\bfseries 6} (2018) 40} [\href{https://arxiv.org/abs/1711.02180}{{\ttfamily
  arXiv:1711.02180}}].

\bibitem{Han:2022qgg}
T.~Han, J.~Liao, H.~Liu, D.~Marfatia and R.~Ruiz, \emph{{BSM $\nu$ physics:
  complementarity across energies -- a white paper for Snowmass 2021}},  in
  \emph{{2022 Snowmass Summer Study}}, 3, 2022
  [\href{https://arxiv.org/abs/2203.06131}{{\ttfamily arXiv:2203.06131}}].

\bibitem{CMS:2018ipl}
{\scshape CMS} Collaboration, A.~M. Sirunyan et~al., \emph{{Combination of
  searches for Higgs boson pair production in proton-proton collisions at
  $\sqrt{s} = $ 13 TeV}},
  \href{http://dx.doi.org/10.1103/PhysRevLett.122.121803}{\emph{Phys. Rev.
  Lett.} {\bfseries 122} (2019) 121803}
  [\href{https://arxiv.org/abs/1811.09689}{{\ttfamily arXiv:1811.09689}}].

\bibitem{ATLAS:2019qdc}
{\scshape ATLAS} Collaboration, G.~Aad et~al., \emph{{Combination of searches
  for Higgs boson pairs in $pp$ collisions at $\sqrt{s} = $13 TeV with the
  ATLAS detector}},
  \href{http://dx.doi.org/10.1016/j.physletb.2019.135103}{\emph{Phys. Lett. B}
  {\bfseries 800} (2020) 135103}
  [\href{https://arxiv.org/abs/1906.02025}{{\ttfamily arXiv:1906.02025}}].

\bibitem{ATLAS-CONF-2021-052}
{\scshape ATLAS Collaboration} Collaboration, \emph{{Combination of searches
  for non-resonant and resonant Higgs boson pair production in the
  $b\bar{b}\gamma\gamma$, $b\bar{b}\tau^{+}\tau^{-}$ and $b\bar{b}b\bar{b}$
  decay channels using $pp$ collisions at $\sqrt{s}$ = 13 TeV with the ATLAS
  detector}},  tech. rep., CERN, Geneva, 2021.

\bibitem{Dorsch:2017nza}
G.~C. Dorsch, S.~J. Huber, K.~Mimasu and J.~M. No, \emph{{The Higgs Vacuum
  Uplifted: Revisiting the Electroweak Phase Transition with a Second Higgs
  Doublet}}, \href{http://dx.doi.org/10.1007/JHEP12(2017)086}{\emph{JHEP}
  {\bfseries 12} (2017) 086}
  [\href{https://arxiv.org/abs/1705.09186}{{\ttfamily arXiv:1705.09186}}].

\bibitem{Carena:2018vpt}
M.~Carena, Z.~Liu and M.~Riembau, \emph{{Probing the electroweak phase
  transition via enhanced di-Higgs boson production}},
  \href{http://dx.doi.org/10.1103/PhysRevD.97.095032}{\emph{Phys. Rev. D}
  {\bfseries 97} (2018) 095032}
  [\href{https://arxiv.org/abs/1801.00794}{{\ttfamily arXiv:1801.00794}}].

\bibitem{Arco:2022lai}
F.~Arco, S.~Heinemeyer, M.~M\"uhlleitner and K.~Radchenko, \emph{{Sensitivity
  to Triple Higgs Couplings via Di-Higgs Production in the 2HDM at the
  (HL-)LHC}},  \href{https://arxiv.org/abs/2212.11242}{{\ttfamily
  arXiv:2212.11242}}.

\bibitem{Blennow:2009pk}
M.~Blennow and E.~Fernandez-Martinez, \emph{{Neutrino oscillation parameter
  sampling with MonteCUBES}},
  \href{http://dx.doi.org/10.1016/j.cpc.2009.09.014}{\emph{Comput. Phys.
  Commun.} {\bfseries 181} (2010) 227--231}
  [\href{https://arxiv.org/abs/0903.3985}{{\ttfamily arXiv:0903.3985}}].

\bibitem{Quiros:1999jp}
M.~Quiros, \emph{{Finite temperature field theory and phase transitions}},  in
  \emph{{ICTP Summer School in High-Energy Physics and Cosmology}},
  pp.~187--259, 1, 1999 [\href{https://arxiv.org/abs/hep-ph/9901312}{{\ttfamily
  arXiv:hep-ph/9901312}}].

\bibitem{Biekotter:2021ysx}
T.~Biek\"otter, S.~Heinemeyer, J.~M. No, M.~O. Olea and G.~Weiglein,
  \emph{{Fate of electroweak symmetry in the early Universe: Non-restoration
  and trapped vacua in the N2HDM}},
  \href{http://dx.doi.org/10.1088/1475-7516/2021/06/018}{\emph{JCAP} {\bfseries
  06} (2021) 018} [\href{https://arxiv.org/abs/2103.12707}{{\ttfamily
  arXiv:2103.12707}}].

\bibitem{Baum:2020vfl}
S.~Baum, M.~Carena, N.~R. Shah, C.~E.~M. Wagner and Y.~Wang, \emph{{Nucleation
  is more than critical: A case study of the electroweak phase transition in
  the NMSSM}}, \href{http://dx.doi.org/10.1007/JHEP03(2021)055}{\emph{JHEP}
  {\bfseries 03} (2021) 055}
  [\href{https://arxiv.org/abs/2009.10743}{{\ttfamily arXiv:2009.10743}}].

\bibitem{Coleman:1980aw}
S.~R. Coleman and F.~De~Luccia, \emph{{Gravitational Effects on and of Vacuum
  Decay}}, \href{http://dx.doi.org/10.1103/PhysRevD.21.3305}{\emph{Phys. Rev.}
  {\bfseries D21} (1980) 3305}.

\bibitem{Ivanov:2022osf}
A.~Ivanov, A.~Ivanov, M.~Matteini, M.~Matteini, M.~Nemev\v{s}ek,
  M.~Nemev\v{s}ek et~al., \emph{{Analytic thin wall false vacuum decay rate}},
  \href{http://dx.doi.org/10.1007/JHEP03(2022)209}{\emph{JHEP} {\bfseries 03}
  (2022) 209} [\href{https://arxiv.org/abs/2202.04498}{{\ttfamily
  arXiv:2202.04498}}].

\bibitem{Wainwright:2011kj}
C.~L. Wainwright, \emph{{CosmoTransitions: Computing Cosmological Phase
  Transition Temperatures and Bubble Profiles with Multiple Fields}},
  \href{http://dx.doi.org/10.1016/j.cpc.2012.04.004}{\emph{Comput. Phys.
  Commun.} {\bfseries 183} (2012) 2006--2013}
  [\href{https://arxiv.org/abs/1109.4189}{{\ttfamily arXiv:1109.4189}}].

\bibitem{Athron:2019nbd}
P.~Athron, C.~Bal\'azs, M.~Bardsley, A.~Fowlie, D.~Harries and G.~White,
  \emph{{BubbleProfiler: finding the field profile and action for cosmological
  phase transitions}},
  \href{http://dx.doi.org/10.1016/j.cpc.2019.05.017}{\emph{Comput. Phys.
  Commun.} {\bfseries 244} (2019) 448--468}
  [\href{https://arxiv.org/abs/1901.03714}{{\ttfamily arXiv:1901.03714}}].

\bibitem{Guada:2020xnz}
V.~Guada, M.~Nemev\v{s}ek and M.~Pintar, \emph{{FindBounce: Package for
  multi-field bounce actions}},
  \href{http://dx.doi.org/10.1016/j.cpc.2020.107480}{\emph{Comput. Phys.
  Commun.} {\bfseries 256} (2020) 107480}
  [\href{https://arxiv.org/abs/2002.00881}{{\ttfamily arXiv:2002.00881}}].

\bibitem{Linde:1980tt}
A.~D. Linde, \emph{{Fate of the False Vacuum at Finite Temperature: Theory and
  Applications}},
  \href{http://dx.doi.org/10.1016/0370-2693(81)90281-1}{\emph{Phys. Lett. B}
  {\bfseries 100} (1981) 37--40}.

\bibitem{Anderson:1991zb}
G.~W. Anderson and L.~J. Hall, \emph{{The Electroweak phase transition and
  baryogenesis}}, \href{http://dx.doi.org/10.1103/PhysRevD.45.2685}{\emph{Phys.
  Rev. D} {\bfseries 45} (1992) 2685--2698}.

\bibitem{McLerran:1990zh}
L.~D. McLerran, M.~E. Shaposhnikov, N.~Turok and M.~B. Voloshin, \emph{{Why the
  baryon asymmetry of the universe is approximately 10**-10}},
  \href{http://dx.doi.org/10.1016/0370-2693(91)91794-V}{\emph{Phys. Lett. B}
  {\bfseries 256} (1991) 451--456}.

\bibitem{Dine:1991ck}
M.~Dine, P.~Huet and R.~L. Singleton, Jr., \emph{{Baryogenesis at the
  electroweak scale}},
  \href{http://dx.doi.org/10.1016/0550-3213(92)90113-P}{\emph{Nucl. Phys.}
  {\bfseries B375} (1992) 625--648}.

\bibitem{Coleman:1987rm}
S.~R. Coleman, \emph{{Quantum Tunneling and Negative Eigenvalues}},
  \href{http://dx.doi.org/10.1016/0550-3213(88)90308-2}{\emph{Nucl. Phys. B}
  {\bfseries 298} (1988) 178--186}.

\bibitem{Cline:1999wi}
J.~M. Cline, G.~D. Moore and G.~Servant, \emph{{Was the electroweak phase
  transition preceded by a color broken phase?}},
  \href{http://dx.doi.org/10.1103/PhysRevD.60.105035}{\emph{Phys. Rev. D}
  {\bfseries 60} (1999) 105035}
  [\href{https://arxiv.org/abs/hep-ph/9902220}{{\ttfamily
  arXiv:hep-ph/9902220}}].

\bibitem{Espinosa:2018szu}
J.~R. Espinosa and T.~Konstandin, \emph{{A Fresh Look at the Calculation of
  Tunneling Actions in Multi-Field Potentials}},
  \href{http://dx.doi.org/10.1088/1475-7516/2019/01/051}{\emph{JCAP} {\bfseries
  01} (2019) 051} [\href{https://arxiv.org/abs/1811.09185}{{\ttfamily
  arXiv:1811.09185}}].

\bibitem{Akula:2016gpl}
S.~Akula, C.~Bal\'azs and G.~A. White, \emph{{Semi-analytic techniques for
  calculating bubble wall profiles}},
  \href{http://dx.doi.org/10.1140/epjc/s10052-016-4519-5}{\emph{Eur. Phys. J.
  C} {\bfseries 76} (2016) 681}
  [\href{https://arxiv.org/abs/1608.00008}{{\ttfamily arXiv:1608.00008}}].

\bibitem{Cepeda:2019klc}
M.~Cepeda et~al., \emph{{Report from Working Group 2}: {Higgs Physics at the
  HL-LHC and HE-LHC}},
  \href{http://dx.doi.org/10.23731/CYRM-2019-007.221}{\emph{CERN Yellow Rep.
  Monogr.} {\bfseries 7} (2019) 221--584}
  [\href{https://arxiv.org/abs/1902.00134}{{\ttfamily arXiv:1902.00134}}].

\end{thebibliography}\endgroup

\end{document}